\documentclass[fleqn,usenatbib]{mnras}

\usepackage[T1]{fontenc}
\usepackage{bbding}
\usepackage{ulem}

\DeclareRobustCommand{\VAN}[3]{#2}
\let\VANthebibliography\thebibliography
\def\thebibliography{\DeclareRobustCommand{\VAN}[3]{##3}\VANthebibliography}

\usepackage{graphicx}	
\usepackage{amsmath}	
\usepackage{amssymb}	

\usepackage{rotating}
\usepackage{pdflscape}
\usepackage{tabulary}                                                           

\usepackage{color}

\usepackage{url}

\usepackage{float}

\usepackage{placeins}

\title[A Study of Nine Triply Eclipsing Triples]{A Study of Nine Compact Triply Eclipsing Triples}

\author[Rappaport et al.]{
S.\,A~Rappaport$^1$ \thanks{E-mail: sar@mit.edu},
T. Borkovits$^{2,3,4,5,6}$,
R.~Gagliano$^{7}$,
T.\,L.~Jacobs$^{8}$,
A.~Tokovinin$^9$,
\newauthor
T.\,Mitnyan$^{3}$,
R.\,Kom\v{z}\'{i}k$^{10}$,
V.\,B. Kostov$^{11,12}$,
B.\,P. Powell$^{11}$,
G.\,Torres$^{13}$, 
I.~Terentev$^{14}$,
\newauthor
M.\,Omohundro$^{14}$, 
T.\,Pribulla$^{10}$,
A.\,Vanderburg$^{1}$,
M.\,H.\,Kristiansen$^{15}$,
D.\,Latham$^{13}$,
\newauthor
H.\,M.\,Schwengeler$^{14}$,
D.\,LaCourse$^{16}$,
I. B. B\'\i r\'o$^{2,3}$, 
I. Cs\'anyi$^2$,
D. R. Czavalinga$^3$,
\newauthor
Z. Garai$^{5,6,10}$,
A.\,P\'al$^{4,17}$,
J.\,E.\,Rodriguez$^{18}$,
D.\,J.\,Stevens$^{19}$ \\
$^1$ Department of Physics, Kavli Institute for Astrophysics and Space Research, M.I.T., Cambridge, MA 02139, USA\\
$^{2}$Baja Astronomical Observatory of University of Szeged, H-6500 Baja, Szegedi \'ut, Kt. 766, Hungary\\
$^{3}$ ELKH--SZTE Stellar Astrophysics Research Group, H-6500 Baja, Szegedi \'ut, Kt. 766, Hungary\\
$^{4}$Konkoly Observatory, Research Centre for Astronomy and Earth Sciences,  H-1121 Budapest, Konkoly Thege Mikl\'os \'ut 15-17, Hungary\\
$^5$ ELTE E{\"o}tv{\"o}s Lor\'and University, Gothard Astrophysical Observatory, Szent Imre h. u. 112, 9700 Szombathely, Hungary \\
$^{6}$ MTA-ELTE Exoplanet Research Group, H-9700 Szombathely, Szent Imre h. u. 112, Hungary \\
$^{7}$ Amateur Astronomer, Glendale, AZ 85308 \\
$^{8}$ Amateur Astronomer, 12812 SE 69th Place Bellevue, WA 98006, USA \\
$^{9}$ Cerro Tololo Inter-American Observatory $|$ NSF’s NOIRLab, Casilla 603, La Serena, Chile \\
$^{10}$ Astronomical Institute, Slovak Academy of Sciences, 05960 Tatransk\'a Lomnica, Slovakia \\ 
$^{11}$ NASA Goddard Space Flight Center, 8800 Greenbelt Road, Greenbelt, MD 20771, USA \\
$^{12}$ SETI Institute, 189 Bernardo Avenue, Suite 200, Mountain View, CA 94043, USA\\
$^{13}$ Center for Astrophysics $|$ Harvard \& Smithsonian, 60 Garden St., Cambridge, MA 02138, USA \\
$^{14}$ Citizen Scientist, c/o Zooniverse, Dept,~of Physics, University of Oxford, Denys Wilkinson Building, Keble Road, Oxford, OX1 3RH, UK \\
$^{15}$ Brorfelde Observatory, Observator Gyldenkernes Vej 7, DK-4340 T\o ll\o se, Denmark \\
$^{16}$ Amateur Astronomer, 7507 52nd Place NE Marysville, WA 98270, USA \\
$^{17}$ Kavli Institute for Astrophysics and Space Research, M.I.T., Cambridge, MA 02139, USA \\ 
$^{18}$ Department of Physics and Astronomy, Michigan State University, East Lansing, MI 48824, USA \\
$^{19}$ University of Minnesota Duluth, Duluth, MN 55812, USA \\ }
\date{Accepted XXX. Received YYY; in original form ZZZ}

\pubyear{2023}

\begin{document}
\label{firstpage}
\pagerange{\pageref{firstpage}--\pageref{lastpage}}
\maketitle

\begin{abstract}
 In this work we report the independent discovery and analysis of nine new compact triply eclipsing triple star systems found with the {\it TESS} mission: TICs 47151245, 81525800, 99013269, 229785001, 276162169, 280883908, 294803663, 332521671, and 356324779.  Each of these nine systems exhibits distinct third-body eclipses where the third (`tertiary') star occults the inner eclipsing binary (EB), or vice versa.  We utilize a photodynamical analysis of the {\it TESS} photometry, archival photometric data, {\it TESS} eclipse timing variations of the EBs, available archival spectral energy distribution curves (SED), and, in some cases, newly acquired radial velocity observations, to solve for the parameters of all three stars, as well as most of the orbital elements. From these analyses we find that the outer orbits of all nine systems are viewed nearly edge on (i.e., within $\lesssim 4^\circ$), and 6 of the systems are coplanar to within $5^\circ$; the others have mutual inclination angles of $20^\circ$, $41^\circ$, and possibly $179^\circ$ (i.e., a retrograde outer orbit).  The outer orbital periods range from 47.8 days to 604 days, with eccentricities spanning 0.004 to 0.61.  The masses of all 18 EB stars are in the range of 0.9--2.6\,M$_\odot$ and are mostly situated near the main sequence.  By contrast, the masses and radii of the tertiary stars range from 1.4--2.8\,M$_\odot$ and 1.5--13\,R$_\odot$, respectively.  We make use of the system parameters from these 9 systems, plus those from a comparable number of compact triply eclipsing triples published previously, to gain some statistical insight into their properties.
\end{abstract}

\begin{keywords}
binaries:eclipsing -- binaries:close -- stars:individual: TIC\,47151245, TIC\,81525800, TIC\,99013269, TIC\,229785001, TIC\,276162169, TIC\,280883908, TIC\,294803663, TIC\,332521671, TIC\,356324779
\end{keywords}



\section{Introduction}
\label{sect_intro}

Triply eclipsing triple star systems have a number of characteristics in common with ordinary eclipsing binaries (EBs), but can be far richer astrophysically.  For example, detection of a sequential set of one type of outer eclipse, such as the primary outer eclipses, immediately determines the outer orbital period, $P_{\rm out}$.  Detection of both types of outer eclipses (primary and secondary) leads to a value of $e_{\rm out} \cos \omega_{\rm out}$ for the outer orbit, by analogy with how this quantity is determined in regular eclipsing binaries (EBs) based on the spacing of the two types of eclipses \citep{stern39}.  As is the case for an ordinary EB, an eccentric outer triple's orbit, coupled with an outer inclination angle $i_{\rm out} < 90^\circ$, may allow for only a single eclipse per orbit, typically nearer to periastron.  For simple EBs there is also a relation between the eclipse durations and $e \sin \omega$ \citep{kopal59}; while for triples this is much more difficult to apply, but should also hold approximately\footnote{Use of the word ``approximately'' in this context acknowledges the fact that the exact shape and duration of outer eclipses do not repeat precisely due to the fact that these depend on the phasing of the inner binary at the times of the outer conjunctions.} when averaged over many outer eclipses. In principle, if the radius of the tertiary star, $R_B$, is a substantial fraction of the outer semi-major axis, $a_{\rm out}$, then this may result in ellipsoidal light variations (ELVs) raised on the tertiary star. Additionally, the tertiary star may be brought into corotation with the orbit, by analogy with ordinary binaries, in which case starspots can corotate with the outer orbit producing additional periodic modulations.  For circular orbits in flat triples (i.e., with zero mutual inclination angle, $i_{\rm mut}$, between the inner EB plane and the outer orbital plane), the eclipse depths are still proportional to the surface brightness of the two stars involved in the eclipses.

Several additional features are also possible in triply eclipsing triples that have no counterpart in ordinary EBs.  The very presence of the tertiary star will cause the orbital period of the inner binary to become longer \citep[see, e.g.][]{rappaport13,borkovitsetal15}.  The time-varying nature of this effect  leads to a very useful `dynamical delay' in the eclipse timing of the EB when at least one of the two orbits is eccentric, or if the two orbits are inclined \citep[see, e.g.,][]{borkovitsetal11}.  The presence of the tertiary can cause driven apsidal motion in the inner binary, and will lead to orbital plane precession -- at least to the extent that the two planes are not aligned. If the mutual inclination angle, $i_{\rm mut}$ is large enough, i.e., $\gtrsim 39^\circ$, this can lead to Zeipel-Lidov-Kozai (ZLK) cycles (\citealt{vonzeipel910,lidov962,kozai962}; see also the review by \citet{borkovits22} for a discussion of these and other dynamical effects in multi-stellar systems).

Another important way that triply eclipsing triples are astrophysically richer than simple EBs is that the dynamical interactions among the stars, as well as the geometrical information from the inner and outer eclipses, are often sufficient to extract the system masses, radii, $T_{\rm eff}$'s, and ages with considerable accuracy, as well as the detailed architecture of the orbits \citep[see, e.g.,][]{borkovitsetal19a,borkovitsetal20b,borkovitsetal22,rappaport22}.  

When one of the stars in a binary evolves, it can begin to transfer mass to the other star.  The effects of the mass loss on the orbit, and on the mass-losing, as well as the mass-accreting, stars have been widely studied for many decades, and are relatively well understood \citep{bhattacharya91,eggleton06,tauris23}.  In the case of triple systems, if the tertiary star evolves first it will start to transfer mass to the binary.  This transfer may be stable or unstable, and the latter might lead to something analogous to a common envelope scenario for binary stars.  In the case where one of the EB stars evolves first, any mass ejected from the binary will encounter the tertiary star on its way out of the system.  These processes are more varied and more complex than in the binary case, and have received considerable scrutiny over the past decade, in particular \citep[see, e.g.,][]{tauris14,devries14,sabach15,hillel17,comerford20,munoz20,soker21,glanz21,hamers22}.   

The {\it Kepler} \citep{borucki10}, {\it K2} \citep{howell14}, and {\it TESS} \citep{ricker15} missions, with their long-term, wide field, precision photometry from space, have made it much easier to discover triply eclipsing triple star systems. They are typically identified when an extra, isolated pair of eclipses appears in the lightcurve of an ordinary eclipsing binary, or a long exotically-shaped extra eclipse appears that cannot be produced in a simple binary \citep[see the recent extensive review of][]{borkovits22}.  

In a number of recent papers \citep[e.g.,][]{kristiansen22,rappaport22} we discussed how we have found 52 new triply eclipsing triple systems in the {\it TESS} data.  Of those systems, we were able to find the outer orbital period of 20 of them using archival ground-based photometry.\footnote{Discovering one outer eclipse with {\it TESS}, or even two of them separated by a couple of years, is insufficient to determine the outer period.}  In \citet{rappaport22} we reported on six of these triply eclipsing triples, and provided a full set of stellar and orbital parameters for each.  In this work we report on 9 of the remaining systems.  

In Section \ref{sec:discovery} we discuss how the third body events are discovered in {\it TESS} data, and we present plots of the third body events with fitted models.  We mention in Section \ref{sec:other}  other contemporaneous detections of the triple-nature of a number of the nine systems which are studied in this work.  In Section \ref{sec:SED} we discuss fits to the archival spectral energy distributions (SED) and make first estimates of the stellar properties -- masses, radii, and effective temperatures.  In Section~\ref{sec:outer_orbit} we demonstrate how archival photometric data from a number of ground-based surveys were utilized to determine the outer orbital period of most of the triples via the detection of third-body eclipses. We discuss in Section \ref{sec:etv_rv_gaia} several sources of input to the system analysis other than the {\it TESS} photometry, archival photometry, and SED curves.  The detailed photodynamical model by which we jointly analyze the photometric lightcurves, eclipse timing variations, and spectral energy distributions, is reviewed in Section \ref{sec:photodynamical}.  Tables of system parameters, including extracted masses, radii, and effective temperatures, as well as the orbital parameters for both the inner and outer orbits, are given in Section \ref{sec:results} for each of the nine triple systems. We summarize our results in Section \ref{sec:summary} where we also discuss some of the more important findings from this and some of our recent related studies of triply eclipsing triples.  

\section{Discovery of Triply Eclipsing Triples with \textit{TESS}}
\label{sec:discovery}

\begin{table*}
\hspace{-20px}
\footnotesize
\centering
\caption{Main properties of the nine triple systems from different catalogs}
\tiny
\begin{tabular}{lccccccccc}
\hline
\hline
Parameter			              & 47151245 	    &  81525800 	    & 99013269	            &  229785001$^i$	&  276162169     &   280883908 	& 	294803663  	&   332521671 		& 356324779 		        \\  
\hline
RA J2000 			               & 17:13:43.81	     &  06:09:18.45	    & 20:21:04.69           & 18:54:44.67       &  20:13:25.79    &  04:29:21.01	        &  14:51:09.97               &    12:00:01.3         &  04:31:15.65	\\
Dec J2000 		               & $-33$:04:06.99     &  27:43:28.55	    &  29:30:38.14          & 68:46:50.90	&   32:55:27.06   &  69:43:12.75 	        &  $-63$:35:08.82          &   $-53$:14:15.74    & 57:45:01.11 \\
$T^a$                                       & $10.27 \pm 0.02$ & $13.10 \pm 0.01$ & $8.90 \pm 0.03$   & $11.34 \pm 0.01$  & $11.68 \pm 0.06$  & $10.26 \pm 0.01$     &  $11.22 \pm 0.02$  &   $9.60 \pm 0.01$  & $12.68 \pm 0.01$ \\
G$^b$ 				       & $10.34 \pm 0.00$  & $13.60 \pm 0.01$ & $9.55 \pm 0.00$        & $11.85 \pm 0.01$   & $12.06 \pm 0.00$ & $11.01 \pm 0.00$       &  $12.02 \pm 0.00$   & $10.15 \pm 0.00$ & $13.01 \pm 0.00$ 	\\ 
G$_{\rm BP}$$^b$ 		       & $10.44 \pm 0.01$  & $13.99 \pm 0.01$ & $10.10 \pm 0.00$       & $12.03 \pm 0.00$   & $12.39 \pm 0.00$  & $11.69 \pm 0.00$       & $12.73 \pm 0.00$   & $10.58 \pm 0.01$  & $13.27 \pm 0.00$ 	 \\
G$_{\rm RP}$$^b$ 		       & $10.15 \pm 0.01$  & $13.04 \pm 0.01$ & $8.81 \pm 0.00$         & $11.27 \pm 0.00$  & $11.56 \pm 0.00$  & $10.19 \pm 0.00$       & $11.18 \pm 0.00$    & $9.55 \pm 0.00$  & $12.58 \pm 0.00$ 		\\
B$^a$                                       & $10.50 \pm 0.07$ & $14.56 \pm 0.17$ & $10.87 \pm 0.05$       & $12.63 \pm 0.36$  & $12.91 \pm 0.14$ & $12.69 \pm 0.17$       & $13.59 \pm 0.08$   & $11.19 \pm 0.07$ & $13.65 \pm 0.11$  \\
V$^c$                                       & $10.33 \pm 0.01$ & $13.81 \pm 0.17$ & $9.92 \pm 0.01$          & $11.95 \pm 0.03$  &  $12.24 \pm 0.09$ & $11.57 \pm 0.02$       & $12.62 \pm 0.06$     & $10.37 \pm 0.01$ & $13.09 \pm 0.08$ \\
J$^d$				       & $10.00 \pm 0.02$  & $12.33 \pm 0.02$ & $7.87 \pm 0.03$               & $10.83 \pm 0.02$   & $10.90 \pm 0.02$ & $9.11 \pm 0.04$          & $9.92 \pm 0.02$       &  $8.80 \pm 0.02$ & $12.04 \pm 0.02$ 	 	\\
H$^d$ 				       & $9.96 \pm 0.02$  & $12.05 \pm 0.02$ & $7.38 \pm 0.03$                 & $10.56 \pm 0.02$  & $10.76 \pm 0.02$ & $8.51 \pm 0.03$          & $9.33 \pm 0.02$      & $8.37 \pm 0.04$ &  $11.95 \pm 0.03$ 	 	\\
K$^d$ 			                & 9.88$ \pm 0.02$  & $11.97 \pm 0.02$ & $7.25 \pm 0.02$                & $10.50 \pm 0.02$   & $10.68 \pm 0.02$ & $8.35 \pm 0.02$          & $9.14 \pm 0.02$     & $8.26 \pm 0.03$ & $11.85 \pm 0.03$ \\
W1$^e$ 				        & $9.74 \pm 0.02$  & $11.97 \pm 0.02$ & $7.10 \pm 0.05$                & $10.38 \pm 0.02$  & $10.68 \pm 0.02$ & $8.25 \pm 0.02$           & $9.01 \pm 0.02$     & $8.19 \pm 0.02$ & $11.81 \pm 0.02$ 	 	\\
W2$^e$ 				        & $9.80 \pm 0.02$  & $11.97 \pm 0.02$ & $7.22 \pm 0.02$                & $10.42 \pm 0.02$   & $10.71 \pm 0.02$ & $8.33 \pm 0.02$           & $9.04 \pm 0.02$     & $8.23 \pm 0.02$ & $11.84 \pm 0.02$ 	 	\\
W3$^e$ 				        & $9.72 \pm 0.08$  & $11.54 \pm NA $    & $7.18 \pm 0.02$                & $10.35 \pm 0.04$   & $10.90 \pm 0.18$ & $8.27 \pm 0.02$           & $8.95 \pm 0.03 $     & $8.16 \pm 0.02$ & $11.68 \pm 0.24$ 	 	\\
W4$^e$				        & $7.96 \pm 0.31$  & ... 	                     & $7.17 \pm 0.08$                 & ...                          & ... & $8.43 \pm 0.31$           & ...   & $8.10 \pm 0.19$ & ... 		\\
$T_{\rm eff}$ [K]$^b$ 	        & $8192 \pm 275$  & $5374 \pm 250$  & $4806 \pm 110$                & $6277 \pm 1300$  & $5750 \pm 180$ &  $4402 \pm 333$         & $4362 \pm 125$            &  $5201 \pm 110$  &  $6362 \pm 780$		\\  
$T_{\rm eff}$ [K]$^a$ 	        & ...                        &  $6340 \pm 124$   & $6249 \pm 122$                & $6016 \pm NA$     & $9660 \pm 900$ &  	...	                          &  $5734 \pm 122$          &   $5455 \pm 123$	&  $8379 \pm 124$			\\  
Radius [$R_\odot$]$^b$ 	        & ...	                      & $4.23 \pm 0.35$    & $17.8 \pm 0.9$                 &           ...  	               &  $5.19 \pm 0.36$   & ... 	                            &  $14.3 \pm 0.8$            &  $7.82 \pm 0.67$  & $8.41 \pm 2.0$	\\
Radius [$R_\odot$]$^a$ 	        & ...                         & $3.75 \pm NA$       &   $16.5 \pm NA $              & $1.23 \pm NA$    & $4.41 \pm 0.20$   & ... 	                            &  $13.5 \pm NA $           &  $7.67 \pm NA$      & $7.36 \pm NA$  	 	\\
Distance [pc]$^f$			& $1023 \pm 20$  & $1979 \pm 81$       &   $805 \pm 31$               & $780 \pm 200$ 	&  $1552 \pm 25$   & $3072^{+2200}_{-900}$ & $1770 \pm 55$   &  $748 \pm 9.5$  & $3245 \pm 150$ 	 \\  
$E(B-V)$$^a$                            & $0.46 \pm 0.02$ & $0.24 \pm 0.02$    & $0.45 \pm 0.04$                  & ...                         & $0.63 \pm 0.08$ & ... & $0.56 \pm 0.02$    & $0.091 \pm 0.001$ & $0.41 \pm 0.02$ \\ 
$\mu_\alpha$ [mas/yr]$^b$	& $-0.35 \pm 0.03$ & $-0.55 \pm 0.02$  & $-0.37 \pm 0.04$                 & $-5.33 \pm 0.38$  & $-0.44 \pm 0.01$  & $+4.31 \pm 0.11$ 	     & $-6.05 \pm 0.01$	&  $-6.66 \pm 0.02$  & $0.97 \pm 0.02$ 	 \\   
$\mu_\delta$ [mas/yr]$^b$  	& $-4.26 \pm 0.02$ & $-3.29 \pm 0.02$  & $-3.58 \pm 0.05$                 &  $-14.07 \pm 0.49$ & $-0.98 \pm 0.01$ & $-2.62 \pm 0.13$ 	     & $-3.61 \pm 0.02$ & $-0.33 \pm 0.02$  & $-1.23 \pm 0.01$ 	  \\  
RUWE$^{b,g}$                          &  0.749 & 0.957  & 4.20 & 29.5 & 0.916 & 9.83 & 1.43 & 1.13 & 0.948 \\
$P_{\rm binary}$$^h$ [d] & 1.202 & 1.649 & 6.534 & 0.929 & 2.550 & 5.248 & 2.246 &1.247 & 3.477 \\
$P_{\rm triple}$$^h$ [d]  & 284.90 & 47.85 & 604.05 & 165.25 & 117.10 & 184.35 & 153.14 &48.50 & 86.65 \\
\hline
\label{tbl:mags9}
\end{tabular}  

\small
\textit{Notes.}  General: ``NA" in this table indicates that the value is not available. (a) {\it TESS} Input Catalog (TIC v8.2) \citep{TIC8}. (b) Gaia EDR3 \citep{GaiaEDR3}; the uncertainty in $T_{\rm eff}$ and $R$ listed here is 1.5 times the geometric mean of the upper and lower error bars cited in DR2. Magnitude uncertainties listed as 0.00 are $\lesssim 0.005$. (c) AAVSO Photometric All Sky Survey (APASS) DR9, \citep{APASS}, \url{http://vizier.u-strasbg.fr/viz-bin/VizieR?-source=II/336/apass9}. (d) 2MASS catalog \citep{2MASS}.  (e) WISE point source catalog \citep{WISE}. (f) \citet{bailer-jonesetal21}. (g) The Gaia Renormalized Unit Weight Error (RUWE) is the square root of the normalized $\chi^2$ of the astrometric fit to the along-scan observations. Values in excess of about unity are sometimes taken to be a sign of stellar multiplicity. (h) Binary and outer orbital periods from this work; given here for reference purposes.  (i) 2+1+1 quadruple.  \\  

\end{table*}

\begin{table}
\centering
\caption{{\it TESS} Observation Sectors for the Triples$^a$}
\small
\begin{tabular}{lcc}
\hline
\hline
Object & Sectors Observed & Third Body Events  \\
\hline
TIC 47151245 & S12 \& S39  & S39  \\
TIC 81525800 & S43-S45 & S43 \& S44 \\ 
TIC 99013269 & S14-S15, S41, S55 & S41  \\
TIC 229785001 & 23 sectors in S14-S55  & S16, S22, S40 \& S52 \\
TIC 276162169 & S14-15, S41, S54-S55 & S14, S55 \\
TIC 280883908 & S25-S26 \& S52-S53 & S25-S26 \& S53  \\
TIC 294803663 & S11-S12 \& S38-S39 & S12 \\
TIC 332521671 & S10 \& S37 & S37  \\
TIC 356324779 & S19 & S19 \\
\hline
\label{tbl:sectors}  
\end{tabular}

\textit{Notes.}  (a) At least six of these sources will also be observed in later {\it TESS} sectors.

\end{table}

Our `Visual Survey Group' \citep[VSG;][]{kristiansen22} has continued its search for multi-stellar systems in the \textit{TESS} photometric lightcurves.  Thus far, we estimate that we have visually surveyed about 9 million lightcurves from {\it TESS} full-frame images (FFI) of anonymous stars down to {\it TESS} magnitude $T \lesssim 15$.  In addition, we have also visually inspected somewhat more than 1 million lightcurves of preselected eclipsing binaries.  The latter were found in the {\it TESS} data via machine learning (ML) searches \citep[see][]{powell21} of some 120 million lightcurves in Sectors 1-48 {\it TESS} FFI lightcurves.  We have found that the visual and ML searches \citep[see, e.g.,][]{powell21,kostovetal21,kostov22} are an excellent complement to each other.  ML is certainly faster, by orders of magnitude, but more limited than the human brain in terms of what unexpected things it can find.

For the bulk of our visual searches, the lightcurves are displayed with Allan Schmitt's {\tt LcTools} and {\tt LcViewer} software \citep{schmitt19}, which allows for an inspection of a typical lightcurve in just few seconds.  The 9 million lightcurves of `anonymous' stars came from the following sources: Science Processing Operations Center (SPOC, \citealt{jenkins16}); the Difference Imaging Pipeline \citep{oelkers18}; the PSF-based Approach to TESS High quality data Of Stellar clusters (PATHOS, \citealt{nardiello19}); the Cluster Difference Imaging Photometric Survey (CDIPS, \citealt{bouma19}); the MIT Quick Look Pipeline (QLP, \citealt{huang20}); the TESS Image CAlibrator Full Frame Images (TICA, \citealt{fausnaugh20}); and the Goddard Space Flight Center (GSFC, see Sect. 2, \citealt{powell21}). 

These visual plus ML searches have led to numerous discoveries which are summarized in \citet{kristiansen22}.  Among other things, this has led to a catalog of 97 quadruple star systems \citep{kostov22}.  We note here, in passing, that about half of the quadruple systems of our catalog were found by visually surveying anonymous FFI lightcurves, while the other half were the result of visual searches through the set of pre-selected EBs (which were themselves found by a ML algorithm).  

In our searches for triply eclipsing triples we look for an eclipsing binary lightcurve with an additional extra eclipse that is typically strangely shaped and of longer duration than the EB eclipses, or a rapid succession of isolated eclipses.  Once such `extra' eclipses, or `third body' events are found, there is usually very little doubt that they are due to a tertiary star eclipsing the EB or vice versa.  

From searches through lightcurves obtained from the first three full years of {\it TESS} observations, we have found more than 52 of these triply eclipsing triples. Twenty of these 52 systems were found in the set of 9 million `anonymous' FFI lightcurves, while the remaining 32 were found among the sample of only 1 million preselected EB lightcurves.  This statistic is discussed in \citet{rappaport22}. 

Of these 52 triply eclipsing triples, we have determined the outer orbital period for 20 of them via a combination of {\it TESS} and archival ground-based photometry. We have previously reported four of these 20 systems in \citet{borkovitsetal20b} and \citet{borkovitsetal22}, and six more in \citet{rappaport22}.  One additional triple among the 20 was found to have an unreliable outer period.  Here we present the discovery and analysis of the remaining 9 new triply eclipsing triples from among this set.  The TIC numbers and basic cataloged data (e.g., coordinates, magnitudes, distance, etc.) are given in Table~\ref{tbl:mags9}.  The sectors in which they were observed and in which third-body events were found are given in Table \ref{tbl:sectors}.  

\begin{figure*}
\begin{center}
\includegraphics[width=0.332 \textwidth]{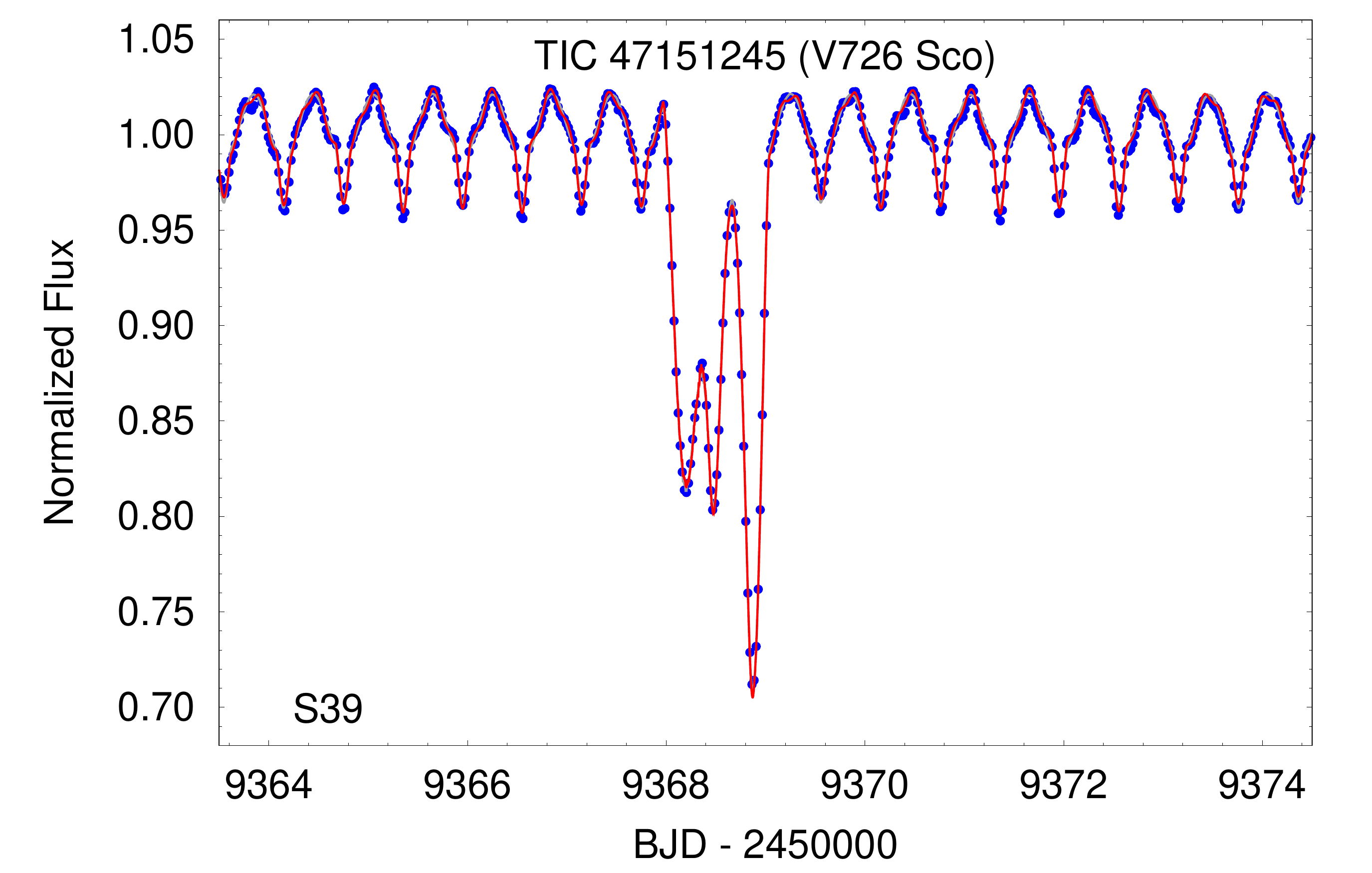} 
\includegraphics[width=0.313 \textwidth]{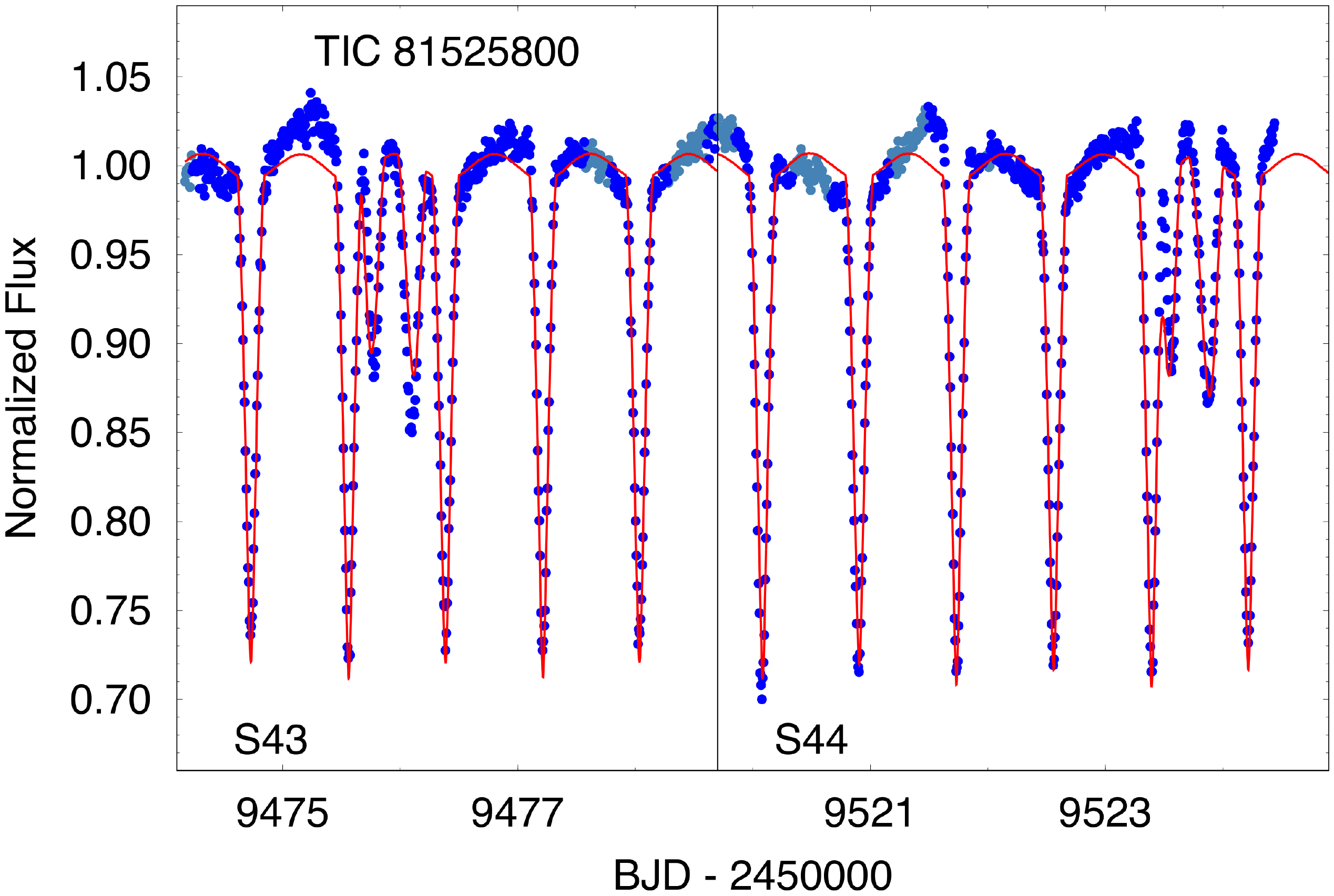}  
\includegraphics[width=0.332 \textwidth]{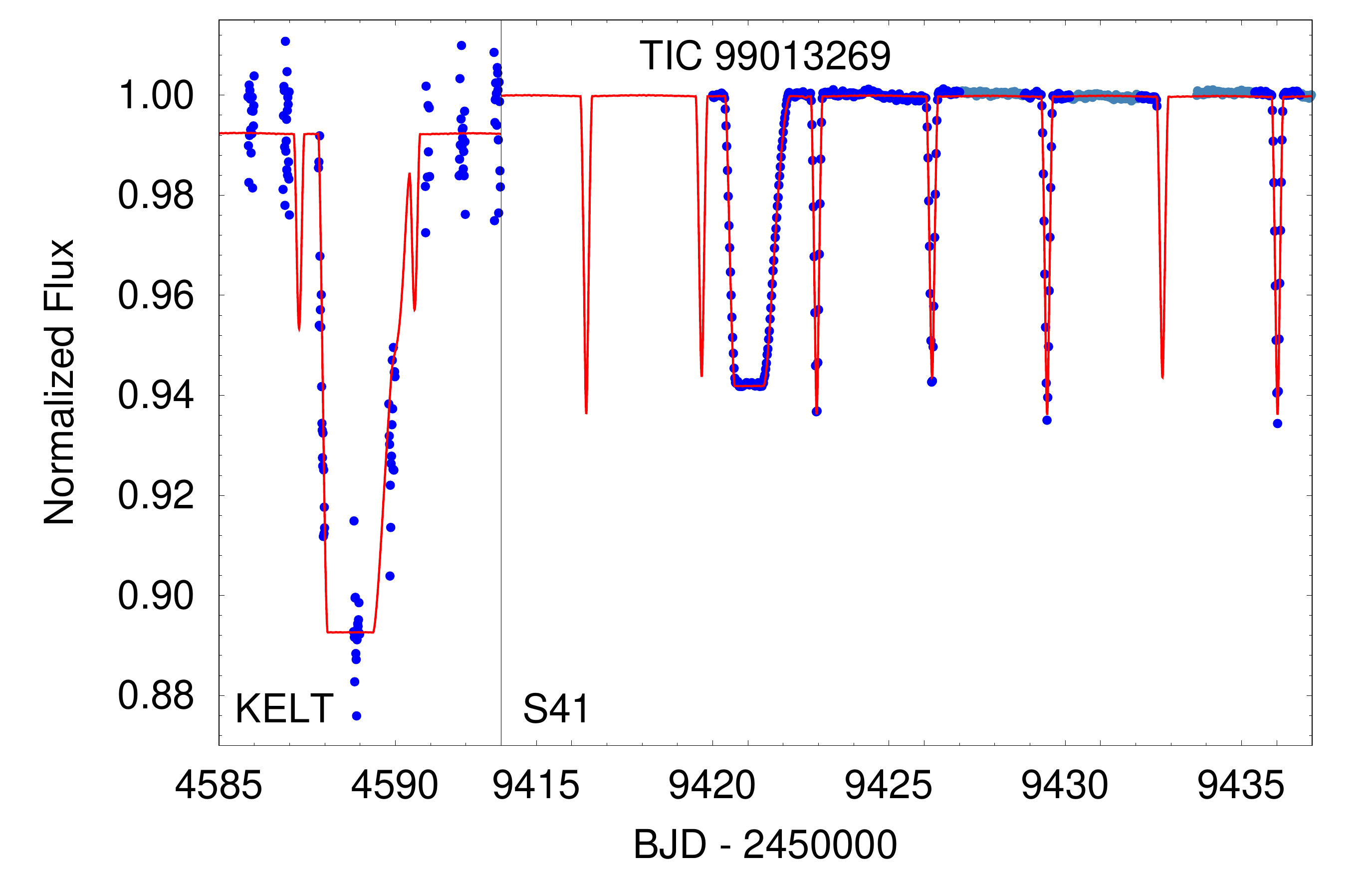}

\includegraphics[width=0.327 \textwidth]{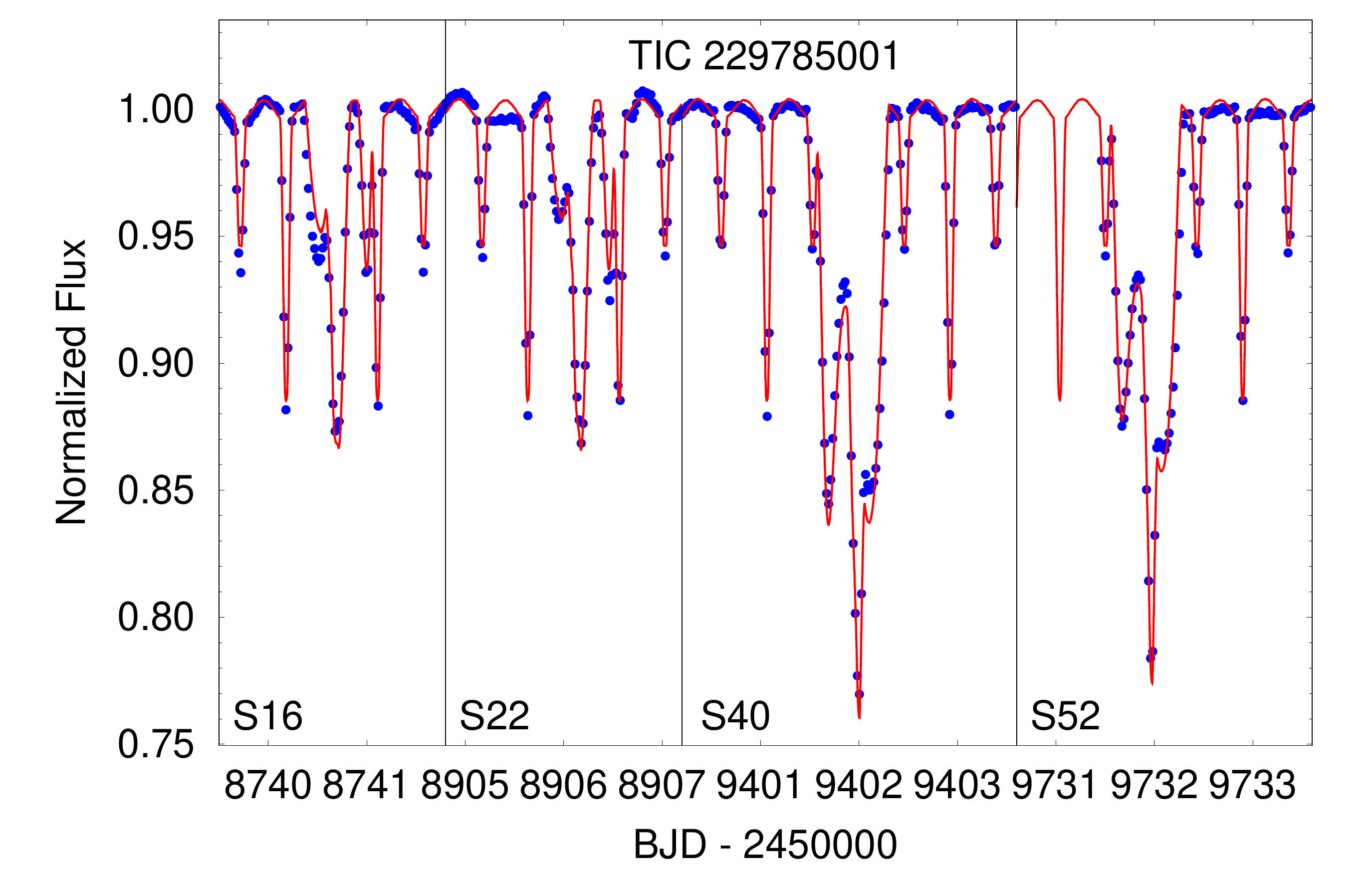}  \hglue-0.09cm
\includegraphics[width=0.337 \textwidth]{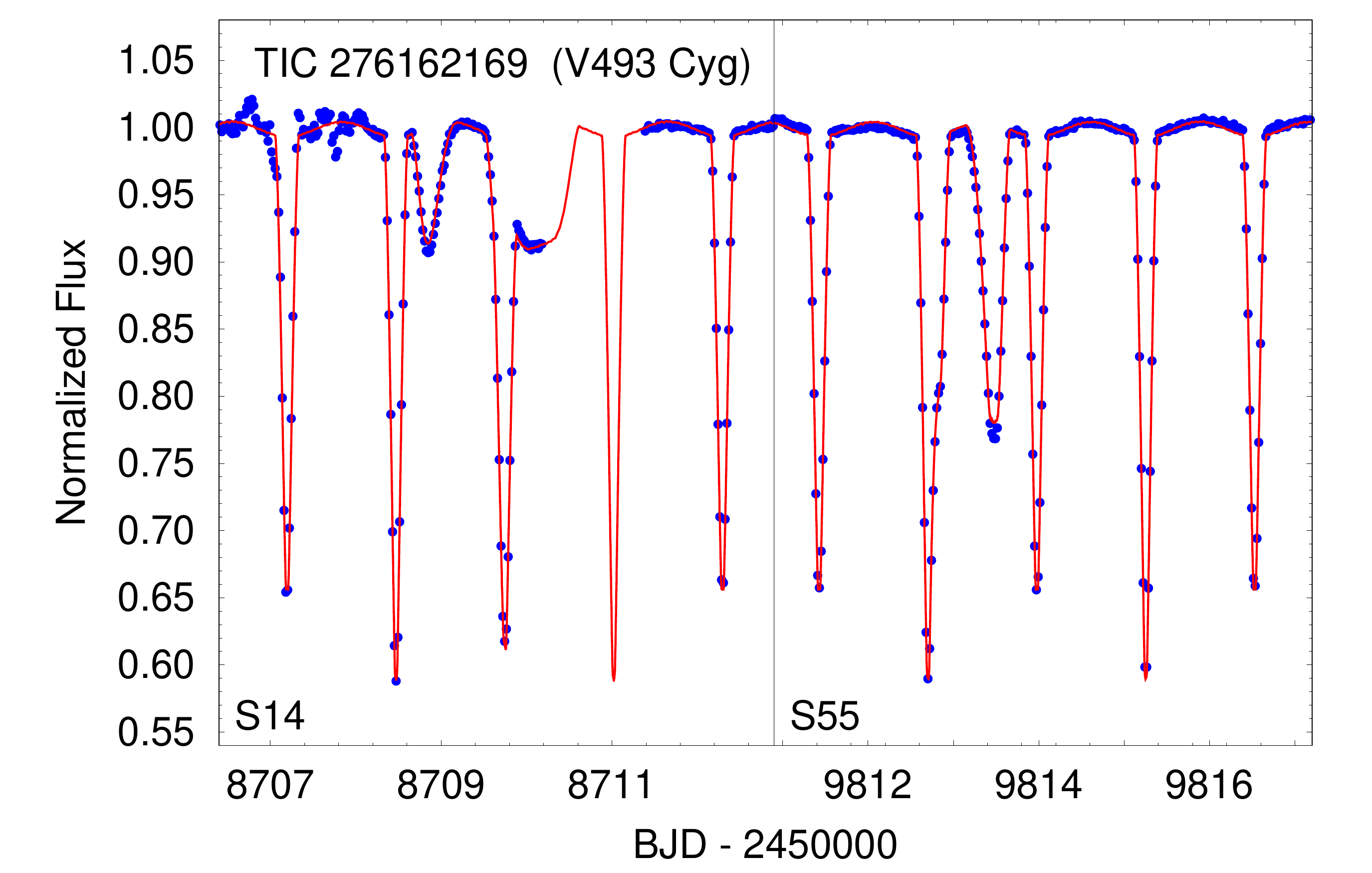}  \hglue-0.09cm
 \includegraphics[width=0.332 \textwidth]{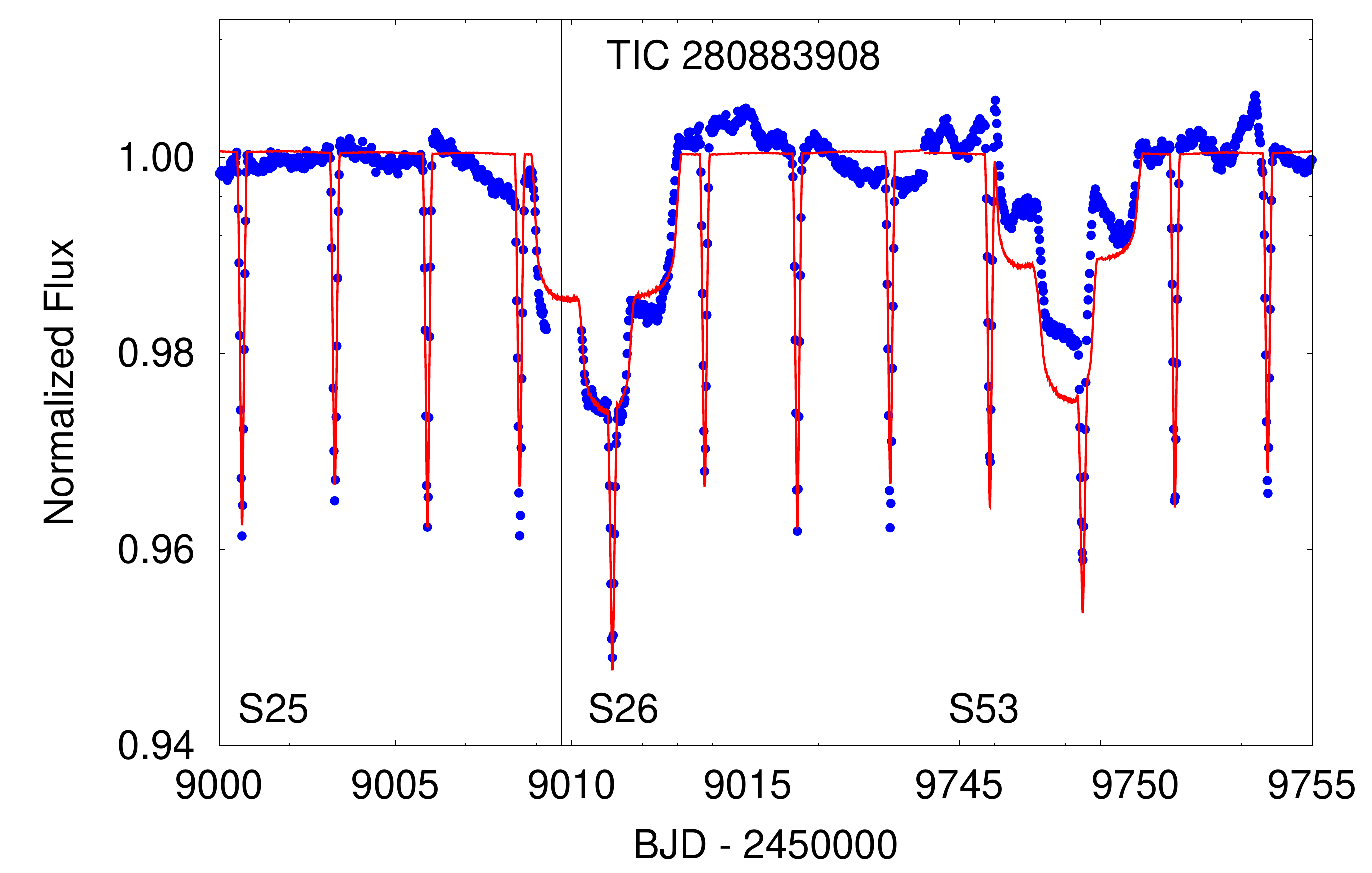}   
 
\includegraphics[width=0.329 \textwidth]{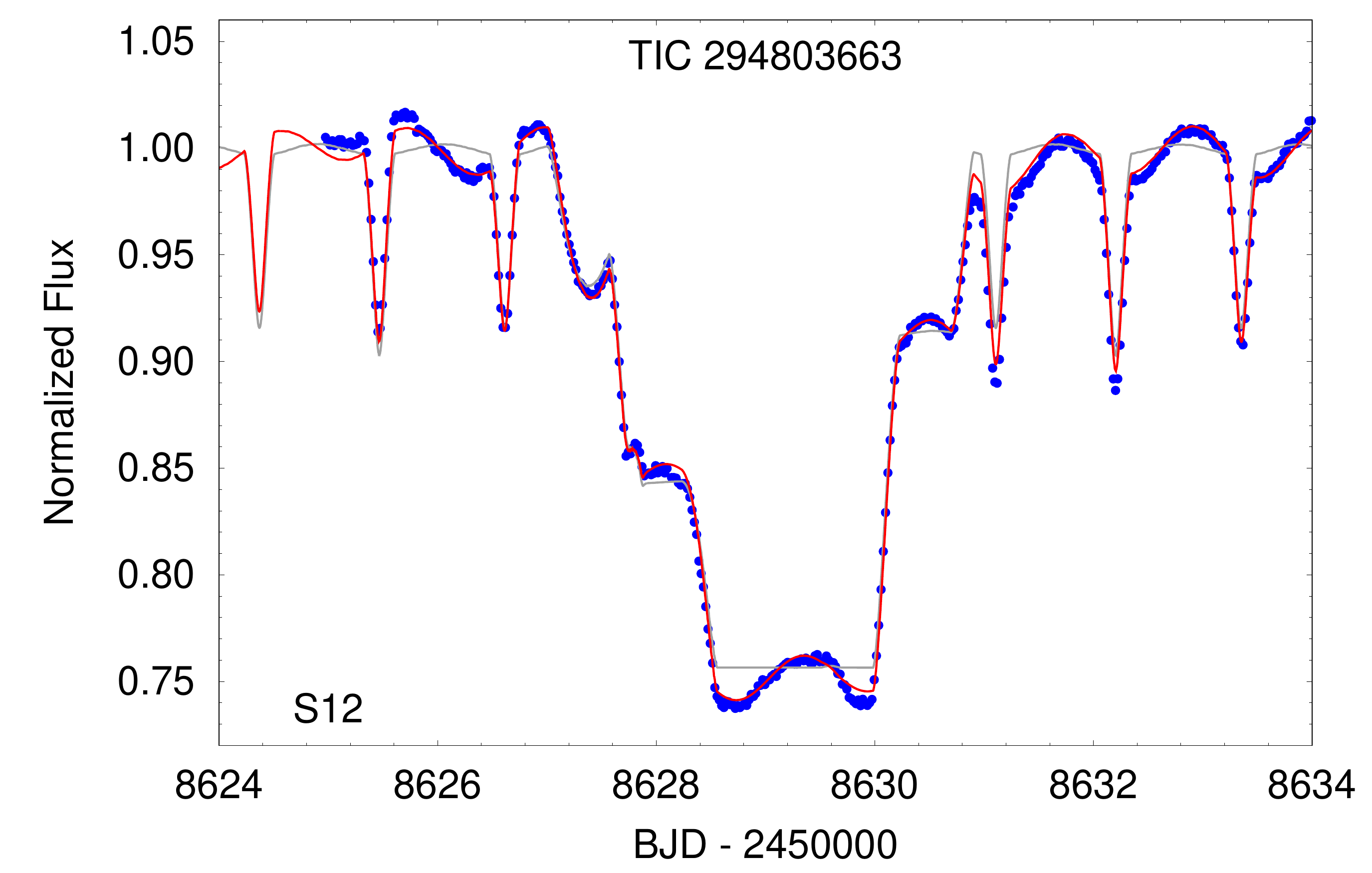} \hglue-0.05cm
\includegraphics[width=0.333 \textwidth]{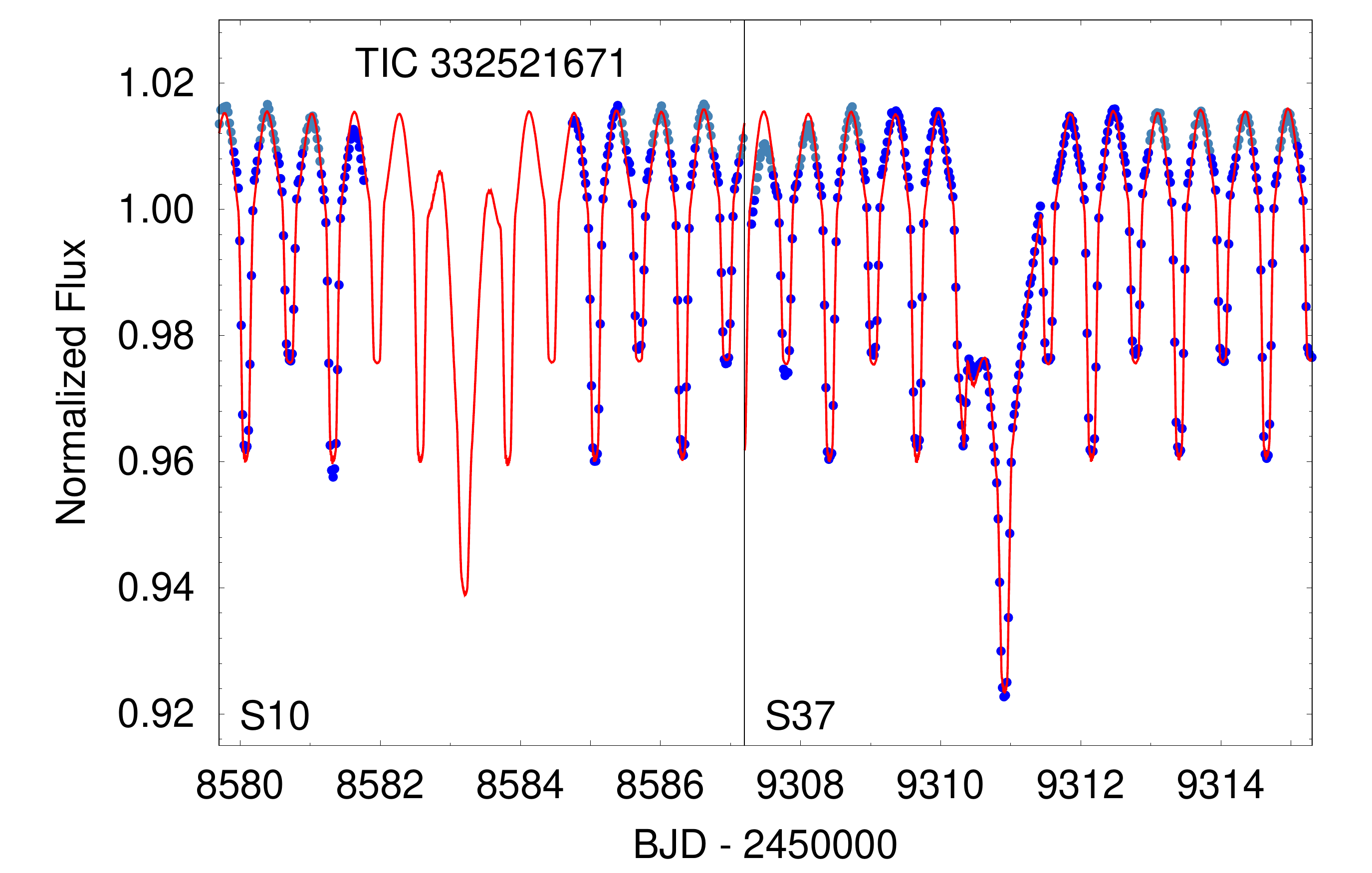} \hglue0.09cm
\includegraphics[width=0.325 \textwidth]{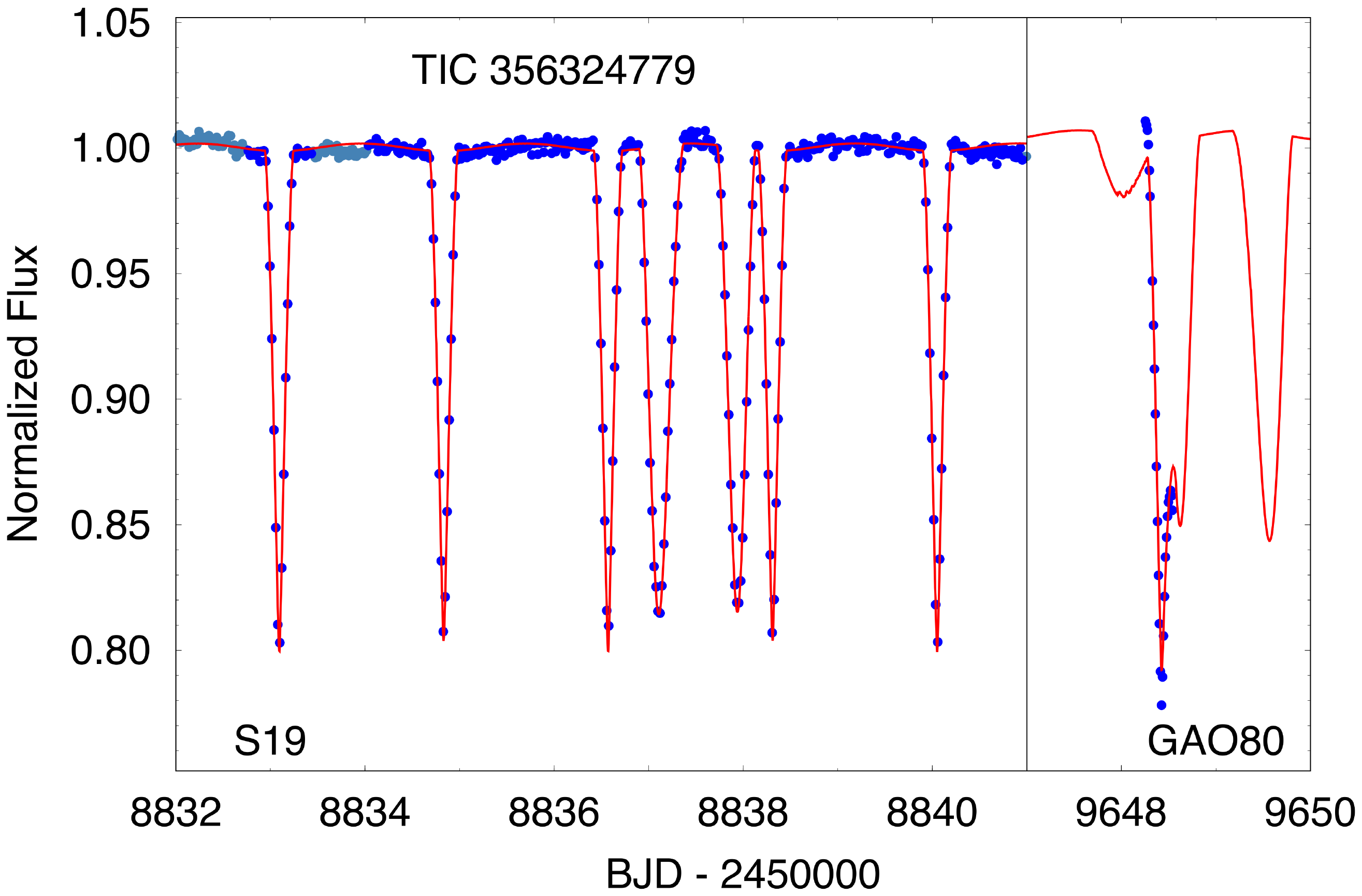}

\caption{{\it TESS} third body lightcurves.  We present a portion of a sector's lightcurve for each source containing the third body event that led to their discoveries.  For most of the sources there is only a single third body event that was detected.  The overplotted model lightcurves are discussed in Sect.~\ref{sec:photodynamical}. The lighter blue points in the out-of-eclipse region were omitted from the photodynamical fits to save computation time.  Note, in the case of TICs\,47151245 and 294803663 the light gray lines represent the pure triple-star model, without the contributions of the stellar oscillations.}
\label{fig:triples}
\end{center}
\end{figure*}

Seven of the nine systems reported in this work were observed only in FFIs from {\it TESS} with either 30-min or 10-min cadence.  The two remaining triples, TICs~99013269 and 276162169 in some particular sectors were also observed in 2-min cadence mode. Segments from the {\it TESS} lightcurves exhibiting third-body eclipses are shown in Fig.~\ref{fig:triples} for all nine sources.  The sources were observed for a minimum of a single sector for one of the triples and a maximum of 23 sectors for another target (which is near the ecliptic pole), with an average of a few sectors per source.  For 5 of the 9 triples only a single third-body eclipse was detected, while at the other extreme 3 and 4 outer eclipses were seen for two other triples.  In most cases, even when more than one outer eclipse were detected, they were in {\it TESS} sectors that were too far apart in time to make a unique determination of the outer period.   

Given that the sectors are only approximately a month long, and the outer orbital periods range from 48 to 604 days, it is not surprising that we observed just 16 third body events out of a total of 48 observing sectors.  Based on the third-body events detected with {\it TESS}, (i) we cannot generally determine the outer orbital period from the {\it TESS} data alone, and (ii) the photodynamical fits to these systems will typically not include both types of outer eclipses (i.e., primary and secondary).  Nonetheless, with the use of ground-based archival data, we were able to determine the outer orbital period for 7 of the 9 systems, as well as the parameter $e_{\rm out} \cos \omega_{\rm out}$ (from the relative separations of the outer eclipses) for 5 of the 9 systems.

\begin{table*}
\centering
\caption{Other Detections of the Triples$^a$}
\small
\begin{tabular}{lccc}
\hline
\hline
Object & This Work$^b$ & Gaia DR3 Orbits$^{b,c}$ & Zasche et al.$^{b,d}$ \\
        & ($P_{\rm out}$, $e_{\rm out}$, $\omega_{\rm out}$)  &   ($P_{\rm out}$, $e_{\rm out}$, $\omega_{\rm out}$) &    \\
\hline
TIC 47151245 & 284.90 d; 0.480; 219$^\circ$  & ... & 285 d\\
TIC 81525800 & 47.85 d; 0.614; 201$^\circ$ & ... & ...\\ 
TIC 99013269 & 604.05 d; 0.463; 271$^\circ$ & Astr.: 609.5 d; 0.458; ... & ... \\
TIC 229785001 & 165.25 d; 0.458; 31$^\circ$ &  Spect.: 166.0 d; 0.493; 16$^\circ$ & 165.3 d \\
TIC 276162169 & 117.10 d; 0.268; 243$^\circ$ & ... & ... \\
TIC 280883908 & 184.35 d; 0.260; 146$^\circ$ & Spect.: 184.1 d; 0.258; 142$^\circ$ & ... \\
TIC 294803663 & 153.14 d; 0.030; 313$^\circ$ & Spect.; 153.1 d; 0.047; 252$^\circ$ & 153.1 d \\
TIC 332521671 & 48.50 d; 0.004; 311$^\circ$ & ... & ... \\
TIC 356324779 & 86.65 d; 0.284; 144$^\circ$ & ... & 86.7 d \\
\hline
\label{tbl:noticed}  
\end{tabular}

\textit{Notes.}  (a) Six of the nine triply eclipsing systems were spotted in prior broad surveys, but no quantitative analysis of the system parameters (especially of the constituent stars) was undertaken.  Four of them have outer orbits reported by Gaia, but no third body eclipses were reported by Gaia.  Four of the systems had third body eclipses detected by \citet{zasche22}, and the outer period could be deduced from those extra eclipses. (b) Where available, we show for each source, the period, eccentricity, and argument of periastron of the outer orbit, separated by semicolons. (c) The Gaia orbital solutions are marked as either astrometric or spectroscopic; \citet{babusiaux22}; \citet{gaia22} (d) \citet{zasche22}.  
\end{table*}

\begin{figure}
\begin{center}
\includegraphics[width=0.99 \columnwidth]{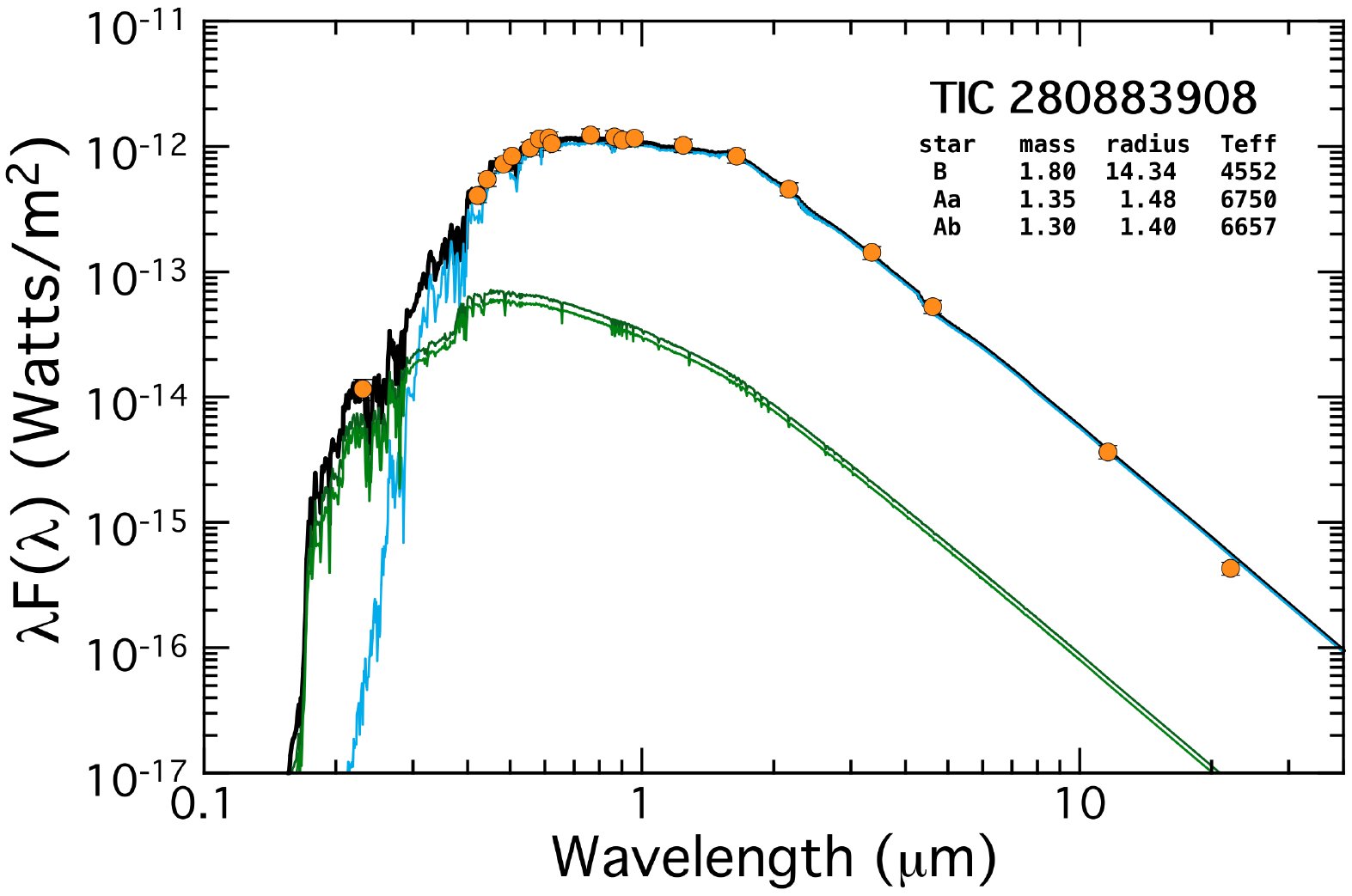}  \hglue0.03cm 
\includegraphics[width= 0.993 \columnwidth]{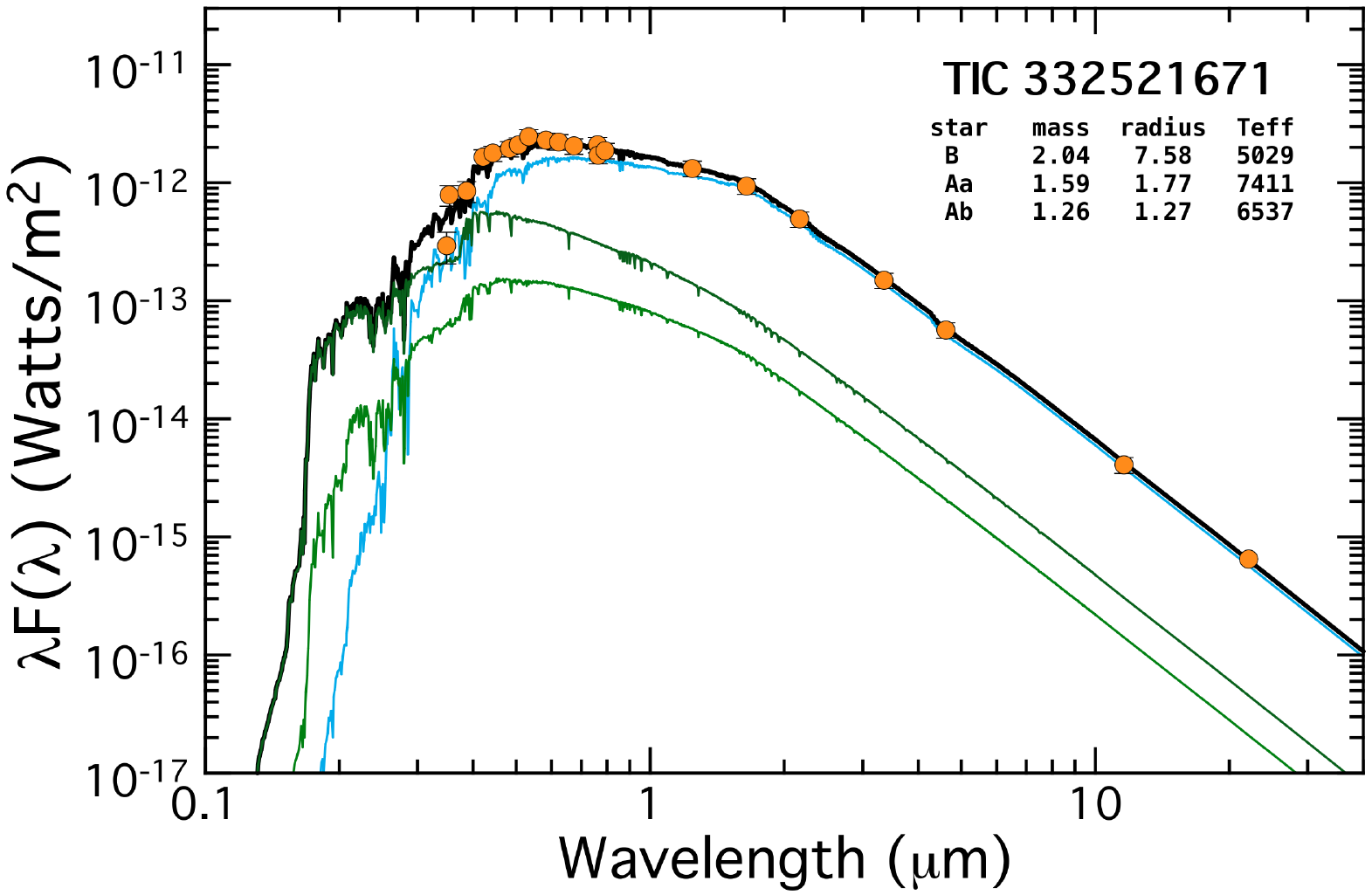}  \hglue0.1cm 
\caption{SED fits for two triply eclipsing triples of the nine discussed in this work. The cyan curve represents the model spectrum of the tertiary star (B) while the green curves are for the EB stars (Aa and Ab).  The sum of the three model spectra is given by the heavy black curve. The fits for the three stellar masses, radii, and $T_{\rm eff}$'s were made using only the measured SED points, and five simple crude constraints enumerated in the text. We make the explicit assumption that the three stars have evolved in a coeval manner without mass transfer.}
\label{fig:SEDs}
\end{center}
\end{figure} 

\begin{figure}
\begin{center}
\includegraphics[width= 0.99 \columnwidth]{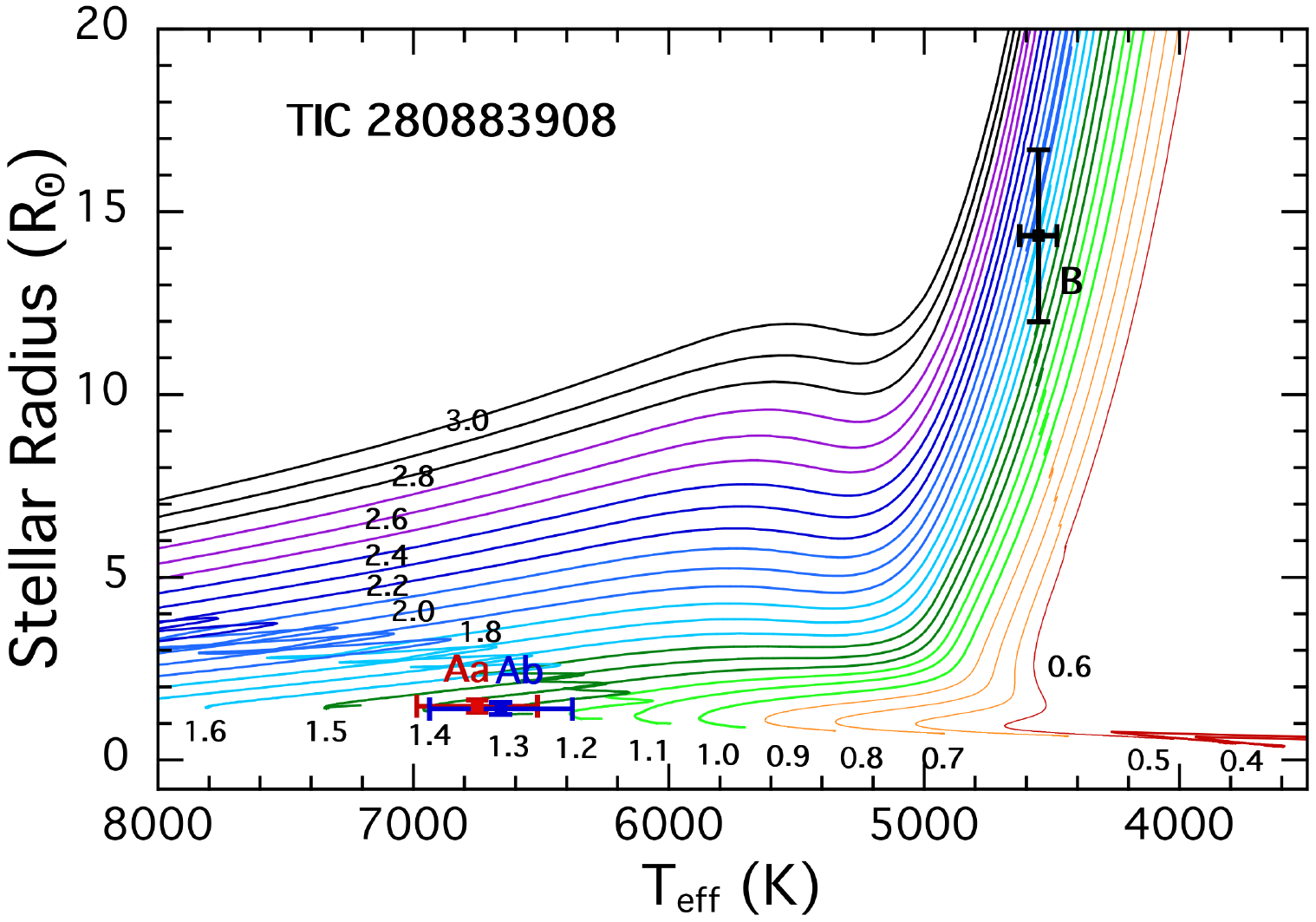} \hglue0.02cm  
\includegraphics[width= 0.994 \columnwidth]{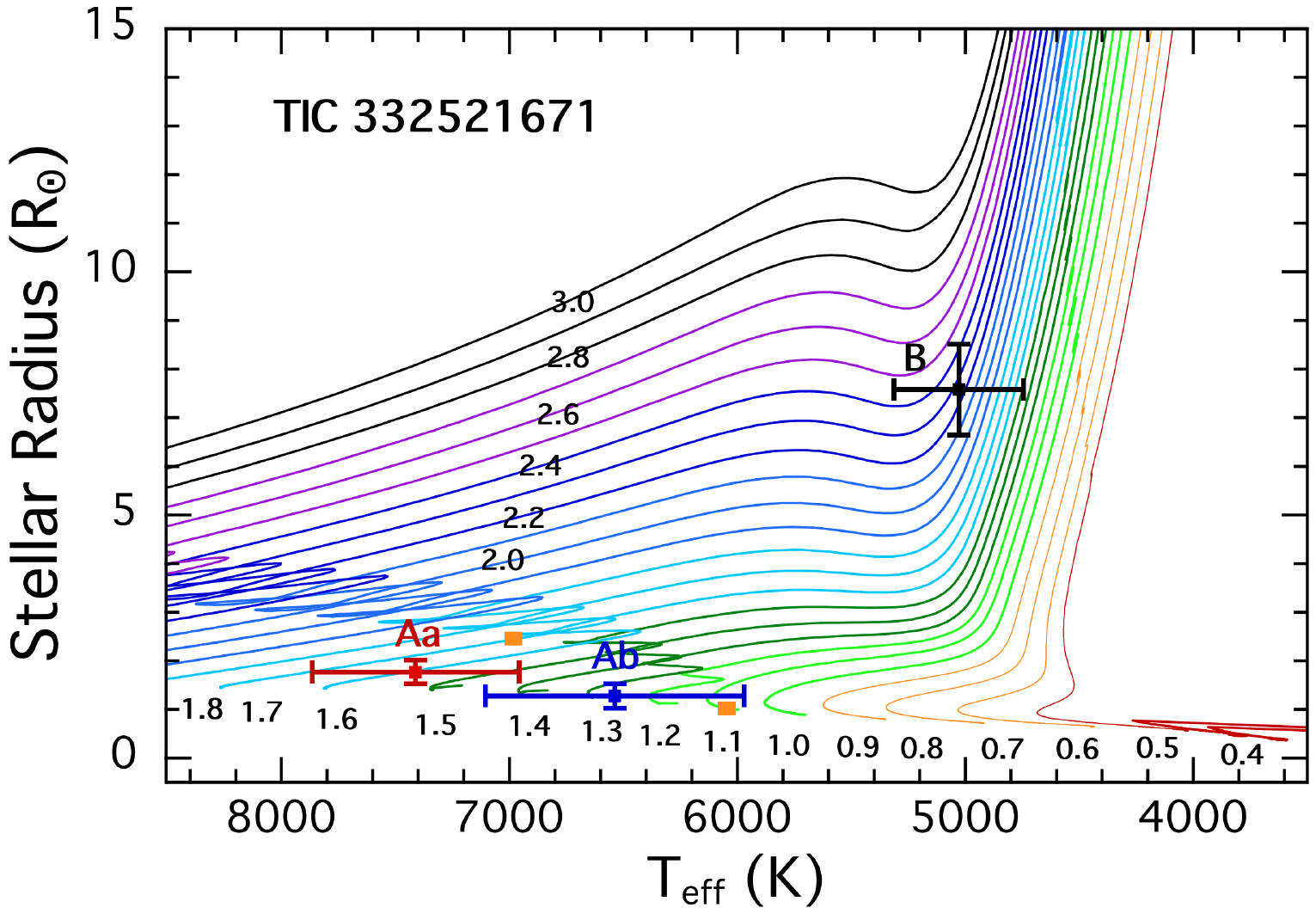} \hglue0.13cm 
\caption{The locations of the three stars in two triple systems of the nine in this work, shown superposed on the {\tt MIST} stellar evolution tracks for stars of solar composition.  The numbers next to the tracks are the stellar masses in M$_\odot$. The locations of the stars in this diagram were taken from the SED fits whose results are given in Fig.~\ref{fig:SEDs}.  More accurate stellar parameters are tabulated in Sect.~\ref{sec:photodynamical} based on the full photodynamical analyses.  The parameters given here for the three stars in TIC 280883908 and for star B in TIC 332521671 are in excellent agreement with those we find from the full photodynamical analysis in Sect.~\ref{sec:photodynamical}.  However, stars Aa and Ab in the latter triple are off by about 1$\sigma$ in $T_{\rm eff}$ and 1-3$\sigma$ in radius from the more comprehensive analysis (shown with small orange markers), which is not atypical when the constraints are limited. }  
\label{fig:evolution}
\end{center}
\end{figure} 

\begin{figure*}
\begin{center}
\includegraphics[width=0.32 \textwidth]{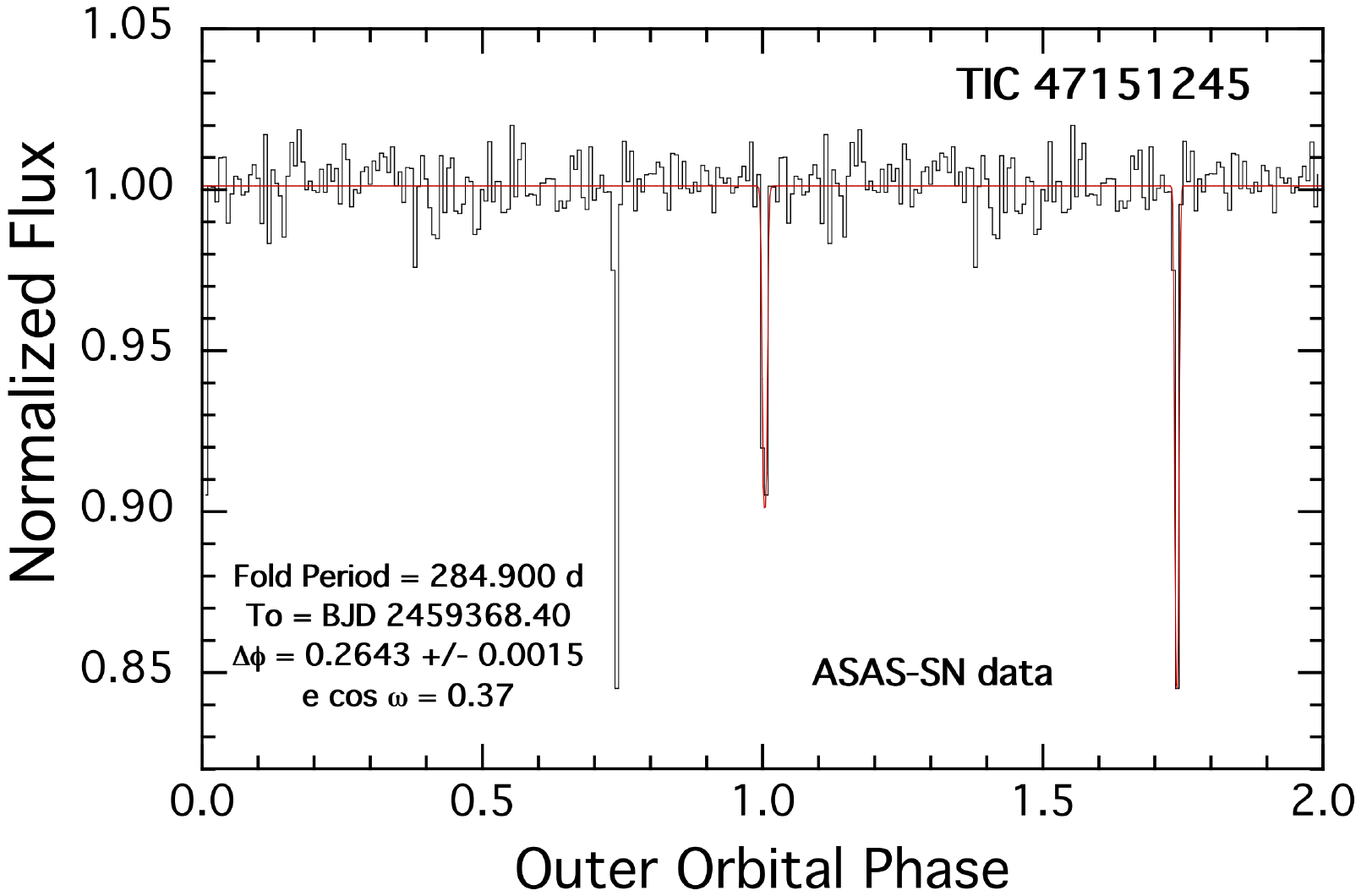} 
\includegraphics[width=0.32 \textwidth]{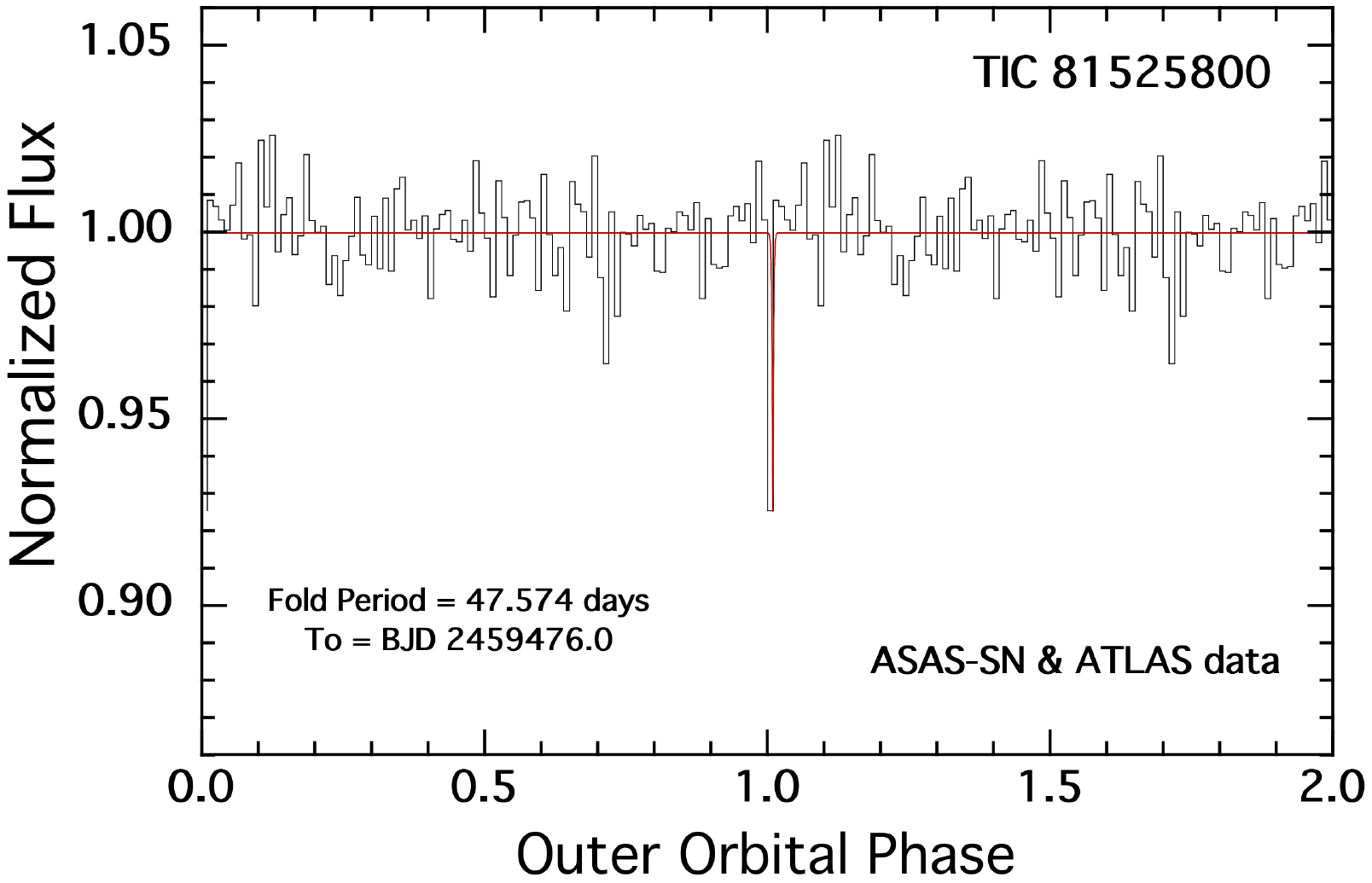}
\includegraphics[width=0.32 \textwidth]{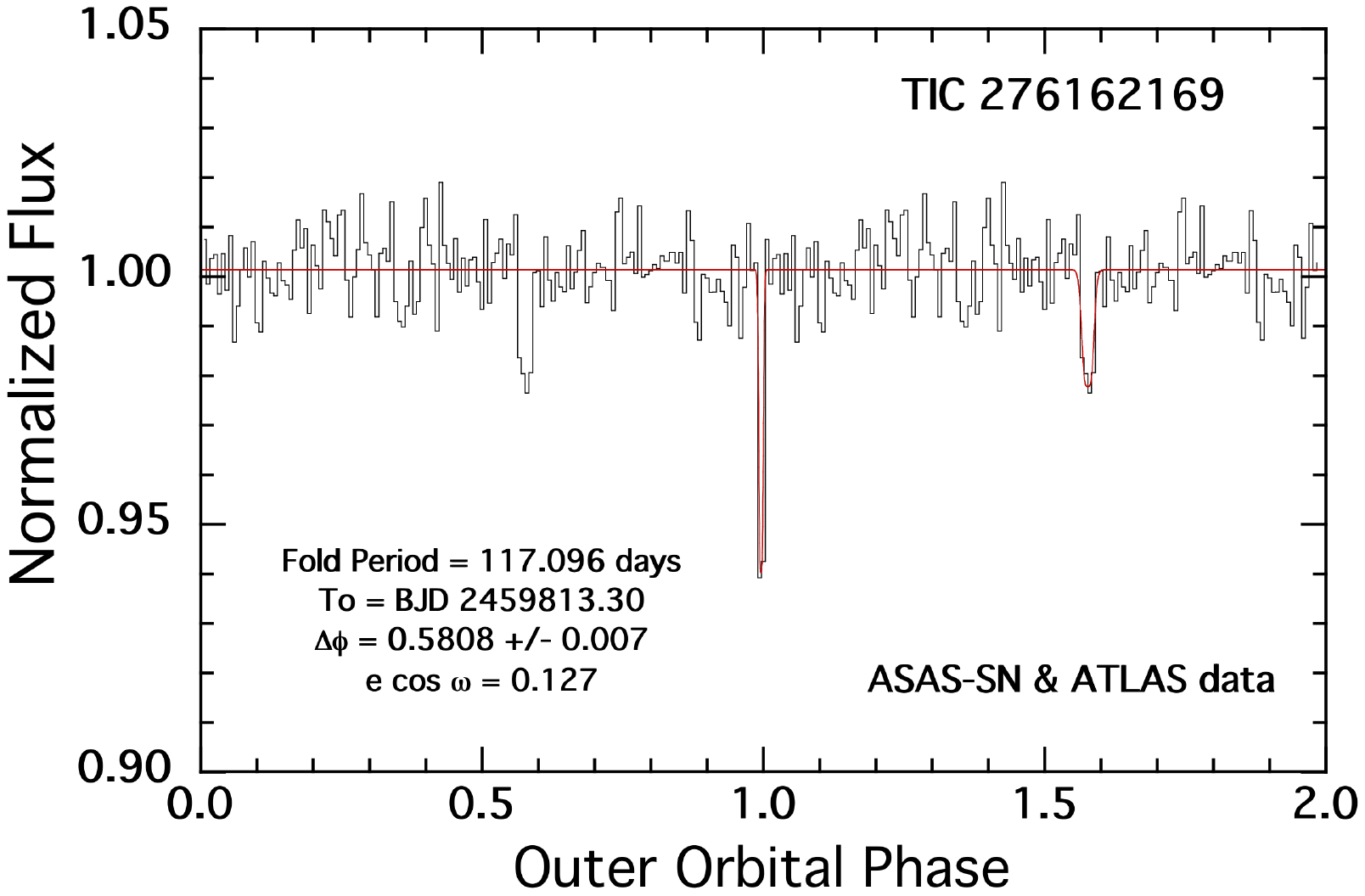}
\includegraphics[width=0.32 \textwidth]{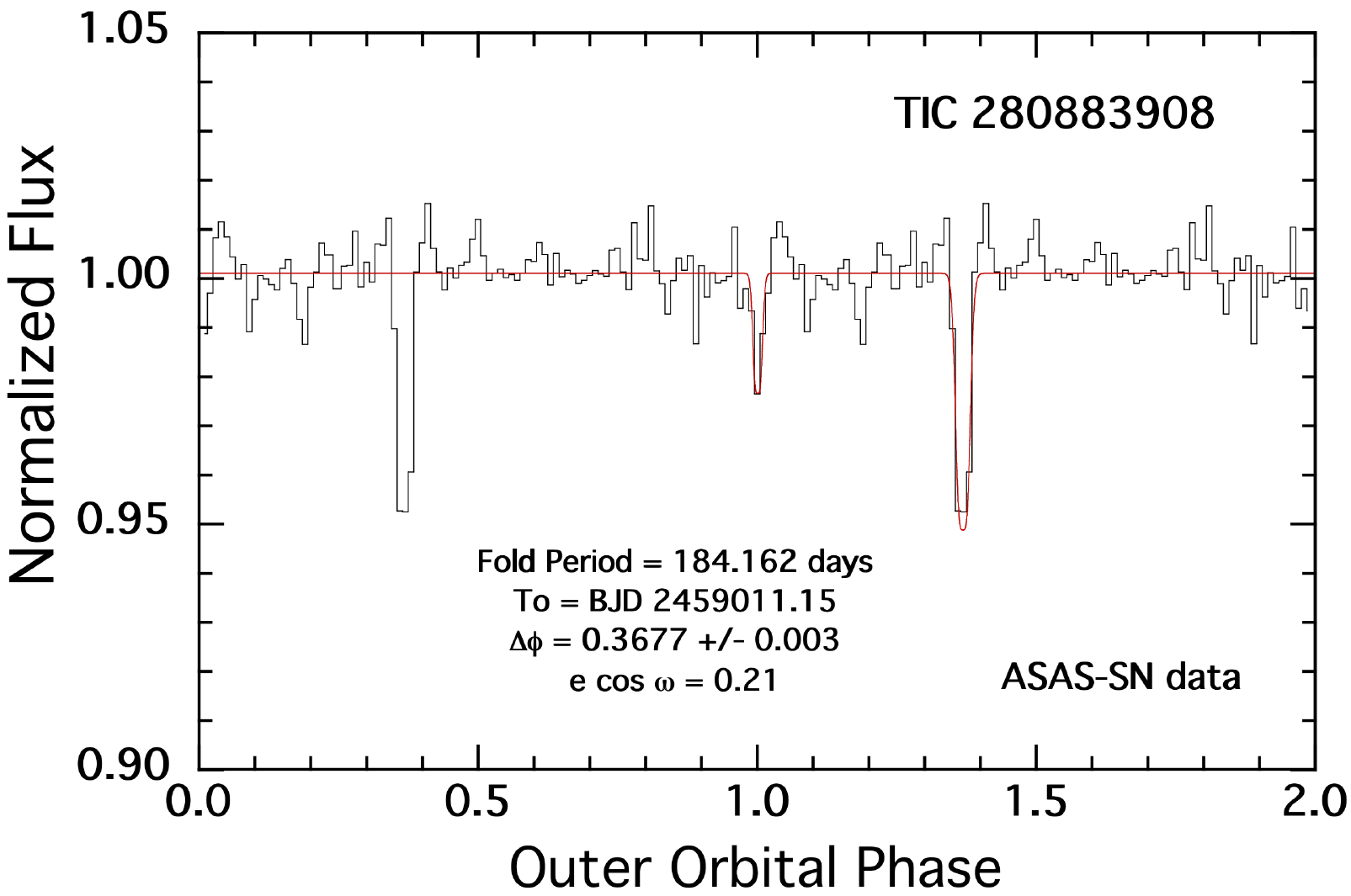} \hglue0.15cm
\includegraphics[width=0.31 \textwidth]{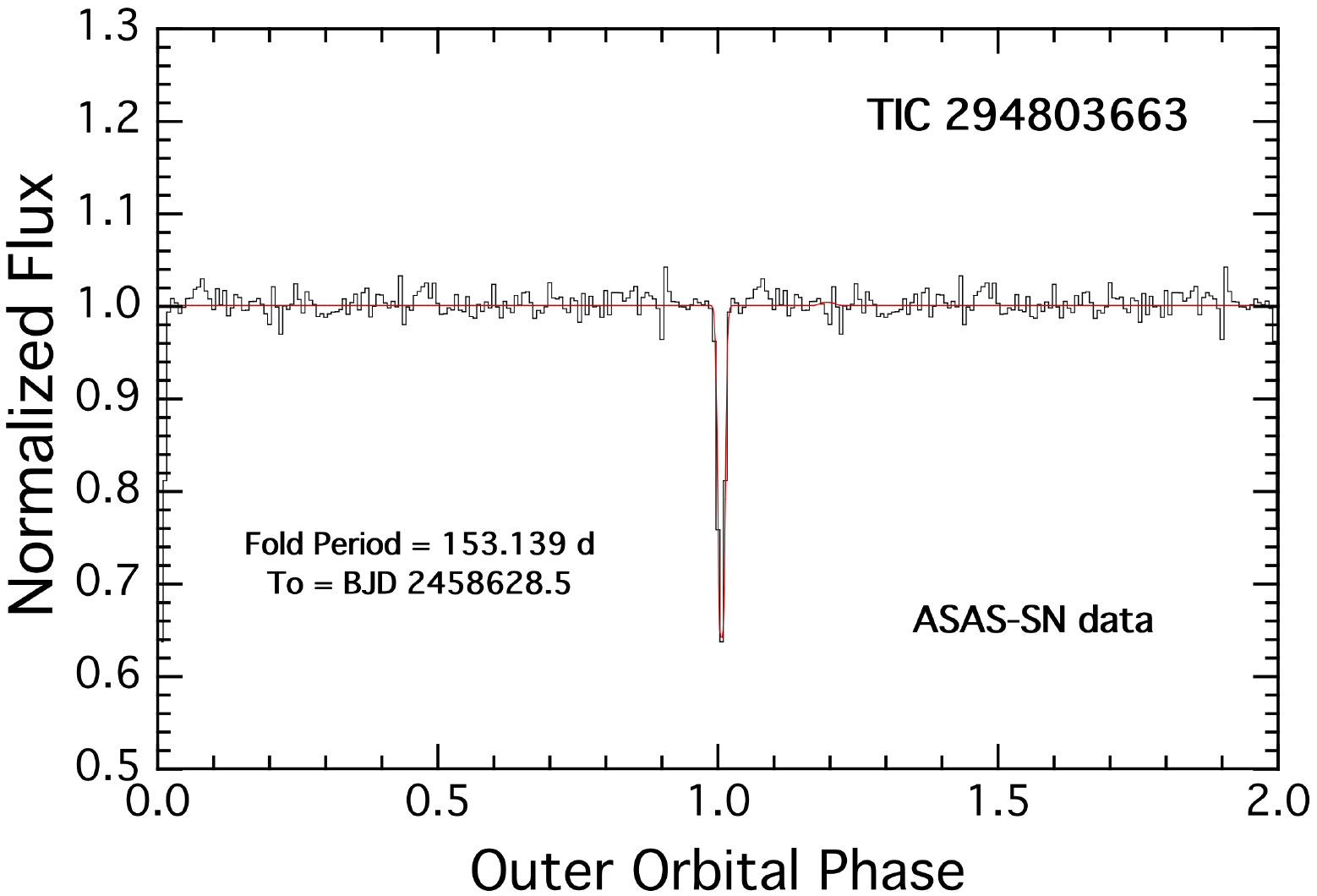}
\includegraphics[width=0.32 \textwidth]{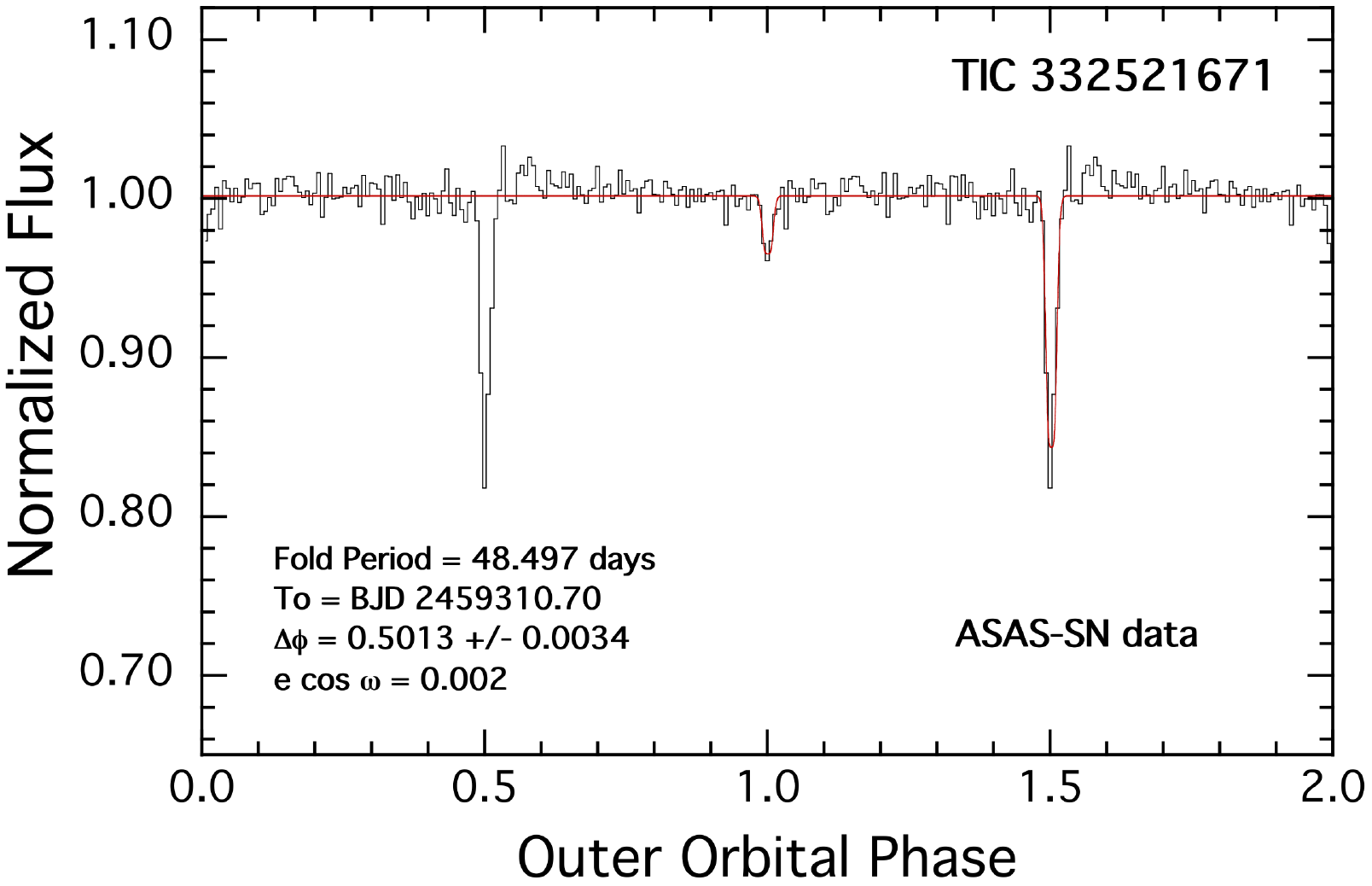} \hglue0.07cm
\includegraphics[width=0.32 \textwidth]{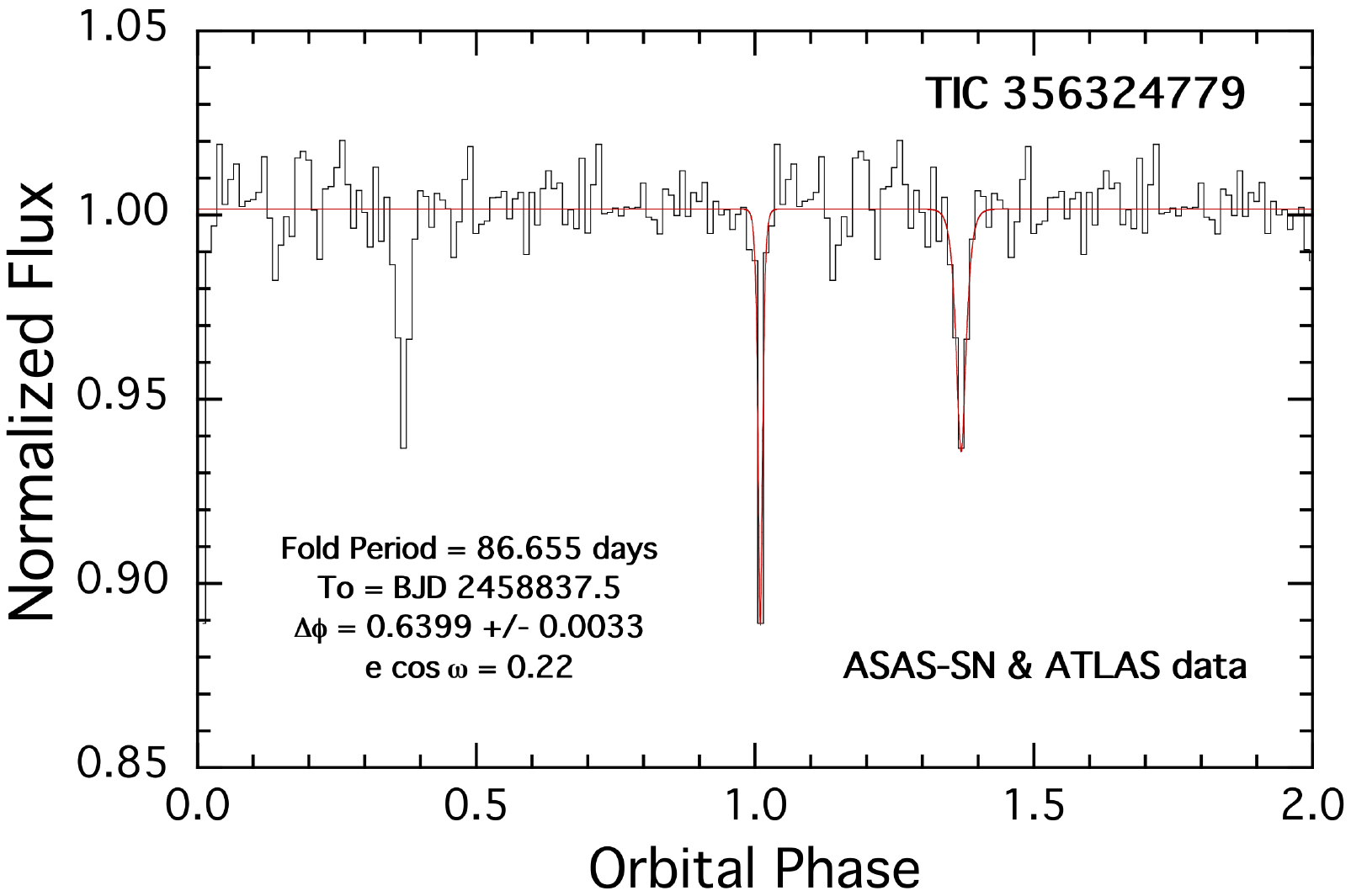}
\caption{Folded, binned, and averaged lightcurves for the outer orbits of seven triply eclipsing triple systems.  These were found using archival data from ASAS-SN and ATLAS (see Sect.~\ref{sec:outer_orbit} for details).  Each panel lists the fold period, the reference time of phase zero (based on the {\it TESS} detection of an outer eclipse), the orbital phase difference between the primary and secondary outer eclipses (for five of the systems), and the inferred value of $e \cos \omega_{\rm out}$ (if both types of outer eclipses are seen).  The red curve is a fit to the one or two outer eclipses during the second plotted orbital cycle only, and are used to determine the orbital phase difference between the two eclipses.}
\label{fig:outer_fold}
\end{center}
\end{figure*} 

\section{Other Detections of These Nine Systems}
\label{sec:other}

Four of the nine systems we report here were noticed previously in the quadruples catalog by \citet{zasche22} as triply eclipsing triples and were correctly reported as such. The outer orbital periods were given but no other information about the systems was presented---either the orbital or stellar parameters.  These sources are itemized in Table \ref{tbl:noticed} along with the outer periods found by \citet{zasche22}.

With the release of Gaia DR3 (\citealt{babusiaux22}; \citealt{gaia22}) are included about half a million orbits measured by Gaia in a number of difference ways (e.g., astrometric, spectroscopic, eclipsing, etc.).  The epochal data on which these orbital determinations were made are not given, nor is any further analysis made of any other system parameters (in particular, the properties of the stars, including the tertiary and those of the inner binary). A check of the Gaia DR3 data set shows that four of the systems we report on have measured orbital parameters, and those are also listed in Table \ref{tbl:noticed}.  For a fifth system (TIC 332521671) Gaia has a completely different outer period listed for nearby TIC 332521670, and we conclude that this latter binary is unrelated to our triple system.   

In the same Table (\ref{tbl:noticed}) we also preview our final results and list the outer period, eccentricity, and argument of periastron, for comparison to the Gaia results and the periods found by \citet{zasche22}.

\section{Preliminary SED Analysis}
\label{sec:SED}

When we first discover a triply eclipsing triple, we make an initial analysis of its composite spectral energy distribution (SED).  This provides us with preliminary estimates of the nature of the three stars in the system.  We make use of the VizieR (\citealt{ochsenbein00}; A.-C.~Simon \& T.~Boch: \url{http://vizier.unistra.fr/vizier/sed/}) SED database which, in turn, utilizes the extensive sky coverage of surveys such as Skymapper \citep{wolf18}, Pan-STARRS \citep{chambers16}, SDSS \citep{gunn98}, 2MASS \citep{2MASS}, WISE \citep{WISE}, and in some cases Galex \citep{bianchi17}. These typically provide $\sim$18-25 flux values over the range 0.12-0.40\,$\mu$m to 21\,$\mu$m.  

For this initial SED analysis, we assume that the three stars in the system have evolved in a coeval manner since their formation as a triple system.  We also assume that no mass transfer has previously occurred among the three stars, in particular between the binary stars.  With these preliminary working hypotheses, there are only four parameters that need to be fitted in a Markov chain Monte Carlo analysis (see, e.g., \citealt{ford05}; \citealt{rappaport21}): $M_{\rm Aa}$, $M_{\rm Ab}$, $M_{\rm B}$, and the age of the system, where $Aa$ and $Ab$ denote the stars in the inner binary, while $B$ refers to the tertiary star in the outer orbit.  We utilize {\it MIST} stellar evolution tracks \citep{paxton11,paxton15,paxton19,dotter16,choi16} for an assumed solar composition of the three stars\footnote{This particular assumption is for the preliminary SED analysis only. See Sect.~\ref{sec:photodynamical} for a description of the more comprehensive photodynamical analysis.}, and the stellar atmosphere models from \citet{castelli03}.  If these four parameters can be determined, then we automatically find the stellar radii and effective temperature of all three stars from evolution tracks.  

To fit the SED curve, we typically need an accurate distance to the source as well as the corresponding interstellar extinction, $A_V$.  For typical objects of comparable brightness and distance to those of our 9 systems Gaia \citep{GaiaEDR3} has a cataloged distance with better than 5\% accuracy.  Information on the extinction to the target can be found from a variety of sources (e.g., Bayestar19; \citealt{green19}).  Thus, in principle, it is not necessary to fit for these parameters.  However, because the systems we are observing are not single stars, and the Gaia distances may in some cases be adversely affected, we add these two quantities (distance and $A_V$) to the fitted MCMC parameters, but with priors that are limited to $\approx \pm$ 4 times their listed uncertainties.

Even though we have 18-25 SED data points, and only $\simeq 6$ MCMC parameters to fit for, that is often insufficient for a robust solution.  The reasons include the facts that (i) the wavelength spacing between SED points is too close to have them represent truly independent wavelengths, and (ii) most SED curves can be fit with a non-physical function containing no more than 4 or 5 parameters which define the characteristic shape. We therefore find it extremely useful to add a number of supplementary constraints in the MCMC fit.  For this set of systems, we used five additional constraints.  The first three of the following stem from the fact that in most cases, the tertiary star tends to dominate the system light either because (in 4 cases) it is substantially evolved off the main sequence, or because (in 7 of 9 cases) it is the most massive star.  In only one case (TIC 276162169) is neither of these true, in which case we forgo the preliminary SED fit.  For the other 8 systems we adopt the following: (i) $M_B$ $> 1.1 \,M_{Aa}$ and $>1.1\,M_{Ab}$, to be certain that the tertiary would be evolved substantially more than the EB stars.  (ii) $R_B$ equals the average of the values given in the {\it TESS} Input Catalog (TIC v8.2) and by Gaia DR2 $\pm 4 \sigma$ of those measurements.  (iii) Same as (ii) but for $T_{\rm eff, B}$.  See Table \ref{tbl:mags9} for the values of $R_B$ and $T_{\rm eff, B}$ (from Gaia and the TIC) based on the composite system light. (iv) The temperature ratio of the two EB stars can be estimated approximately from the ratio of their eclipse depths.  Finally, (v) we take the luminosity of the primary EB star ($L_{\rm Aa}$) to have a fraction of the system luminosity that is greater than the primary eclipse depth.

With just this basic information, we fit the SED for all the properties of the stars in 8 of our 9 systems.  
Here, since there are nine of them, and plots of the fits are not highly illuminating, we show only two illustrative examples---for TIC 280883908 and TIC 332521671 (see Fig.~\ref{fig:SEDs}).  The measured SED points (orange circles) have been corrected for interstellar extinction.  The continuous cyan curve represents the model fit for the tertiary while the green curves are for the stars in the EB.  The total flux from all three stars is represented by the heavy black curve.  For these two triples the tertiary dominates the system light, contributing 91\% and 72\%, respectively. These first estimates of the stellar parameters provide some very quick insight into what kinds of stars we are working with, and can be utilized as the initial input guesses to the photodynamical analysis (see Sect.~\ref{sec:photodynamical}).  

We can also utilize the system parameters for the stars found from this preliminary SED analysis to show the locations of the stars superposed on the {\tt MIST} evolution tracks.  Again, we show only two illustrative examples of this type of visualization of the stars' evolutionary state in Fig.~\ref{fig:evolution} (also for TIC 280883908 and TIC 332521671). Note how both tertiaries are evolved substantially off the main sequence while the EB stars are both still quite near the MS.   

\section{Outer Orbital Periods Found From Archival Data}
\label{sec:outer_orbit}

Following the discovery of a triply eclipsing triple system with \textit{TESS}, and after determining the basic parameters of the constituent stars via an SED fit, we would most like to know the nature of its outer orbit, especially $P_{\rm out}$ and, if possible, the eccentricity, $e_{\rm out}$.  In this aspect of our work, we have most often made use of the ASAS-SN \citep{shappee14,kochanek17} and ATLAS \citep{tonry18,smith20} archival data sets. The ASAS-SN data covers the entire sky and typically contains some $\sim$1800-3000 photometric measurements of a given star, while the ATLAS archives (covering above declinations of $-30^\circ$) often have approximately 2000 photometric measurements.  The ATLAS data go somewhat deeper than the ASAS-SN data, but they tend to saturate more quickly on brighter stars.  The two data sets complement each other nicely, and can quite often be median normalized to each other and co-added for greater statistical precision.   We also routinely check for KELT \citep{pepperetal07,pepperetal12}, WASP \citep{2006PASP..118.1407P}, HAT \citep{bakosetal02} and MASCARA data \citep{talens17} to see whether they are available for a particular system.  

For all of the 9 sources, there were ASAS-SN archival data available.  For four of the sources there were also non-saturated ATLAS data.   With these data we were able to obtain a robust period for the outer orbit for 7 of the 9 systems, and they were also very useful in determining the long-term average EB periods. 
We also obtained KELT data for one of our targets, TIC\,99013269.  However, opposite to the other archival data, whose use is described in this section, the KELT dataset was found to be extremely helpful at the final stage of the photodynamical analysis, hence, it will be discussed there in Sect.~\ref{sec:info_99013269}.

We analyze the archival photometric data by doing a Box Least Squares transform \citep{kovacs02} (BLS) in a search for the outer eclipses (either outer primary or outer secondary eclipse, or both).   We prepare the archival data for a BLS search by (i) removing obvious outlier points, (ii) median normalizing data from different filter bands, and (iii) subtracting out the lightcurve of the inner EB by Fourier means (as described in \citealt{powell21}).  In the process of the latter operation we remove between 5 and 100 orbital harmonics depending on the sharpness of the eclipses in the lightcurve of the EB.  This cleaning process requires knowing the orbital period of the EB very accurately.  In turn, we determine the long-term average binary period from the {\it TESS} data and/or from the archival data.  

The results of the outer period search for 7 of our 9 triply eclipsing triple stars are shown in Fig.~\ref{fig:outer_fold}.  Each panel shows a folded, binned, and averaged lightcurve of the archival data about the period which corresponds to the most significant peak in the BLS transform. The time of the third-body eclipse observed in the {\it TESS} data is taken to be zero phase for the outer orbit fold in each panel.  For 5 of these triples, both the primary and secondary outer eclipses are securely detected. For those systems, the outer period of the triple as well as the value of $e_{\rm out} \cos \omega_{\rm out}$ are accurately determined.  For the remaining two triples only one of the outer eclipses is determined, thereby leading to an accurate measure of $P_{\rm out}$, but not for $e_{\rm out} \cos \omega_{\rm out}$.  Two of the systems have quite short outer periods of 47.6 days (TIC 81525800) and 48.5 days (TIC 332521671).


\section{Other sources of information}
\label{sec:etv_rv_gaia}

In addition to the {\it TESS} photometry and archival photometry (typically from ASAS-SN and ATLAS), for our analysis we used other kinds of photometric and spectroscopic datasets, as well.  First of all we derived eclipse timing data from the {\it TESS} observations.  We list these eclipse timing variation (`ETV') curves as distinct from the {\it TESS} photometric lightcurves even though the ETVs are, in fact, derived from the {\it TESS} photometry.  In the photodynamics code (see Sect.~\ref{sec:photodynamical}) the $\chi^2$ contribution found from fitting the modeled eclipse times to the measured ones is separate and distinct from the contribution computed from fitting the modeled lightcurves to the measured {\it TESS} photometry. All of the systems have ETV curves that the photodynamical analyses utilize.  For all but one of the systems, the ETV data carried useful auxiliary information and, moreover, for at least 3 of the systems the ETV points were particularly helpful in constraining the system parameters.  

To supplement the ETVs that were derived from the {\it TESS} data, we also organized ground-based photometric and/or spectroscopic follow up observations for some of our systems.  In this way, we observed photometrically a total of eight further regular eclipses from three of the nine EBs, and also a part of an additional third body eclipse in one case.  Importantly, we also obtained radial velocity (RV) data for five of the nine systems.  Note that we incorporated into the joint photodynamical analyses not only the ETV points determined from these follow up observations, but also the eclipsing lightcurves themselves. Finally, note that Gaia orbital parameters are also available for four of our systems. In some cases we also used Gaia periods to find reliable initial-guess parameters for the photodynamical fit.

We tabulate all the RV and ETV data in Tables~\ref{tab:TIC99013269_RVdata}--\ref{tab:TIC332521671_RVdata} and \ref{Tab:V726_Sco_(TIC_047151245)_ToM}--\ref{Tab:TIC_356324779_ToM} for RVs and ETVs, respectively. Moreover, below we show separately the fits to the RV and ETV points for those systems where it provides substantial supplemental information, and mention when there are also Gaia solutions.

\begin{figure}
\begin{center}
\includegraphics[width=0.99 \columnwidth]{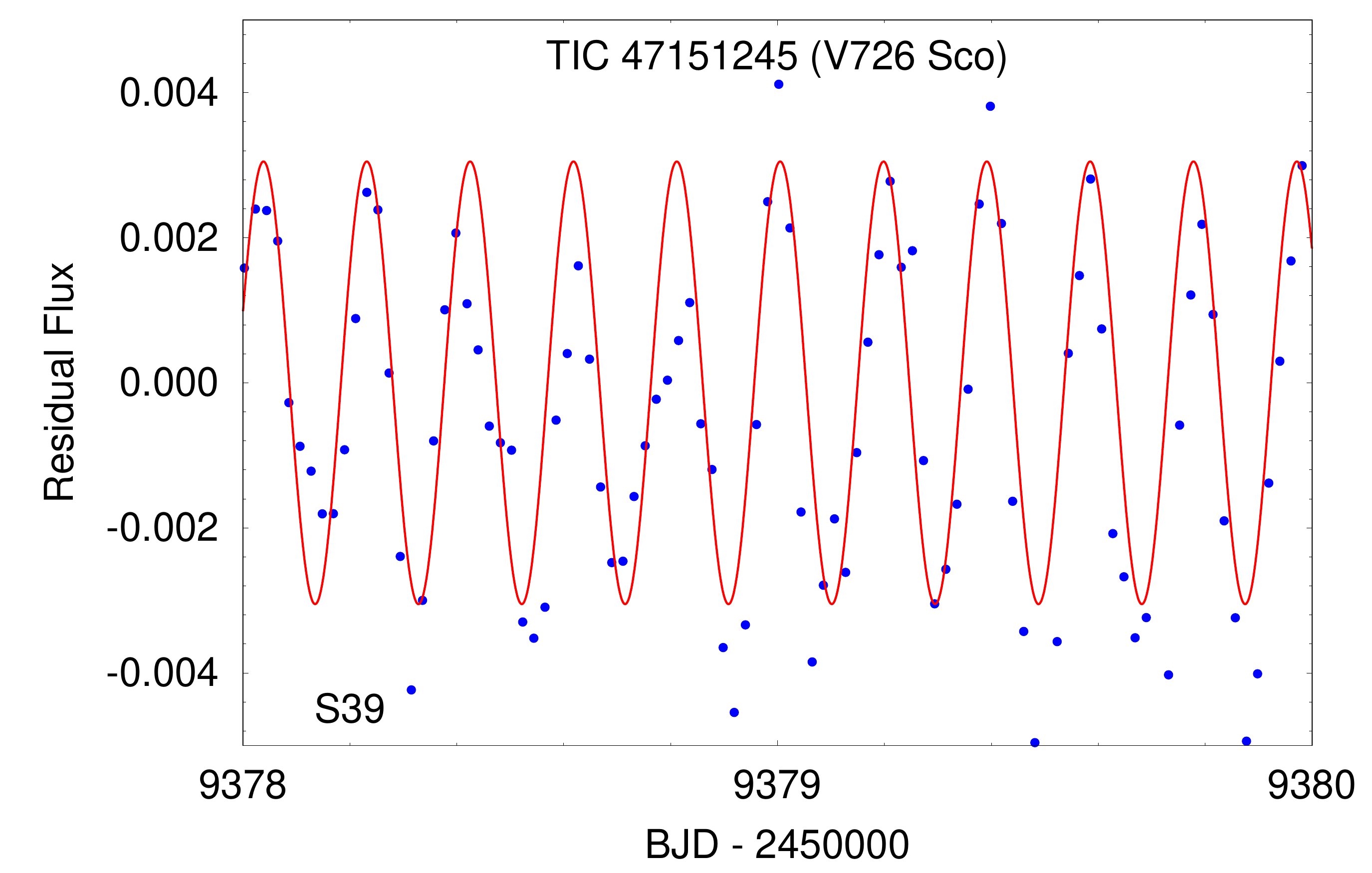}
\caption{Stellar oscillations in TIC 47151245, and the Fourier fit. The dominant pulsation period is 0.1933 d.}
\label{fig:47151245_oscillation}
\end{center}
\end{figure} 

\subsection{TIC 47151245: Archival Data}

TIC 47151245 is the only one of our nine triples which has no Gaia orbit, no useful ETV curve, and no RV data.  In spite of this we have derived a quite meaningful system solution.  Part of the reason for this success is the fact that this system has arguably one of our two best outer orbital folds from the ASAS-SN archival data.  In this lightcurve we detect clearly the primary and secondary outer eclipses, and thereby measure precisely both the outer period and the value of $e_{\rm out} \cos \omega_{\rm out}$ (see Fig.~\ref{fig:outer_fold}).  

Interestingly, we also see a strong oscillation from one of the three hot stars ($\sim$8500 K).  The residuals to the model fit are shown in Fig.~\ref{fig:47151245_oscillation}, along with a Fourier fit which yields only a single period of 0.1933 days.  Unfortunately, we cannot use this pulsation to aid in the mass determination, in part because we do not know which star it comes from.

\subsection{TIC 81525800: ETVs}

\begin{figure}
\begin{center}
\includegraphics[width=0.99 \columnwidth]{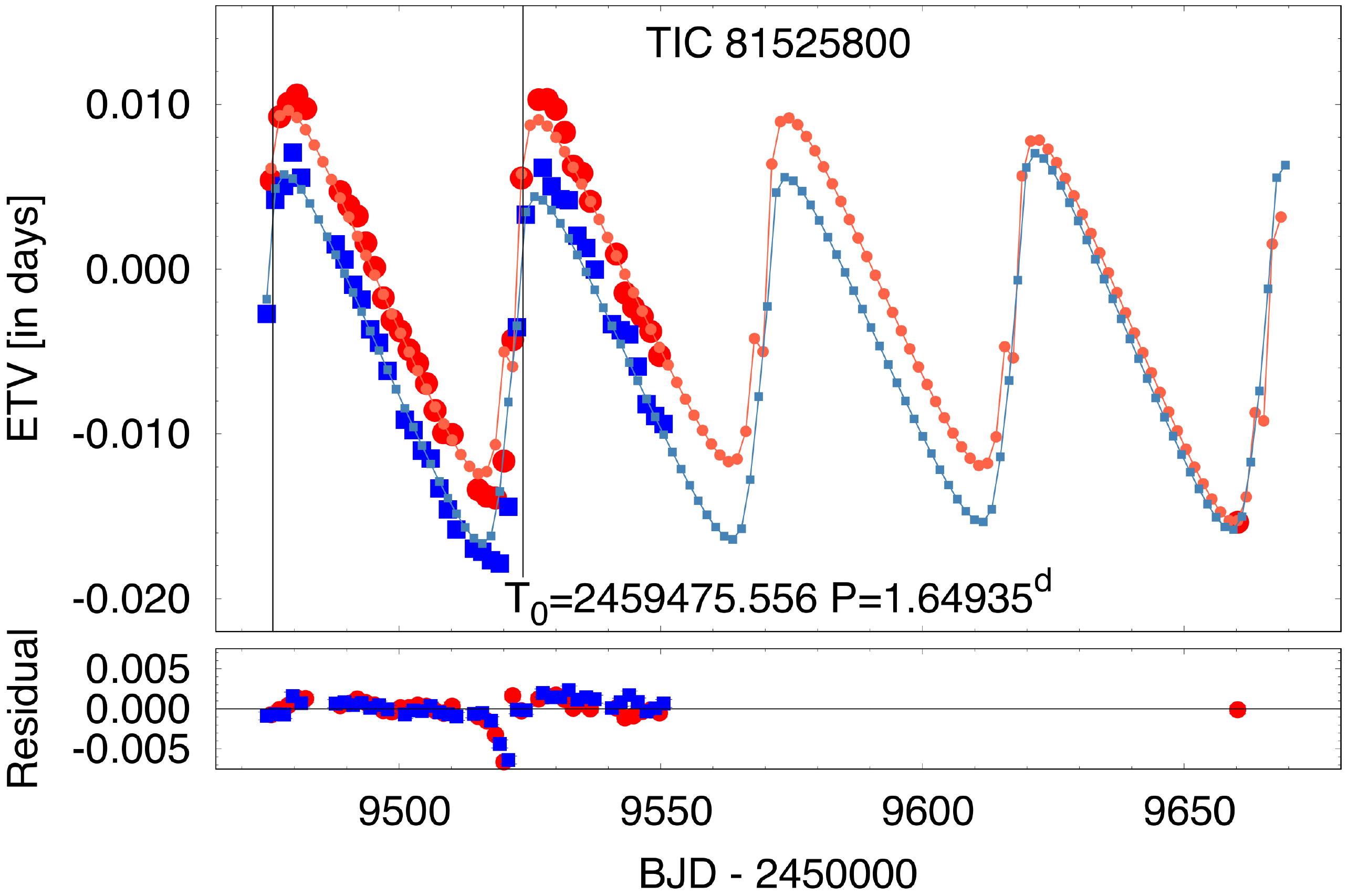}
\caption{Eclipse timing data and its photodynamical model for TIC 81525800. Here, and also for the forthcoming ETV curves, larger red and blue filled circles represent observed primary and secondary eclipses, respectively, while the smaller symbols stand for the model ETV points. These latter points have been connected with straight line segments. Small apparent discontinuities in the model ETV points around the periastron passages in this highly eccentric outer orbit are noticeable.  They arise from the discontinuous sampling of the continuous period variations which are dominated by dynamical, third-body perturbations instead of the pure and much smoother LTTE.  The vertical lines show the locations of the third-body eclipses observed with \textit{TESS}. The earlier points are from Sectors 43, 44, and 45 of the {\it TESS} observations, while the last point is from a ground-based follow-up measurement with the BAO80 instrument.}
\label{fig:TIC81525800}
\end{center}
\end{figure}

The ETV curve for TIC 81525800 is shown in Figure \ref{fig:TIC81525800}.  It covers Sectors 43, 44 and 45 of the {\it TESS} data as well as followup ground-based observations made by the 80-cm RC telescope of Baja Astronomical Observatory (BAO80 -- see \citealp{borkovitsetal22} for technical details), Hungary on the night of 21/22 March 2022.  The ETV curve is dominated by the dynamical delay due to the short outer orbital period of only 47.8 days\footnote{Note that the outer period of this triple is listed in Table \ref{tab:syntheticfit_TIC47151245+81525800} as 49.75 d.  The latter is the instantaneous osculating period in this very close and highly eccentric binary. Hence the apparent discrepancy with the eclipsing period cited here and listed in Fig.~\ref{fig:outer_fold}} and the high outer eccentricity of 0.61.  The outer orbit of 47.8 days is well mapped out by the ETV points after only two sectors of {\it TESS} data.  

\subsection{TIC 99013269: RVs, ETVs, and Gaia results}

This system exhibited only one, but quite remarkable, extra eclipsing event at the very beginning of the Sector 41 observations. Despite the brightness of the EB, the archival data were found to be unusable for determining the outer orbital period. There was weak evidence for a $\sim303$-day outer period, but with the assumption of that period, we were unable to find any realistic and physically stable solution. Therefore, we initiated a long-term spectroscopic follow up for obtaining RV data, and planned to postpone the detailed study of this EB to a later time. 

For the spectroscopic follow up we used three different instruments. 16 of the 20 RV points were obtained with the fibre-fed ACE spectrograph attached to the 1-m RCC telescope of Konkoly Observatory at Piszk\'es-tet\H o, Hungary. Three points were acquired with a fibre-fed \'Echelle spectrograph mounted on the 1.3-m Nasmyth-Cassegrain telescope of Skalnat\'e Pleso Observatory, Slovakia. Finally, one RV measurement was taken with the ESpeRo spectrograph \citep{bonevetal17} mounted on the 2m RCC telescope of the Rozhen National Astronomical Observatory (NAO), Bulgaria.

\begin{figure}
\begin{center}
\includegraphics[width=0.99 \columnwidth]{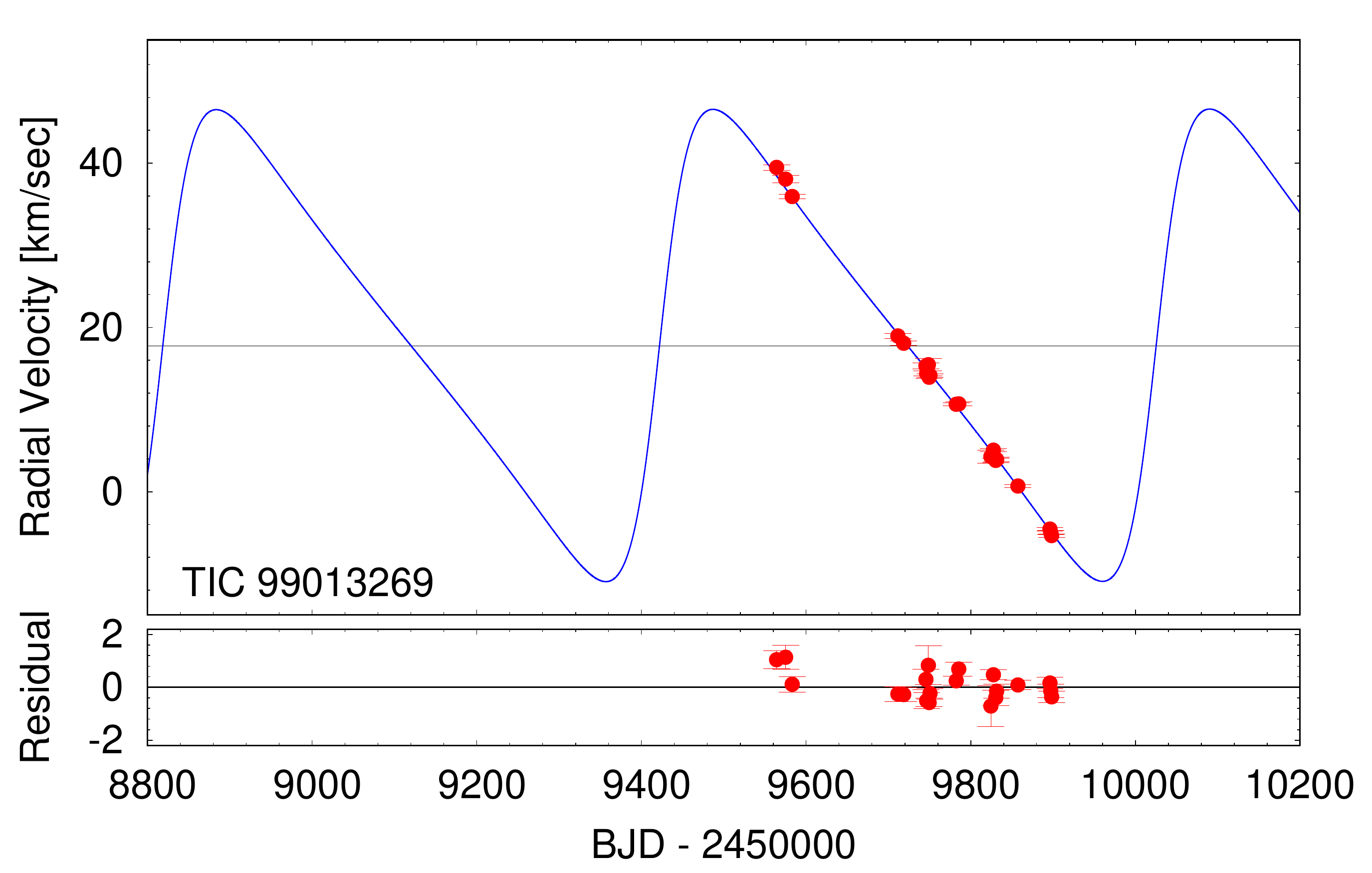}
\includegraphics[width=0.99 \columnwidth]{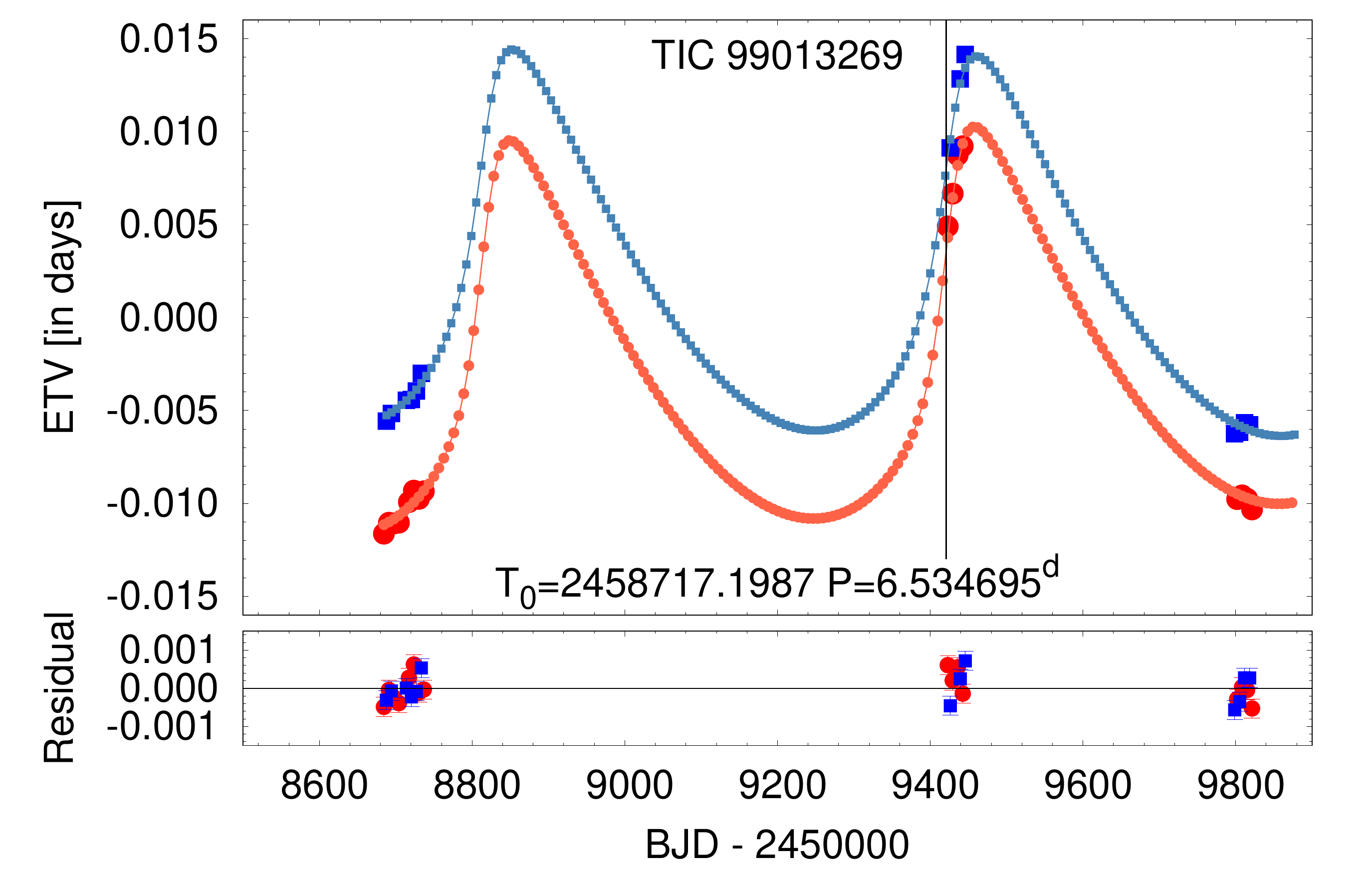}
\caption{Radial velocity and eclipse timing data for TIC 99013269.  The RV data were gathered with the 1-m Piszk\'estet\H o (Hungary), 1.3-m Skalnat\'e Pleso (Slovakia) and 2m Rozhen (Bulgaria) telescopes, while the ETV data are from \textit{TESS} Sectors 14, 15, 41 and 55.  The outer period is $604.05 \pm 0.01$ d.}
\label{fig:TIC99013269}
\end{center}
\end{figure}

After Gaia's third data release, however, we immediately realized that this system is tabulated in the Gaia DR3 non-single stars (NSS) catalog (\citealt{babusiaux22}; \citealt{gaia22}) as an astrometric binary with a period of $P_\mathrm{astr}=609.5$\,d.  Hence, we initiated some preliminary photodynamical runs, using \textit{TESS} Sector 14, 15 and 41 data, the ETV curves determined from these data, and the first set of our RV data that was available at that time.  From those inputs we found a consistent and astrophysically reliable solution.  This photodynamical run was reiterated and refined when the Sector 55 \textit{TESS} observations were released. The RV and ETV curves with the best-fit joint photodynamical solution are plotted in Fig.~\ref{fig:TIC99013269}, top and bottom panels, respectively.  Our outer period and eccentricity are $604.05 \pm 0.01$ days and $0.463 \pm 0.005$, compared to Gaia's values of $609.5 \pm 2$ days and $0.46 \pm 0.03$, so in fairly decent agreement to within the larger Gaia uncertainties.  Of course, Gaia provides no information about the stellar constituents of the triple.

\subsection{TIC 229785001: ETVs and Gaia results}

\begin{figure}
\begin{center}
\includegraphics[width=0.99 \columnwidth]{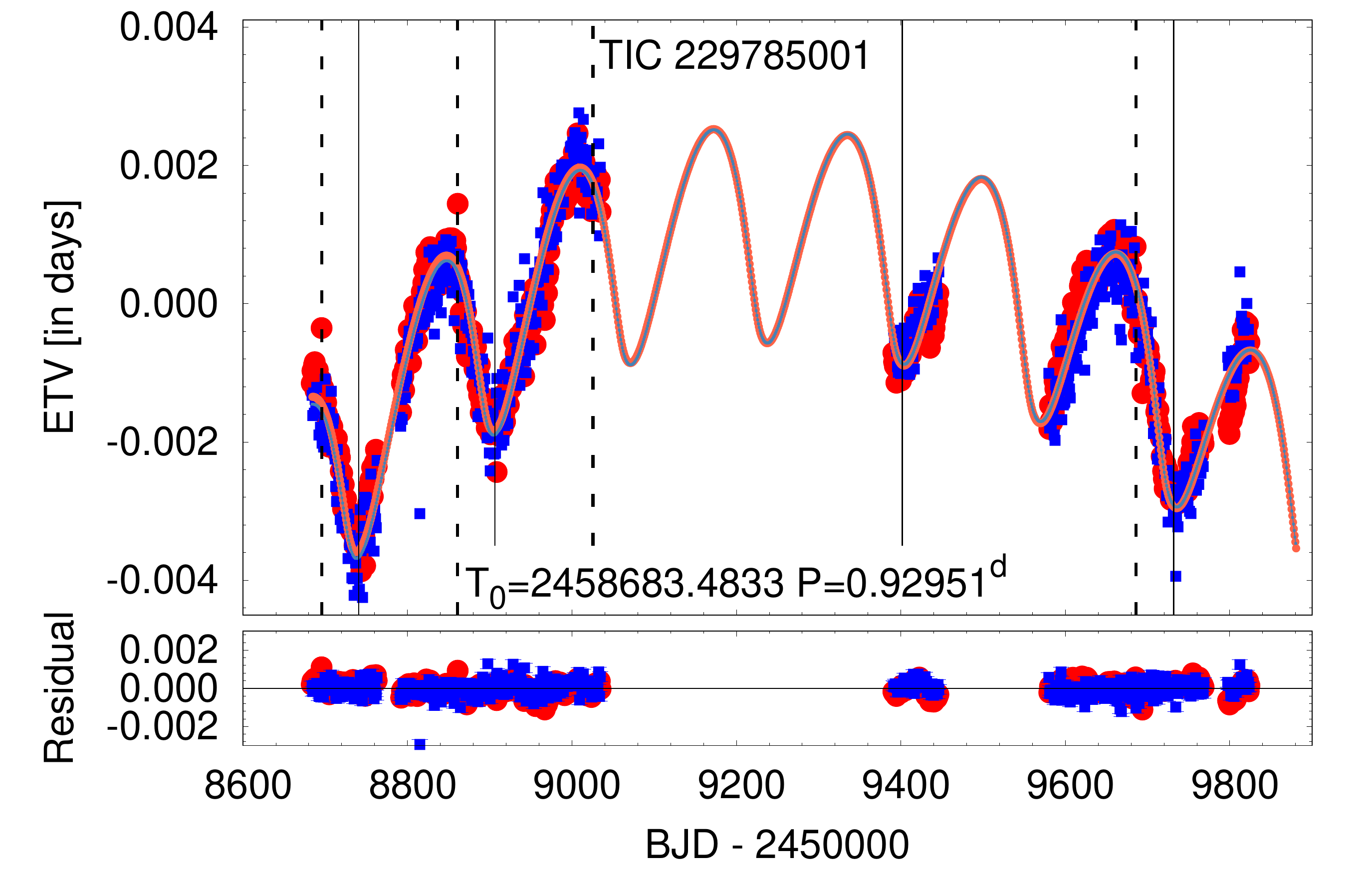}
\caption{Eclipse timing data for TIC 229785001.  The ETV data are from \textit{TESS} Sectors 14-16, 17-26, 40-41, 47-53, and 55.  Solid and dashed vertical lines represent the locations of the primary and secondary third-body eclipses, respectively.}
\label{fig:TIC229785001}
\end{center}
\end{figure}

This system is located in the northern continuous viewing zone (NCVZ) of \textit{TESS} and, hence, it was observed almost continuously during years 2 and 4 of the prime and extended missions of the spacecraft.  The ETVs in Fig.\,\ref{fig:TIC229785001} clearly indicate a longer time-scale variation in addition to the 165-day outer orbit of the triple. Therefore, for a satisfactory modeling of the lightcurve features (mainly the locations and morphologies of the third-body eclipses) we assumed that TIC 229785001 is indeed at least a quadruple stellar system with a hierarchy type of (2+1)+1. Hence, we added a fourth, distant star to the triple system revolving on a wider orbit about the center of mass of the inner triple subsystem. The relatively short observing window of {\it TESS} did not make it possible to determine the parameters of this outermost orbit with satisfactory precision or, even, uniquely. Thus, the parameters of the outmost orbit given later in Table~\ref{tab:syntheticfit_TIC99013269+229785001} should not to be taken too seriously, and may be substantially revised after future observations.

Gaia gives an outer orbital period of 166.0 days compared to our value of 165.25 days, and the corresponding outer eccentricities are 0.49 vs.~0.46, respectively.  Thus, these are in basic agreement.

\subsection{TIC 276162169: ETVs}

\begin{figure}
\begin{center}
\includegraphics[width=1.02 \columnwidth]{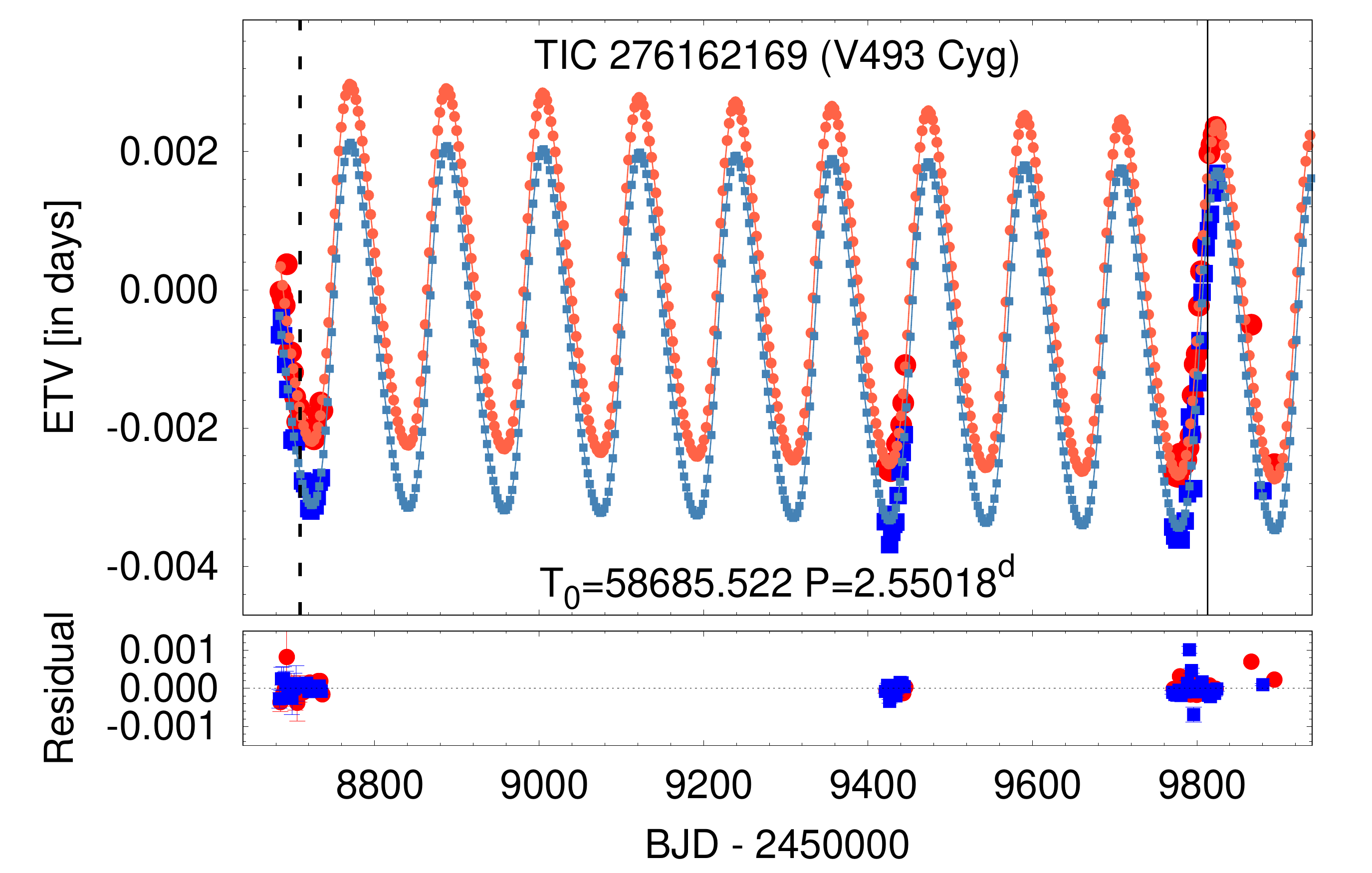}
\caption{Eclipse timing data for TIC 276162169.  The majority of the ETV data are from \textit{TESS} Sectors 14-15, 41, 54-55, while the last three points are obtained during our follow up observations with the GAO80 instrument.}
\label{fig:TIC276162169}
\end{center}
\end{figure}

The ETV curve shown in Figure \ref{fig:TIC276162169} samples the outer orbit of 117 days relatively well over {\it TESS} sectors 14-15, 41, and 54-55, and also includes three points from ground-based observations with the 80-cm RC telescope of Gothard Astrophysical Observatory, Szombathely, Hungary (GAO80 -- see, again, \citealp{borkovitsetal22} for technical details). In this case the dynamical and light-travel time effect (LTTE) delays give quite similar contributions to the ETV curve, which in fact strongly constrains the orbital and dynamical properties of the triple system.

Note, this is the only system in our sample for which former, ground-based eclipse-timing observations are available. The Lichtenknecker Database of the BAV\footnote{{Bundesdeutsche Arbeitsgemeinschaft f\"ur Ver\"andliche Sterne} -- \url{https://www.bav-astro.eu/index.php/veroeffentlichungen/lichtenknecker-database/lkdb-b-r}} lists 26 times of eclipse between 1935 and 2017. However, due to their large uncertainties (which substantially exceed the full amplitude of the ETVs induced by the third-body), we have decided not to use these data.

\subsection{TIC 280883908: RVs, ETVs, and Gaia results} 

\begin{figure}
\begin{center}
\includegraphics[width=0.99 \columnwidth]{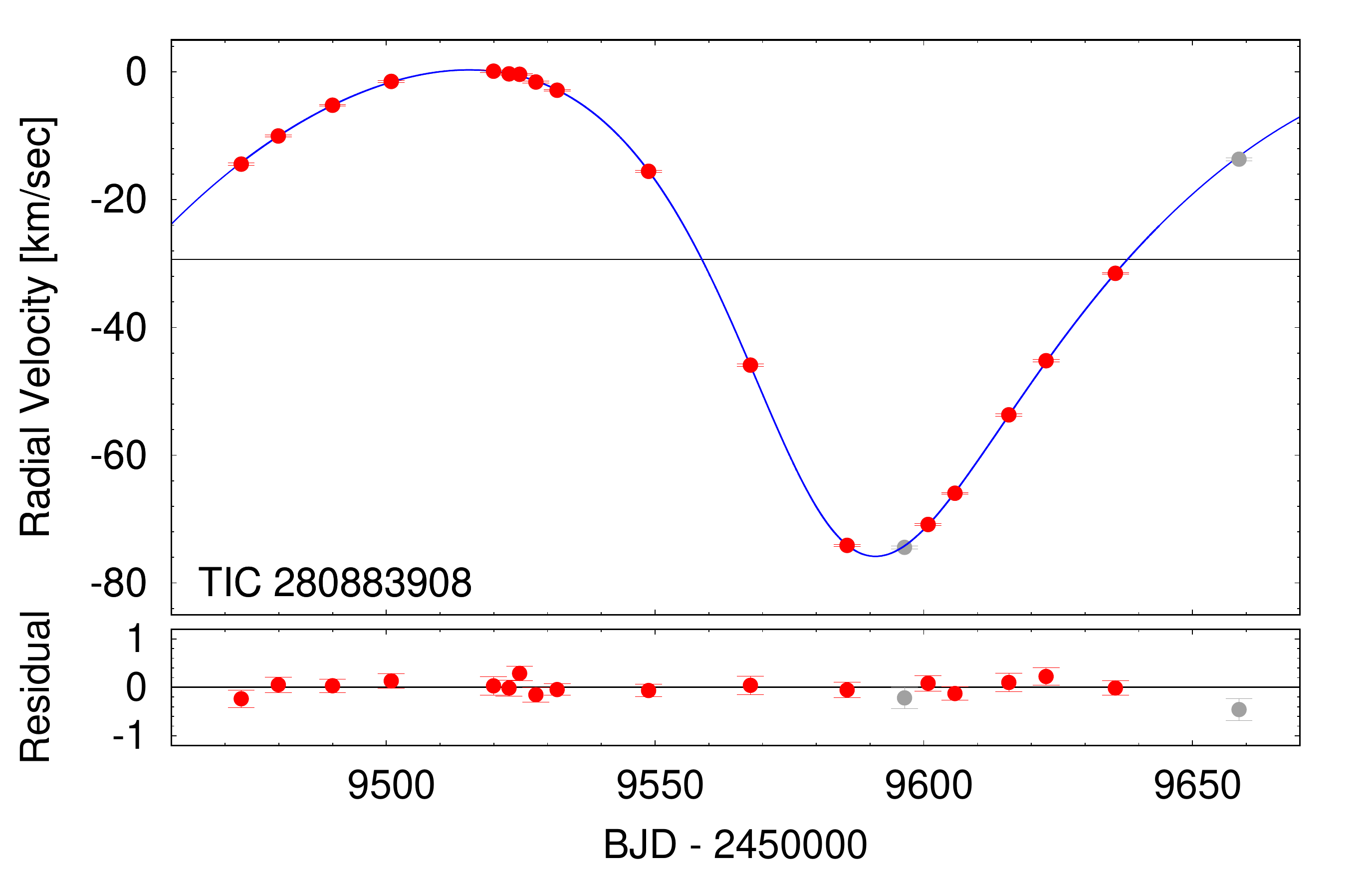}
\includegraphics[width=0.99 \columnwidth]{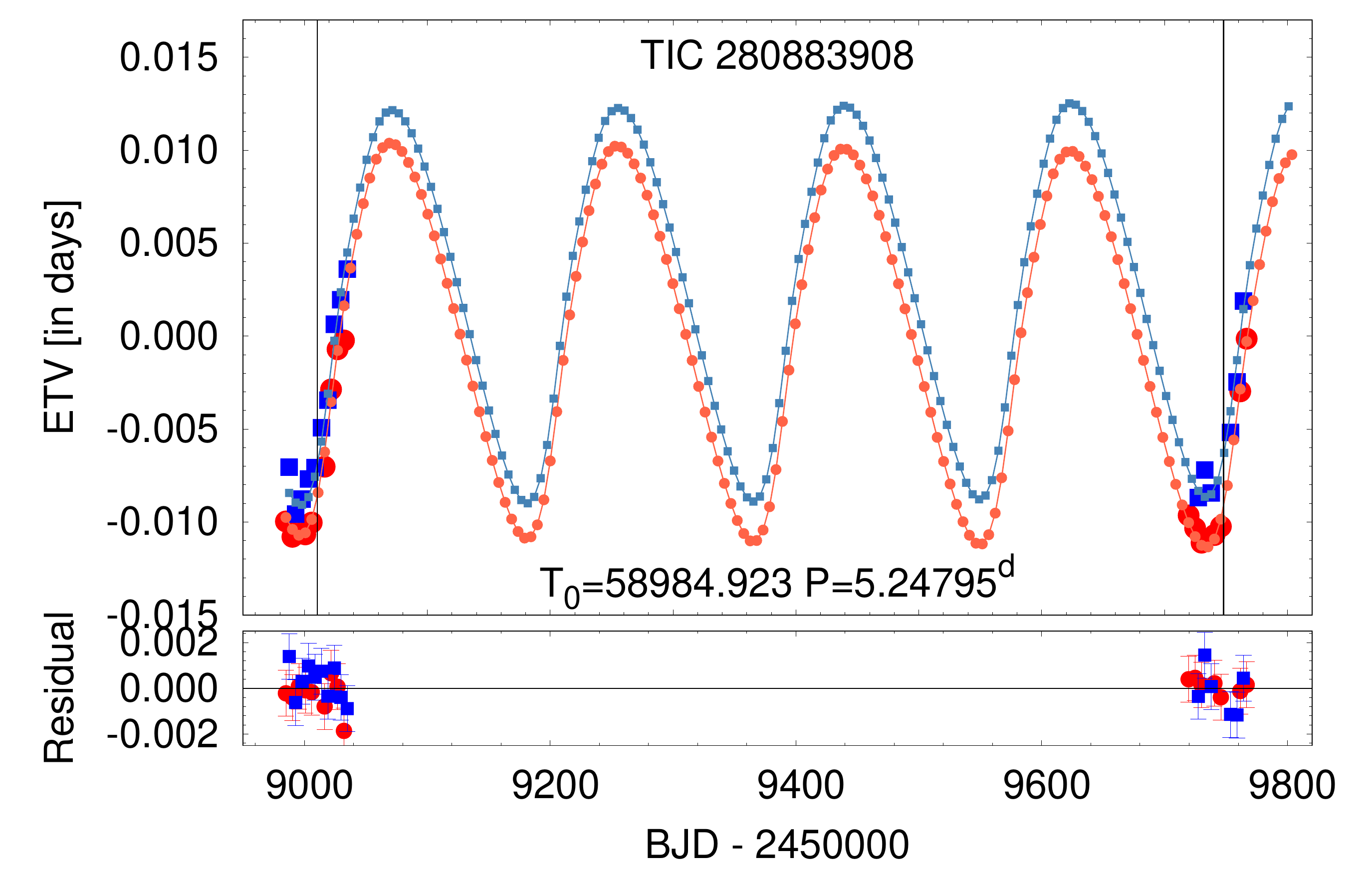}
\caption{Radial velocity and eclipse timing data for TIC 280883908.  All but one RV data points are from TRES and make this the best sampled of any of our RV curves.  The two RV points shown as gray dots were not used for the photodynamical analysis (see text). Furthermore, note that the very last TRES RV point, obtained about 260 days later than the previous one, was not plotted to avoid a large decrease in the horizontal resolution of the plot. The ETV data are from two pairs of {\it TESS} Sectors 25-26 and 52-53, which are separated by two years.  Each pair of {\it TESS} observations samples nearly the same part of the outer orbit of 184 days.}
\label{fig:TIC280883908}
\end{center}
\end{figure}

We obtained 19 RV points with the Tillinghast Reflector Echelle Spectrograph (TRES) on the 1.5m reflector at the Fred Lawrence Whipple Observatory, Arizona, USA.  The two most recent points were obtained after our final photodynamical analysis and, hence, were not used in the model fit.  We obtained one additional point with the 2m Rozhen telescope. In order to keep the homogeneity of the RV measurements, however, we did not use this latter point for the analysis. Figure \ref{fig:TIC280883908} (top panel) shows all but the latest RV point for TIC 280883908 together with the photodynamical solution curve. 

The ETV points taken from the {\it TESS} observations are shown in the bottom panel of Figure \ref{fig:TIC280883908}.  In spite of the fact that the ETV points cover just two {\it TESS} sectors of $\sim$50 days, or a little more than a quarter of the outer orbit, they are a useful complement to the RV data since the ETV points measure the motion of the EB, while the RVs follow the tertiary star.

Gaia gives an outer orbital period of 184.1 days compared to our value of 184.35 days, and the corresponding outer eccentricities are essentially the same at 0.26.  So, again, these are in good agreement.

\subsection{TIC 294803663: RVs, ETVs, Gaia results, and pulsations}

We obtained six RV data points from the CHIRON spectrometer mounted on the 1.5 m SMARTS telescope at CTIO in Chile \citep{tokovinin13}. Spectra with a resolution of 25,000 were taken in the service mode and processed by the standard pipeline \citep{paredes21}. RVs of the slowly-rotating tertiary component were derived by cross-correlation with a binary mask based on the solar spectrum. Broad lines of the EB components are not detectable.

\begin{figure}
\begin{center}
\includegraphics[width=0.99 \columnwidth]{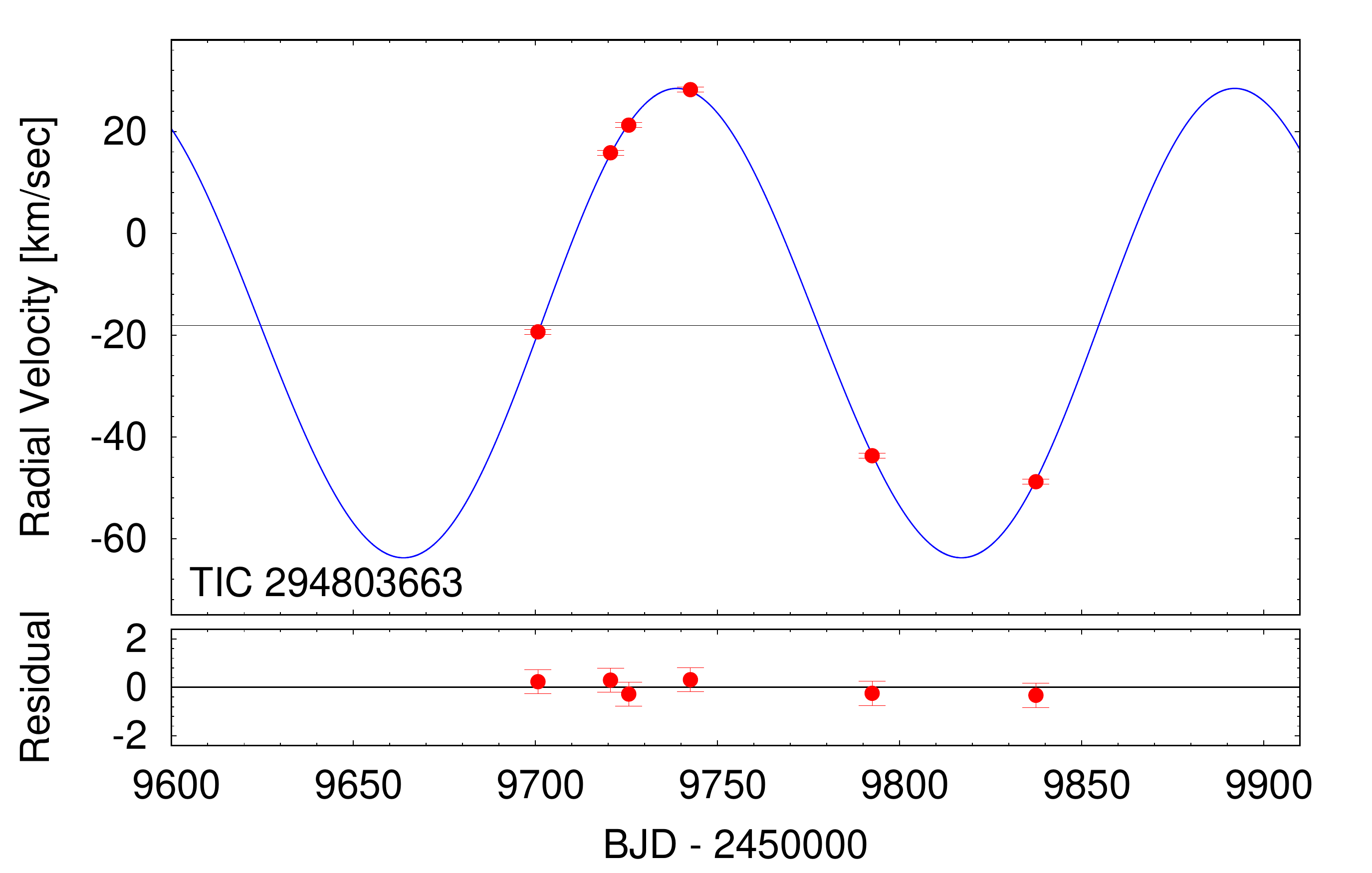}
\includegraphics[width=0.99 \columnwidth]{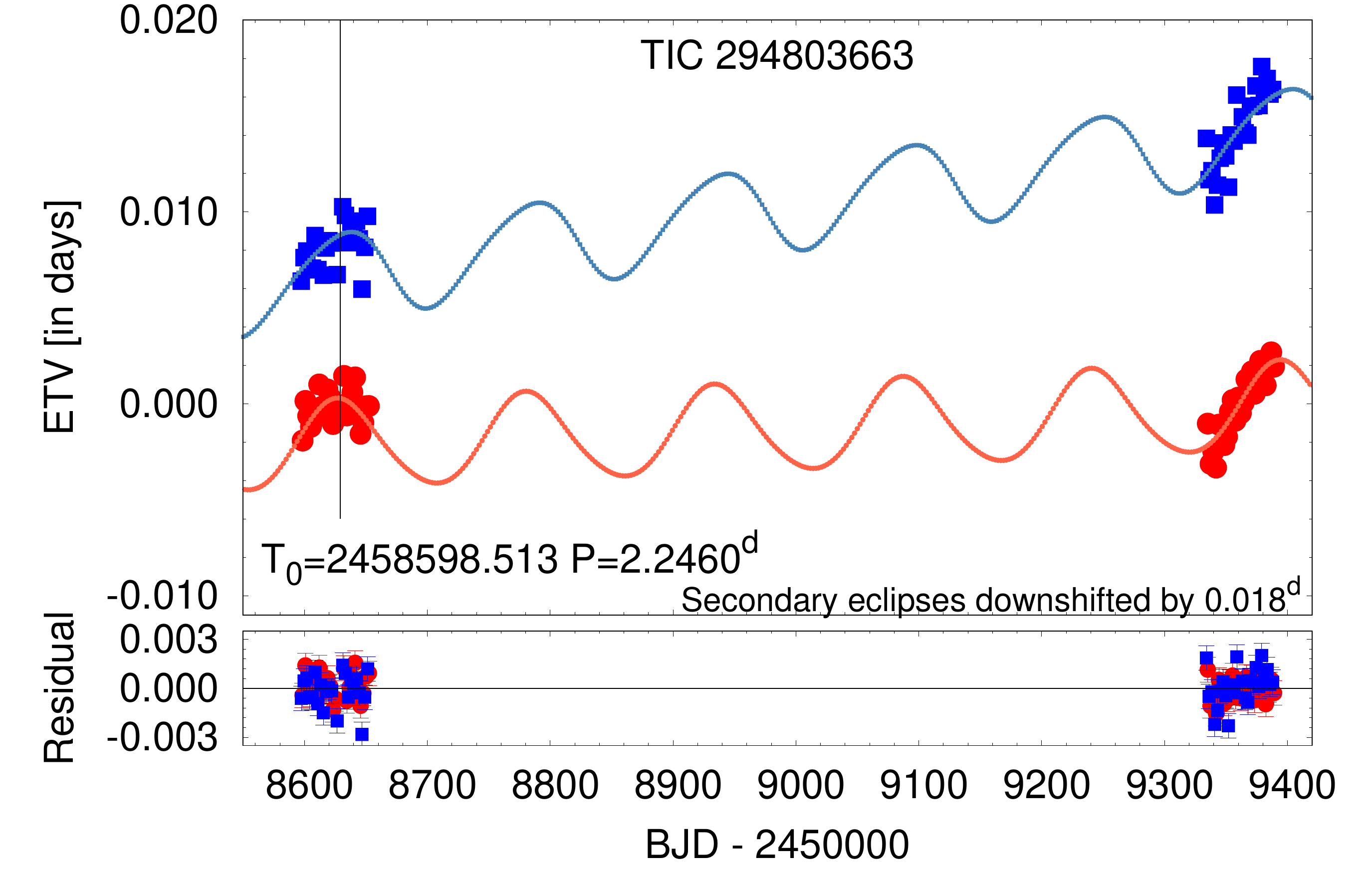}
\caption{Radial velocity and eclipse timing data for TIC 294803663. The RV data are from CHIRON at CTIO.  The model fits are the superposed smooth curves.  Red and blues points (and curves) represent primary and secondary eclipses of the inner eclipsing binary, respectively.  Vertical line represents the location of the third-body eclipse event observed by \textit{TESS}. Note, for a better visibility the secondary ETV curve is downshifted by $0\fd018$.}
\label{fig:294803663}
\end{center}
\end{figure}

This target was observed by {\it TESS} during two pairs of consecutive sectors (11-12 \& 38-39), separated by two years.  The ETV curve (Fig.~\ref{fig:294803663}) during each of the two $\sim$50-day intervals covers two different phases of the outer orbit. While this ETV curve would be insufficient to determine a unique orbit by itself, it is valuable as a supplement to the full orbital solution.

\begin{figure}
\begin{center}
\includegraphics[width=0.99 \columnwidth]{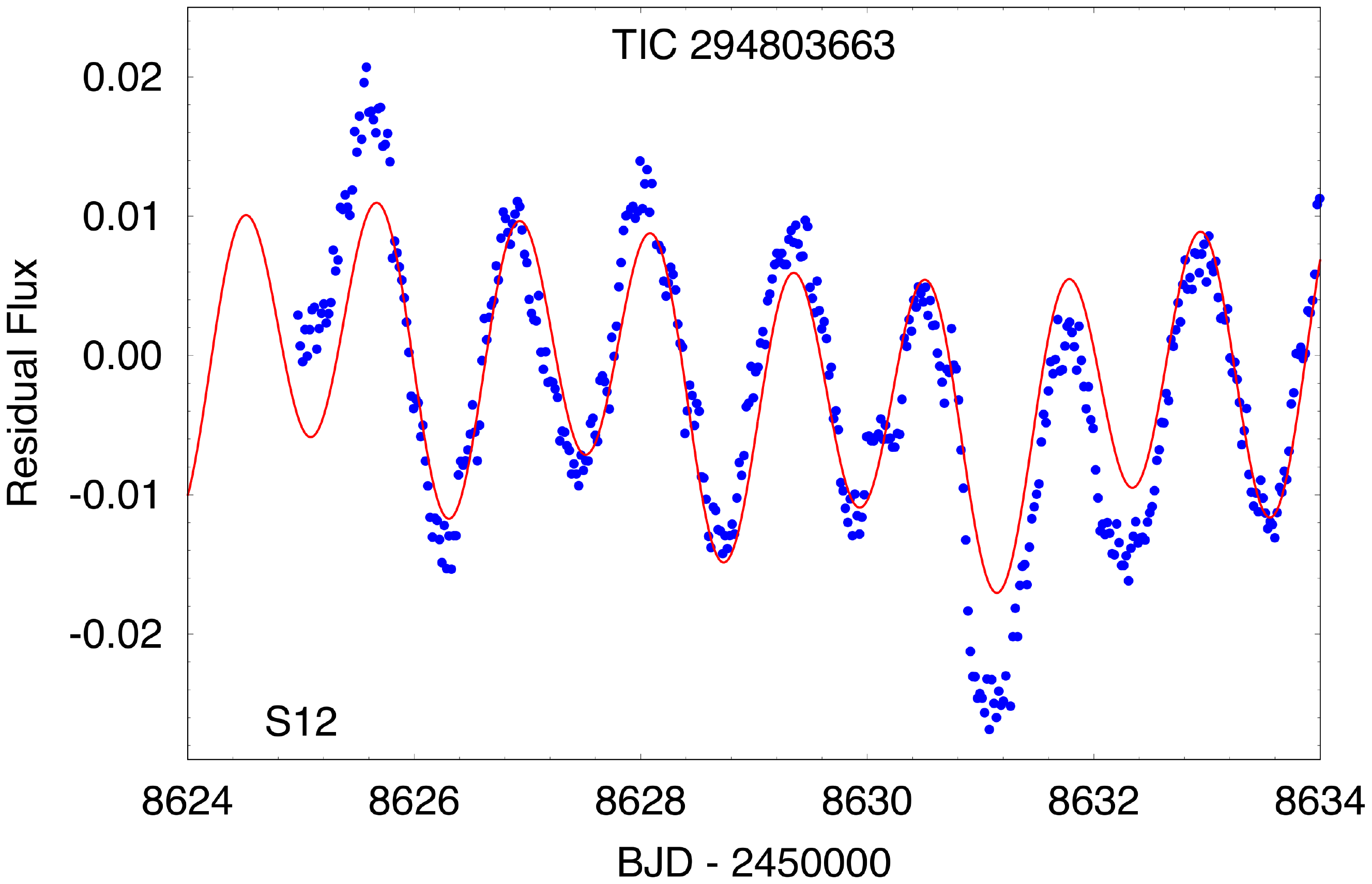}
\caption{Oscillations from the giant tertiary star in TIC 294803663, and Fourier fit.}
\label{fig:294803663_oscillation}
\end{center}
\end{figure}

In addition to the eclipses in this system, we find a strong and nearly periodic signal (see Fig.~\ref{fig:294803663_oscillation}), which we attribute to stellar oscillations in the giant tertiary star.  Four periods are required to fit the oscillations: 1.211 d, 2.427 d, 9.98 d, and 13.56 d.  The latter two periods are likely associated with the {\it TESS} instrument. while the first is just the higher harmonic of the second.


\subsection{TIC 332521671: RVs and ETVs} 

\begin{figure}
\begin{center}
\includegraphics[width=0.99 \columnwidth]{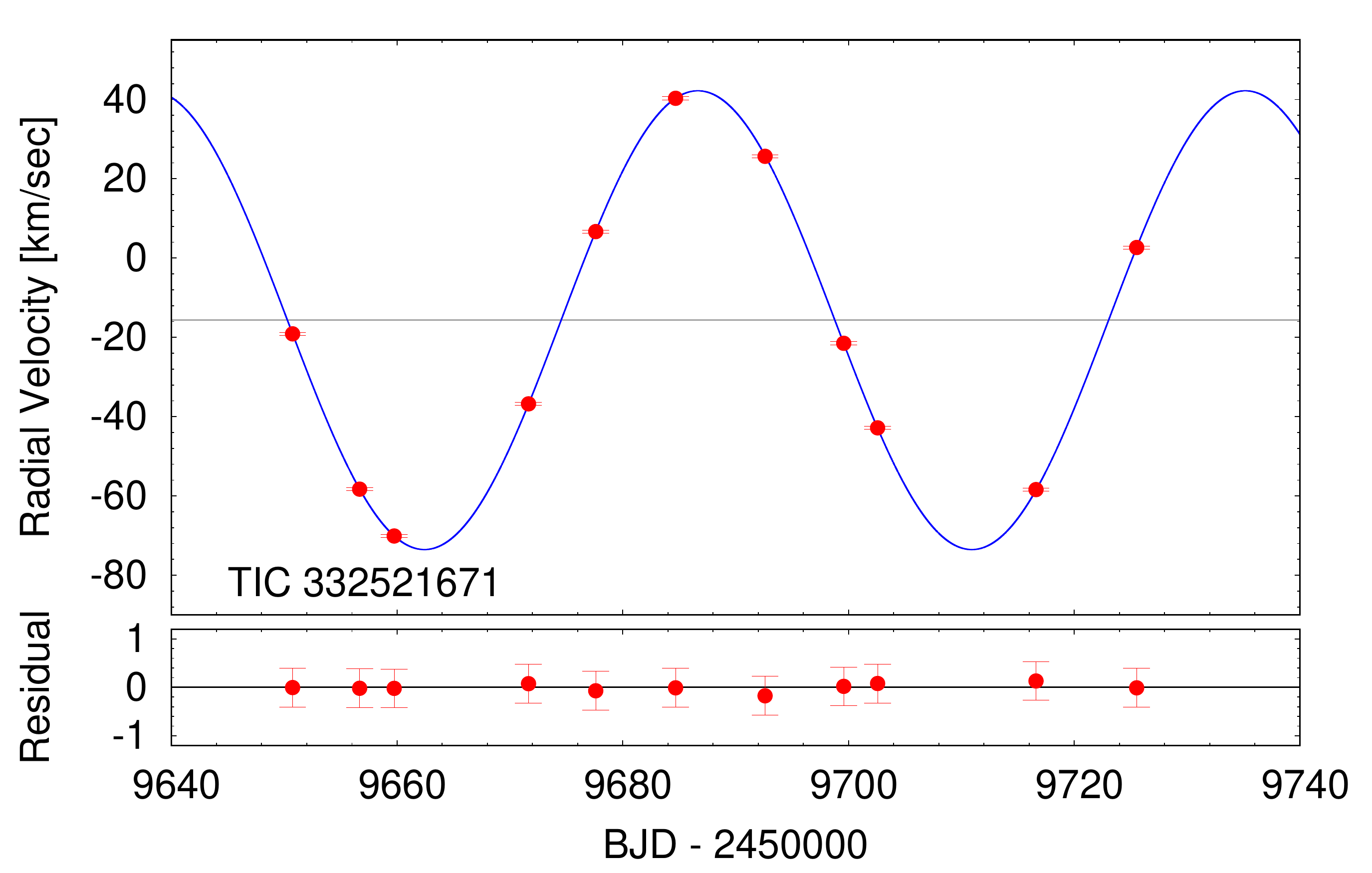} \hglue-0.25cm
\includegraphics[width=1.00 \columnwidth]{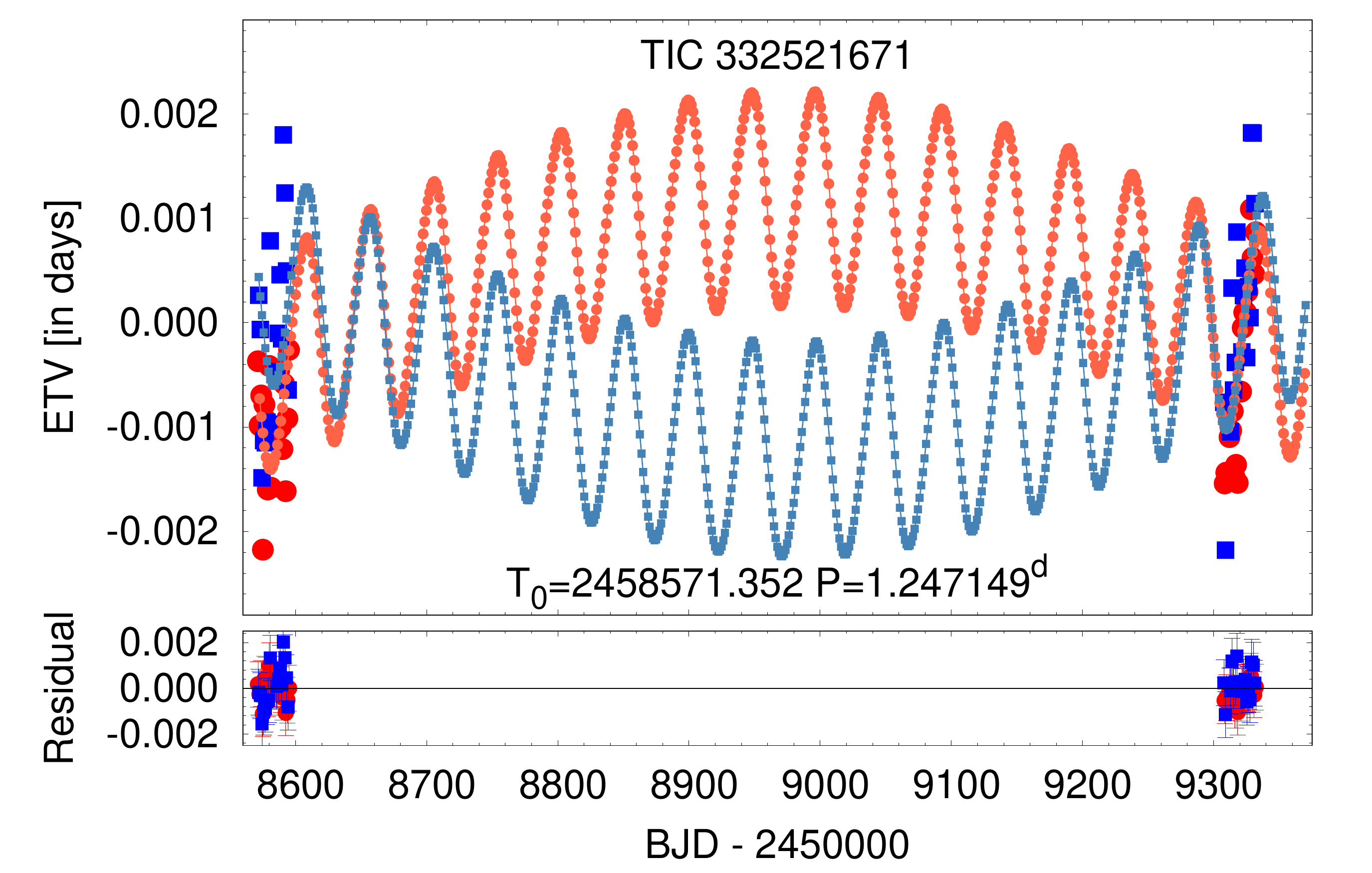}
\caption{Radial velocity and eclipse timing data for TIC332521671.  The RV data are from CHIRON at CTIO, while the ETV data are from {\it TESS} Sectors 10 and 37.}
\label{fig:TIC332521671}
\end{center}
\end{figure}

The radial velocity and eclipse timing data for TIC~332521671 are shown in Fig.~\ref{fig:TIC332521671}.  The RV data are from the CHIRON spectrometer \citep{tokovinin13}.   The complementary ETV data are from {\it TESS} Sectors 10 and 37.  The RV curve hints at a rather circular outer orbit, which indeed turns out to be the case for the 48.6-day outer orbit system.  The ETV curve is rather incomplete, but the full photodynamical model suggests that the inner binary might be undergoing forced apsidal precession with a rather short period of $\sim$4 years. We will discuss, however, in Sect.~\ref{Sect:TIC 332521671} that this effect most probably is not real, but only a pure numerical consequence of the two year-long gap between the Sector 10 and 37 ETV points, where nothing constrains our integrator.

\subsection{TIC 356324779: ETVs}

\begin{figure}
\begin{center}
\includegraphics[width=0.99 \columnwidth]{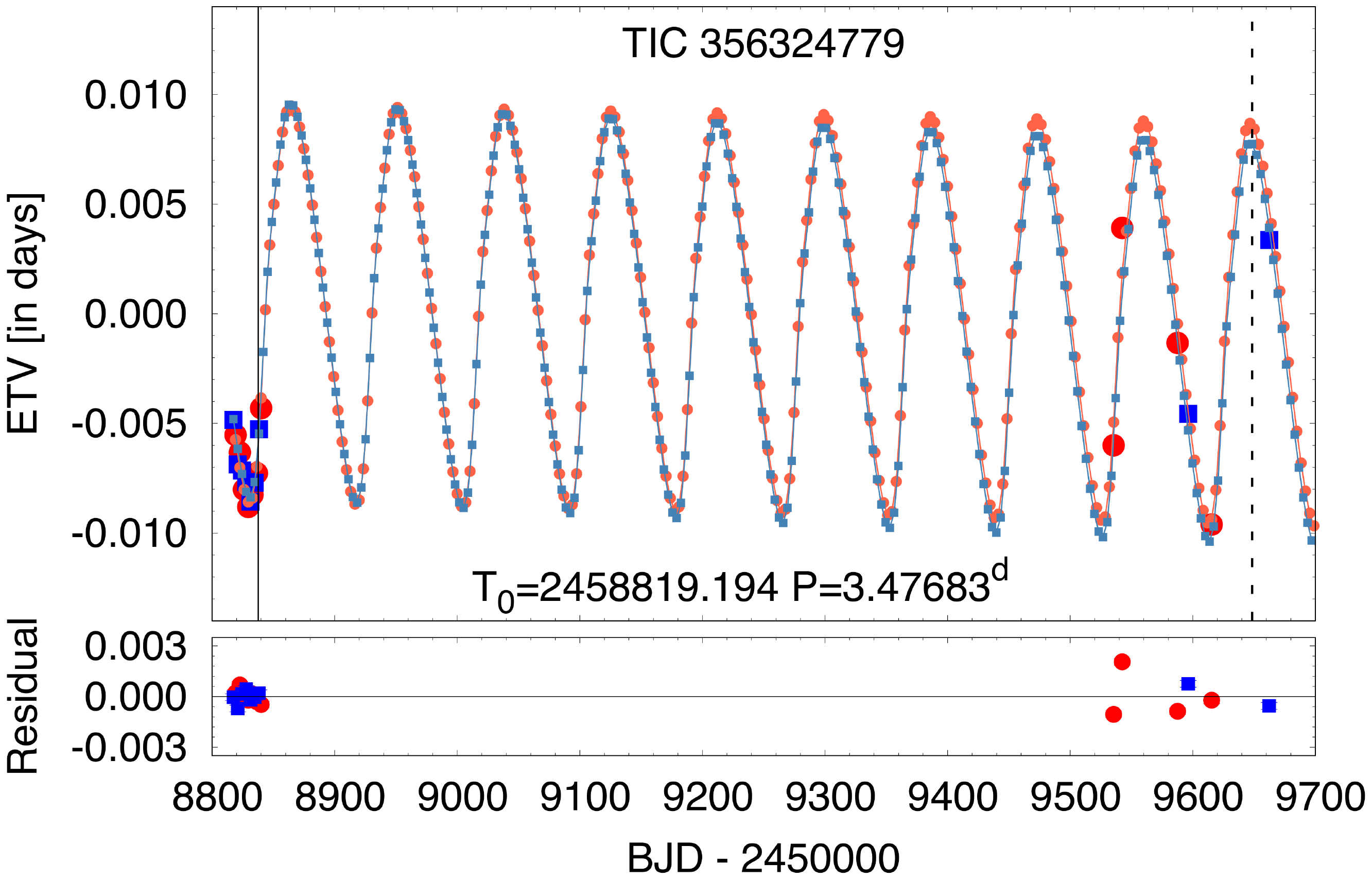}
\caption{Eclipse timing data for TIC 356324779. The earlier points are from Sector 19 of the {\it TESS} observations, while the later points are ground-based followup measurements with BAO80 and GAO80 telescopes.}
\label{fig:TIC356324779}
\end{center}
\end{figure}

The ETV curve for TIC 356324779 is shown in Figure \ref{fig:TIC356324779}.  It covers the Sector 19 {\it TESS} data as well as six followup ground-based inner EB eclipse observations with the BAO80 and GAO80 instruments. Note, besides the regular inner EB eclipses, one additional GAO80 observation serendipitously caught a section of a third-body eclipse (see the very last panel of Fig.~\ref{fig:triples}).

\section{Photodynamical Analysis for the System Parameters}
\label{sec:photodynamical}

We have carried out a photodynamical analysis of all nine of our triply eclipsing triples.  For this phase of the analysis of the system parameters, we utilize the software package {\sc Lightcurvefactory} \citep[see, e.g.][and references therein]{borkovitsetal19a,borkovitsetal20a}. As described in a number of previous papers, the code contains (i) emulators for the {\it TESS} lightcurve, the corresponding ETVs, and radial velocity curve (where available), (ii) a built-in numerical integrator to calculate the perturbed three-body coordinates and velocities of the three stars in the triple;  and (iii) an MCMC-based search routine for finding the best-fit system parameters.  The latter utilizes an implementation of the generic Metropolis-Hastings algorithm \citep[see, e.g., ][]{ford05}. The development, implementation, and use of {\sc Lightcurvefactory}, as well as the steps involved in the analysis process, have been explained in detail as the code was applied to a wide range of multistellar systems \citep{borkovitsetal18,borkovitsetal19a,borkovitsetal19b,borkovitsetal20a,borkovitsetal20b,borkovitsetal21,mitnyanetal20}.  {\sc Lightcurvefactory} has been used successfully to study compact and wider triple systems (with and without outer eclipses),  as well as quadruple systems with either a 2+2 or 2+1+1 configuration.

\begin{table*}
\centering
\caption{Input Information for the System Analysis$^a$}
\small
\begin{tabular}{lccccccc}
\hline
\hline
Object & {\it TESS} 3rd-Body & {\it TESS} EB & Archival Outer & SED Points$^c$ & ETV Curve$^d$ & RV Data$^e$ & Gaia Orbit$^f$ \\
 & Eclipse(s)$^a$ & Lightcurve$^a$ & Eclipses$^b$ &  &  & &  \\
\hline
TIC 47151245 & $\checkmark$  &$\checkmark$ & $\checkmark$ & $\checkmark$ & & \\
TIC 81525800 & $\checkmark$ & $\checkmark$ &$\checkmark$ & $\checkmark$ & $\checkmark$ & & \\ 
TIC 99013269 & $\checkmark$ & $\checkmark$ &  & $\checkmark$ & $\checkmark$ & $\checkmark$ & $\checkmark$ \\
TIC 229785001 & $\checkmark$ &  $\checkmark$ &  & $\checkmark$ & $\checkmark$ & & $\checkmark$\\
TIC 276162169 & $\checkmark$ & $\checkmark$ & $\checkmark$ & $\checkmark$ & $\checkmark$ & &\\
TIC 280883908 & $\checkmark$ & $\checkmark$ & $\checkmark$ & $\checkmark$ &$\checkmark$ & $\checkmark$ & $\checkmark$\\
TIC 294803663 & $\checkmark$ & $\checkmark$ & $\checkmark$ & $\checkmark$ & $\checkmark$ & $\checkmark$ & $\checkmark$\\
TIC 332521671 & $\checkmark$ & $\checkmark$ & $\checkmark$ & $\checkmark$ & $\checkmark$ & $\checkmark$ & \\
TIC 356324779 & $\checkmark$ & $\checkmark$ & $\checkmark$ & $\checkmark$ & $\checkmark$ & & \\
\hline
\label{tbl:input}  
\end{tabular}

\textit{Notes.}  (a) See Fig.~\ref{fig:triples}. (b) See, e.g., Fig.~\ref{fig:outer_fold}. (c) See, e.g., Fig.~\ref{fig:SEDs}. (d) See Figs.~\ref{fig:TIC81525800}, \ref{fig:TIC99013269}, \ref{fig:TIC229785001}, \ref{fig:TIC276162169}, \ref{fig:TIC280883908}, \ref{fig:294803663}, \ref{fig:TIC332521671}, and \ref{fig:TIC356324779}. (e) See Fig.~\ref{fig:TIC99013269}, \ref{fig:TIC280883908}, \ref{fig:294803663}, and \ref{fig:TIC332521671}. (f) \citet{babusiaux22}; \citet{gaia22}
\end{table*}

\begin{figure*}
\begin{center}
\includegraphics[width=0.32 \textwidth]{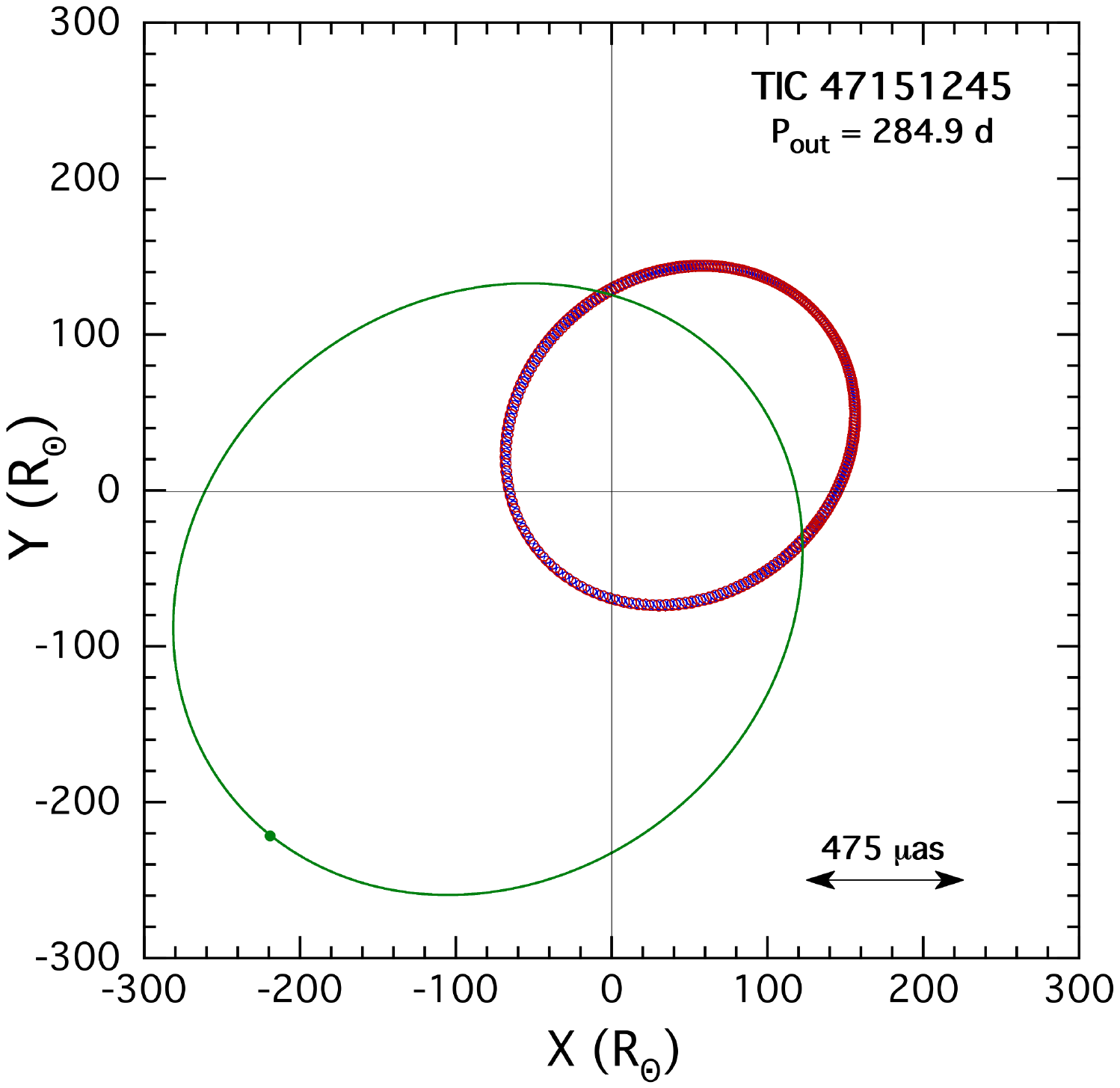}
\includegraphics[width=0.32 \textwidth]{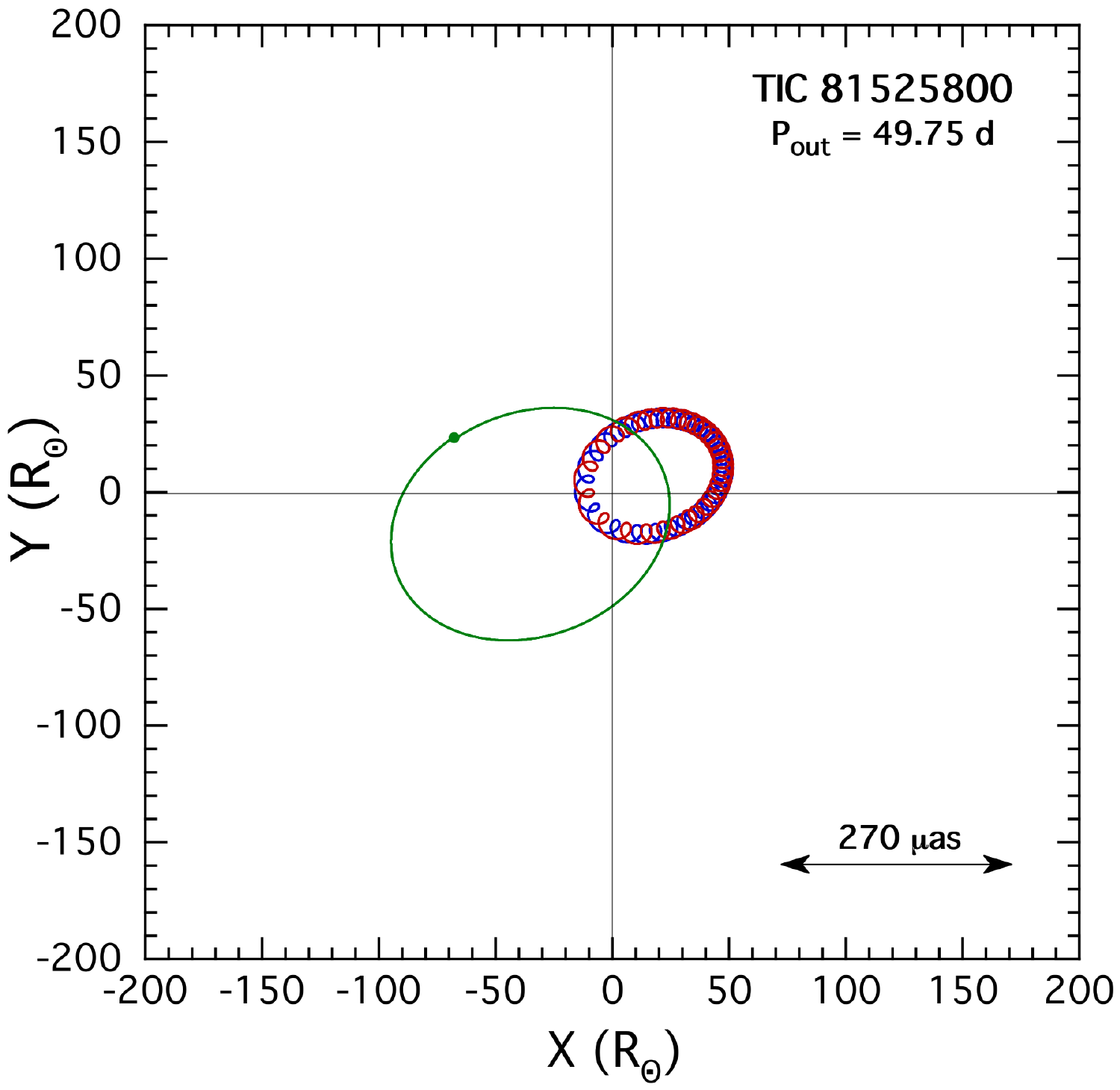} \hglue-0.1cm
\includegraphics[width=0.308 \textwidth]{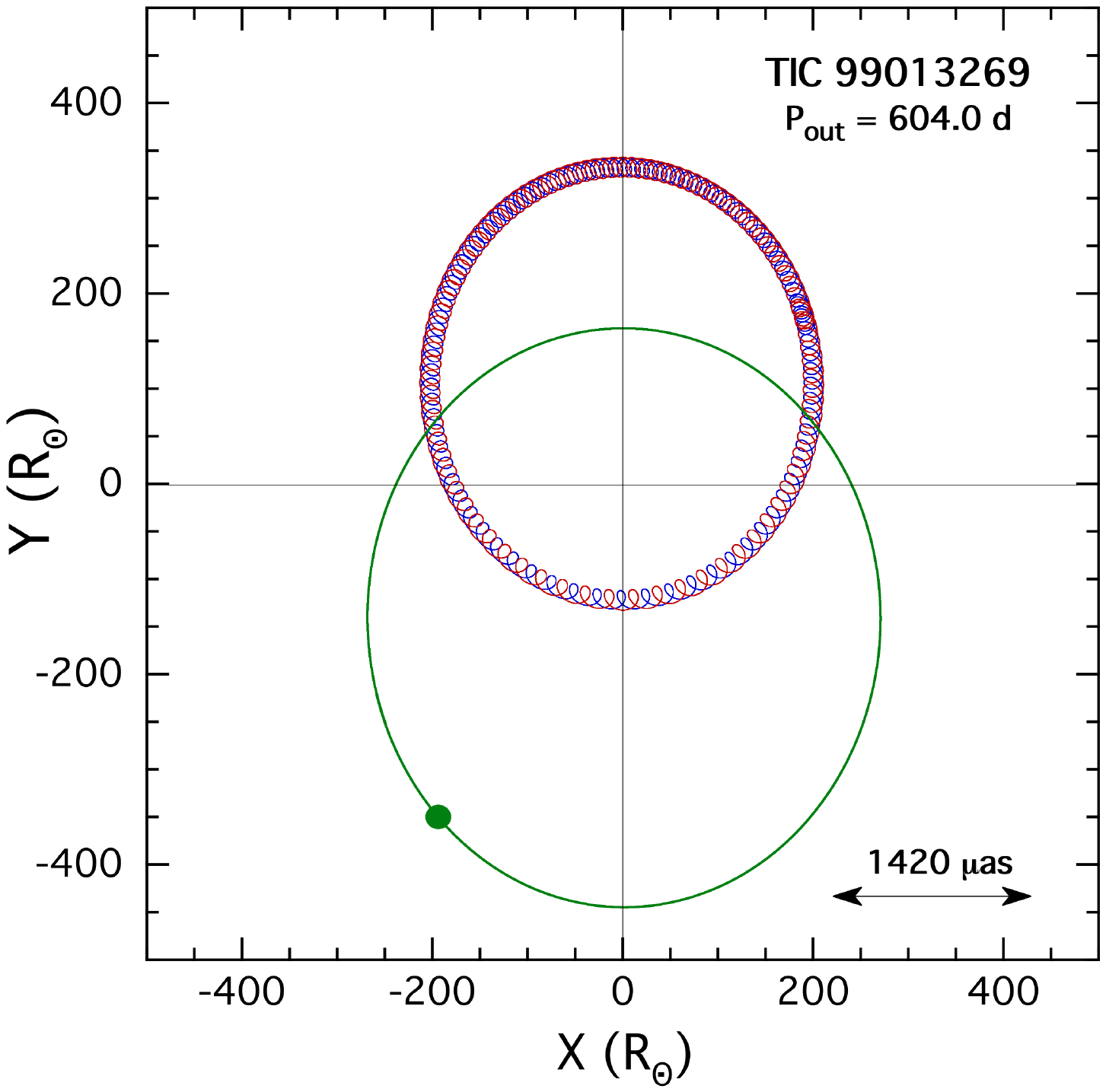} \hglue0.17cm
\includegraphics[width=0.322 \textwidth]{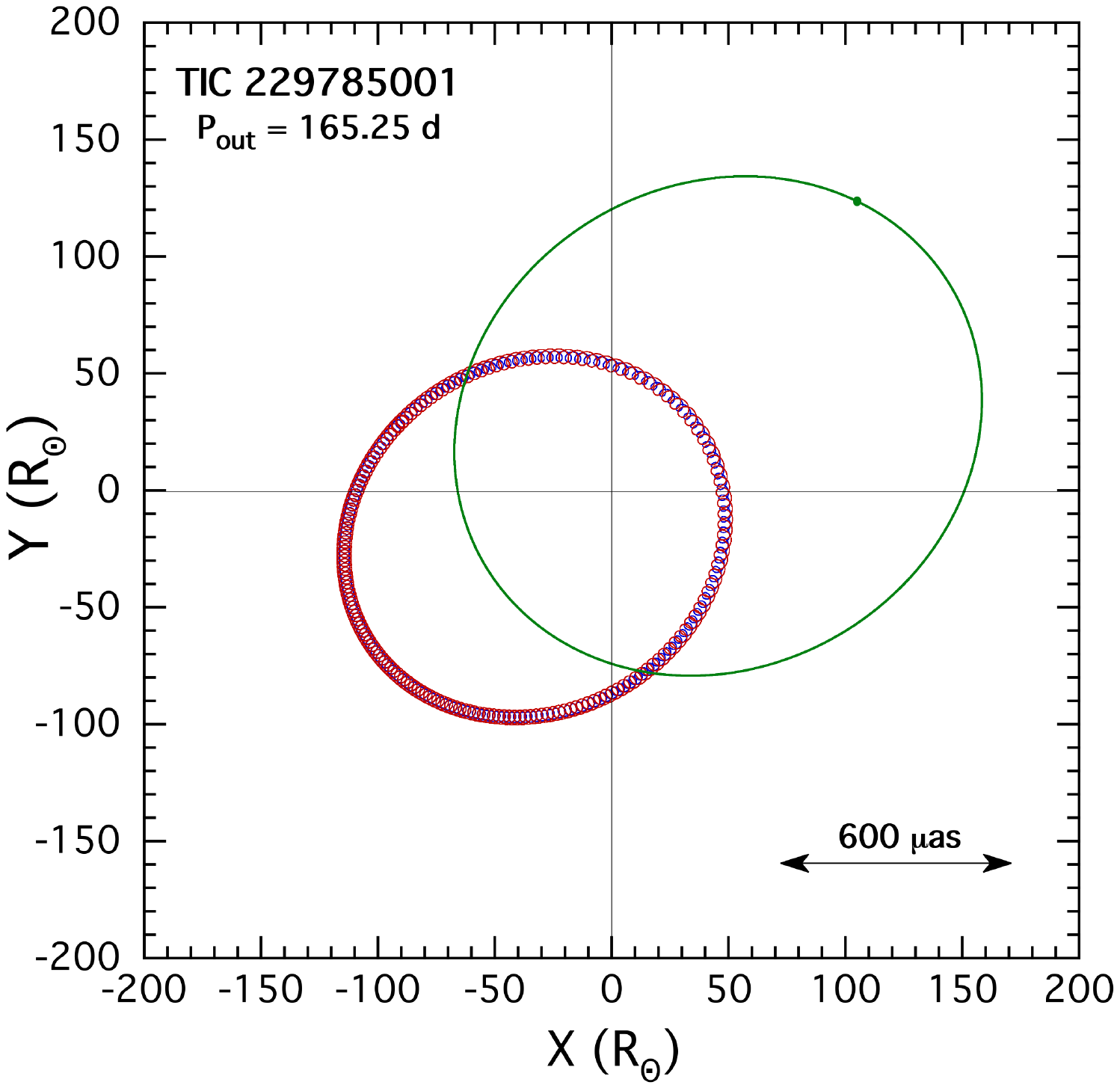} \hglue0.07cm
\includegraphics[width=0.32 \textwidth]{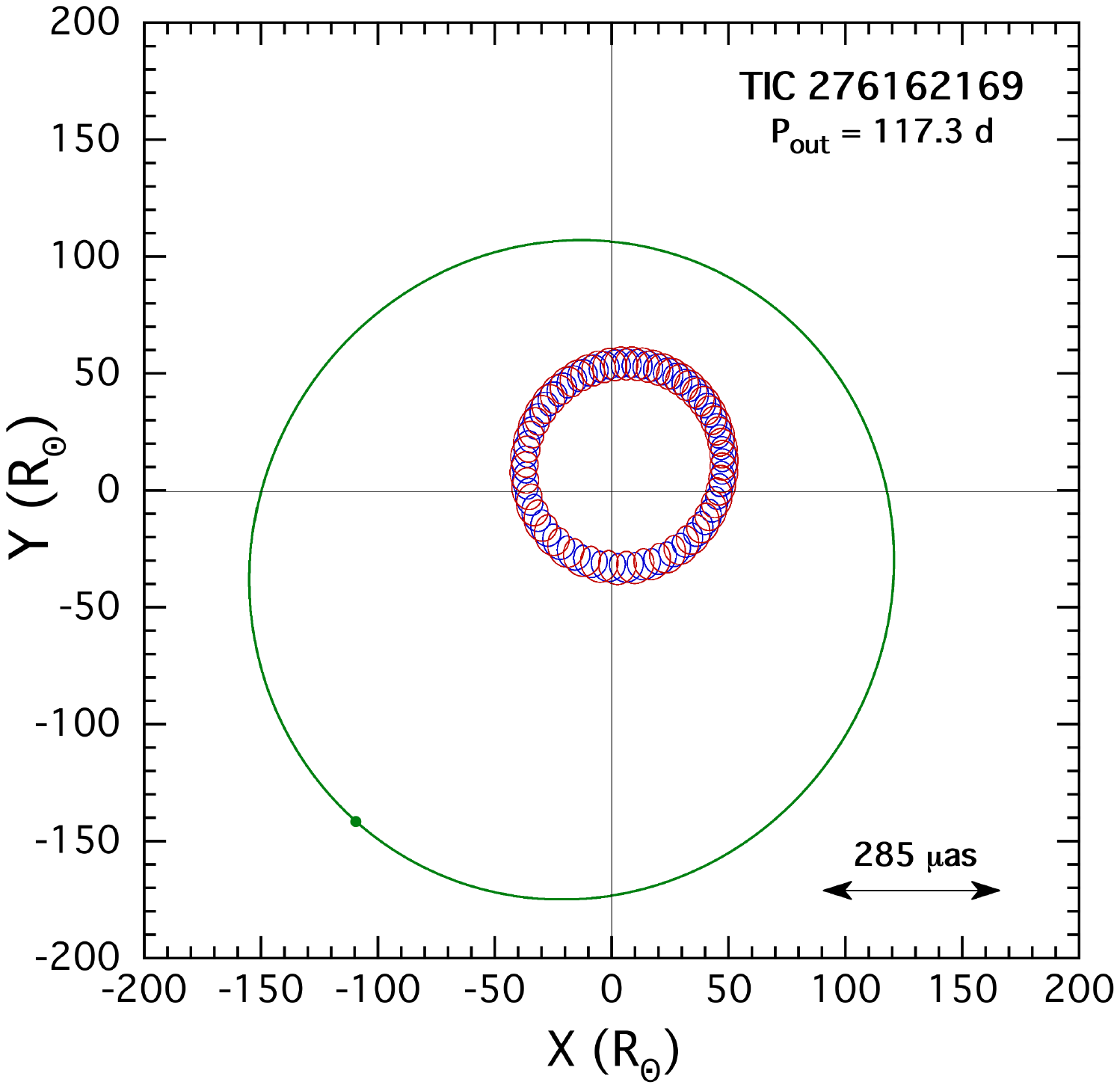} \hglue-0.08cm
\includegraphics[width=0.317 \textwidth]{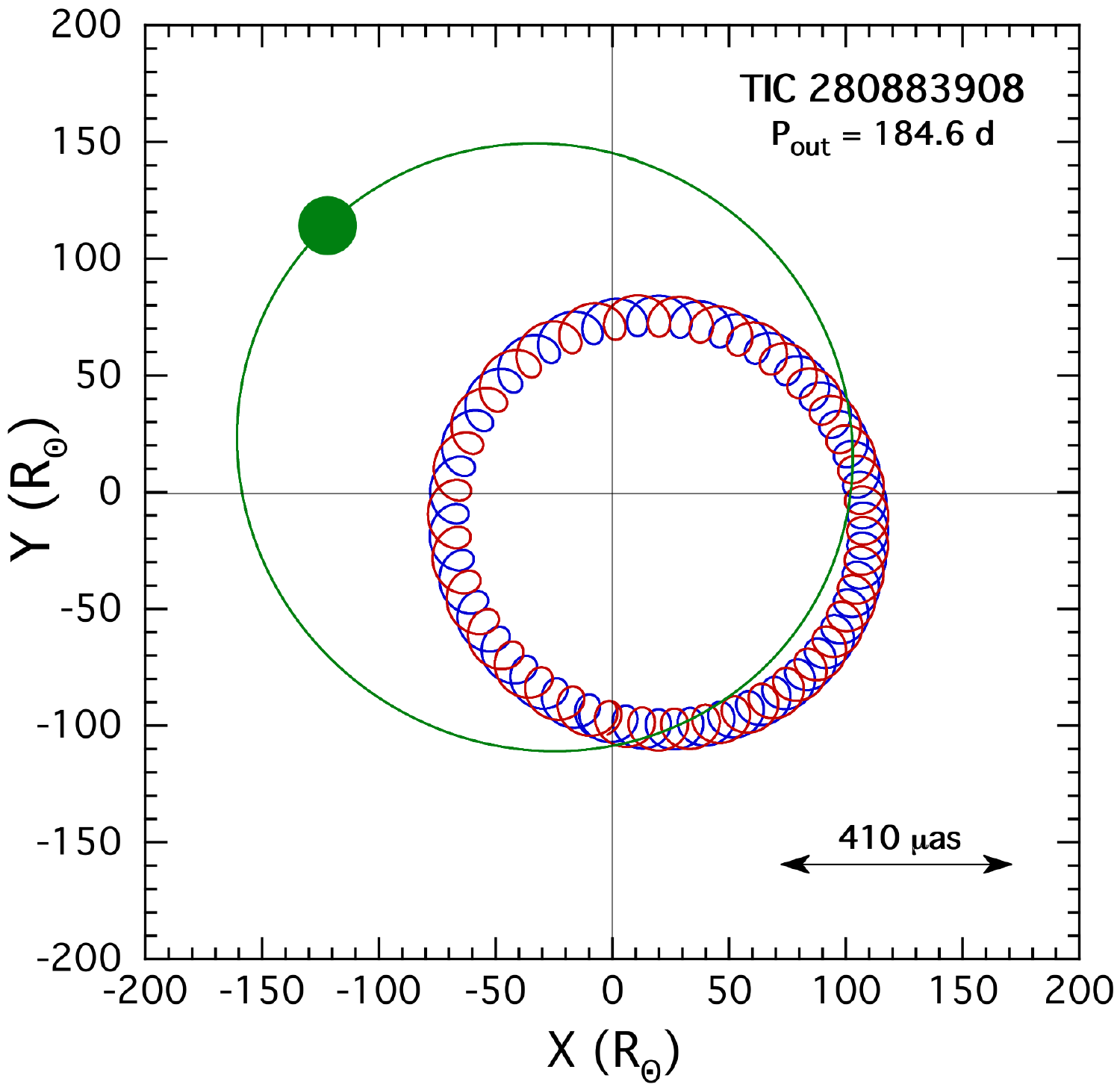} \hglue0.08cm
\includegraphics[width=0.322 \textwidth]{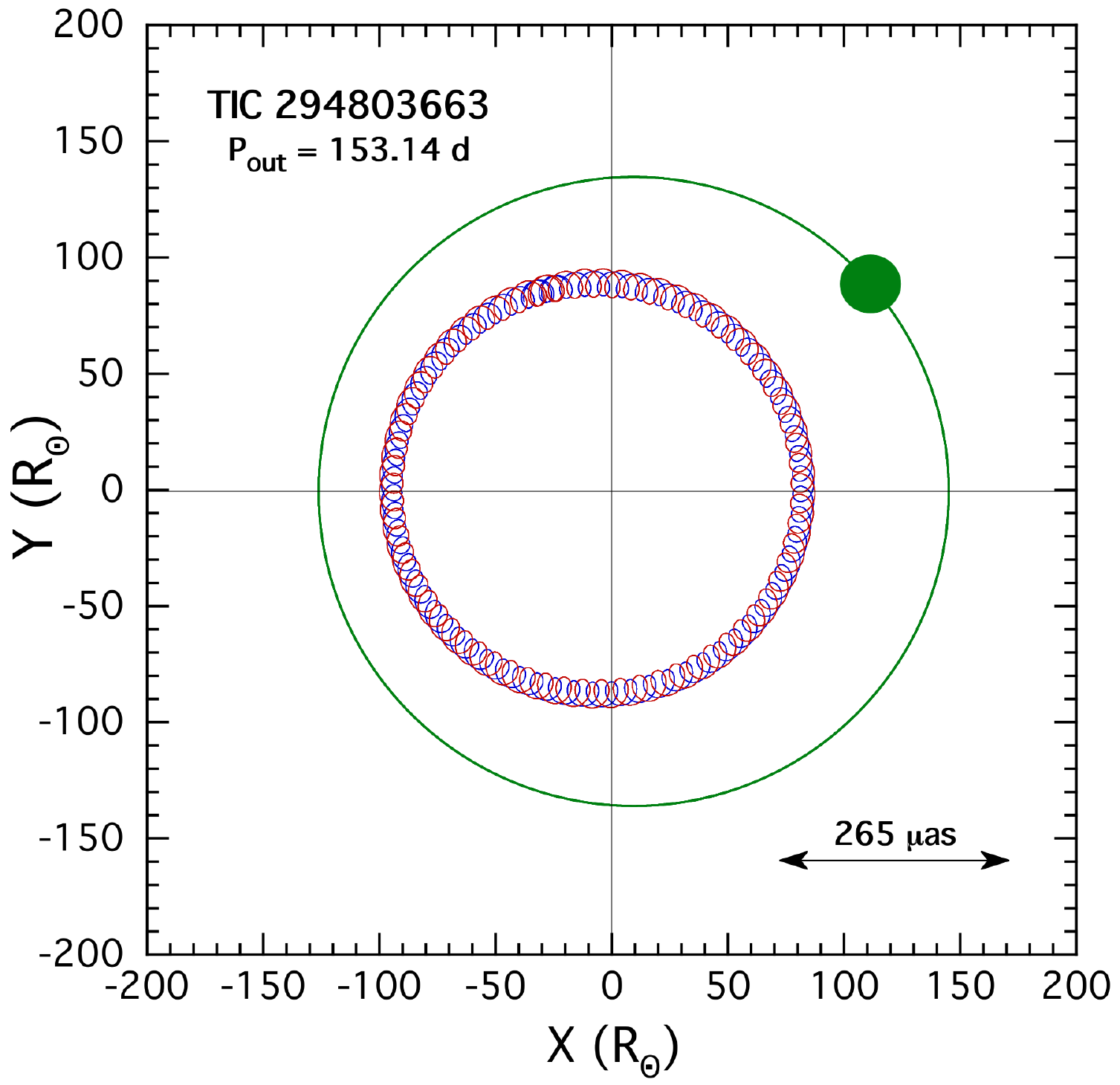} \hglue0.11cm
\includegraphics[width=0.322 \textwidth]{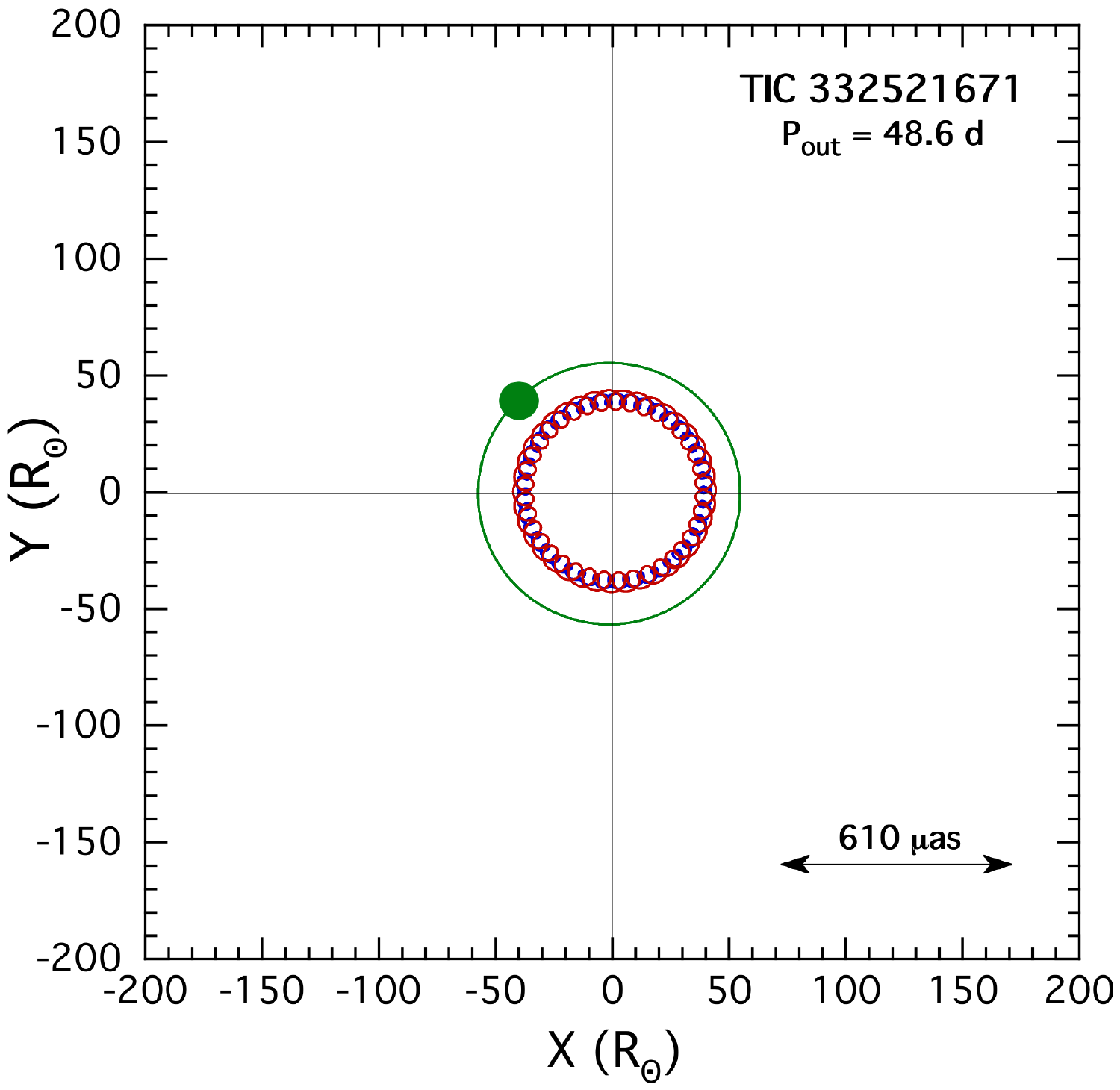} \hglue-0.08cm
\includegraphics[width=0.32 \textwidth]{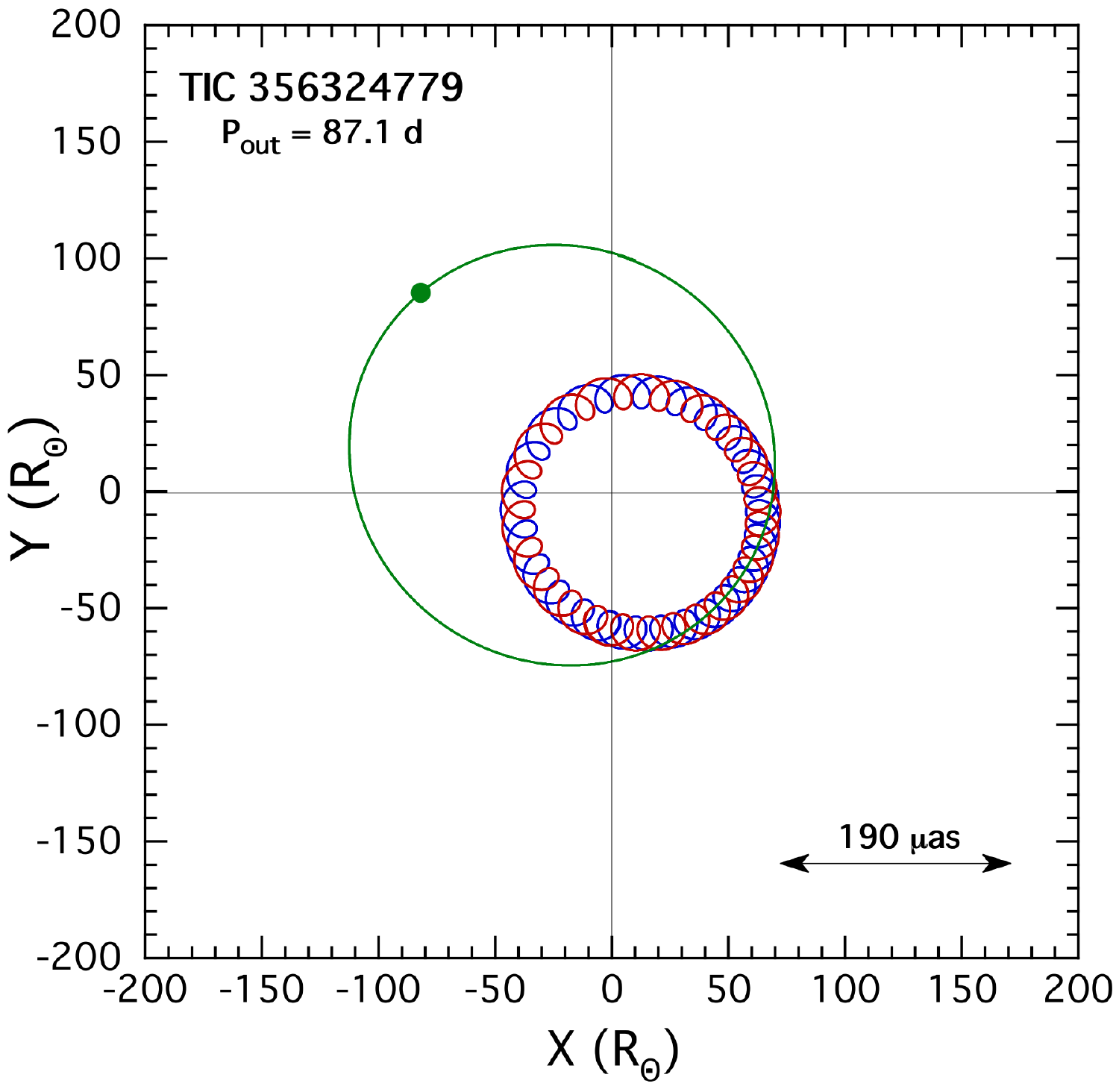}

\caption{The outer orbits of the nine triply eclipsing systems seen from above the orbital plane.  The stars are all moving counter-clockwise except for the tertiary in TIC 276162169 which is moving clockwise (i.e., in a retrograde fashion).  The observer is at $y \rightarrow -\infty$.  Red and blue tracks are for the primary and secondary stars in the EB, respectively, while the green track is that of the tertiary star.  The heavy filled green circle represents the size of the tertiary to scale on the plot.}
\label{fig:orbits}
\end{center}
\end{figure*} 

\begin{table}
\centering
\caption{Definitions of Triple System Parameters in Tables~\ref{tab:syntheticfit_TIC47151245+81525800}--\ref{tab:syntheticfit_TIC356324779}}
\label{tbl:definitions}
\small
\begin{tabular}{lc}
\hline
\hline
Parameter$^a$ & Definition   \\
\hline
$t_0$ & Epoch time for osculating elements    \\
$P$ & Orbital period  \\ 
$a$ & Orbital semimajor axis  \\
$e$ & Orbital eccentricity \\
$\omega$ & Argument of periastron (of secondary) \\
$i$ & Orbital inclination angle \\
$\mathcal{T}_0^\mathrm{inf/sup}$ & Time of conjunction of the secondary$^b$ \\
$\tau$ & Time of periastron passage  \\
$\Omega$ & Longitude of the node relative to inner orbit \\
$i_{\rm mut}$ & Mutual inclination angle$^c$   \\
$q$ & Mass ratio (secondary/primary)  \\ 
$K_\mathrm{pri}$ & ``K'' velocity amplitude of primary \\
$K_\mathrm{sec}$ & ``K'' velocity amplitude of secondary \\
$R/a$ & Stellar radius divided by semimajor axis \\
$T_{\rm eff}/T_{\rm eff,Aa}$ & Temperature relative to EB primary \\
fractional flux  & Stellar contribution in the given band \\
$m$ & Stellar mass  \\
$R$ & Stellar radius   \\
$T_\mathrm{eff}$ & Stellar effective temperature  \\ 
$L_\mathrm{bol}$ & Stellar bolometric luminosity  \\
$M_\mathrm{bol}$ & Stellar absolute bolometric magnitude \\
$M_V$ & Stellar absolute visual magnitude \\
$\log g$ & log surface gravity (cgs units) \\
$[M/H]$ & log metallicity abundance to H, by mass \\
$E(B-V)$ & Color excess in B-V bands  \\
extra light, $\ell_4$  & Contanimating flux in the given band   \\
$(M_V)_\mathrm{tot}$ & System absolute visual magnitude   \\ 
distance & Distance to the source  \\
\hline   
\end{tabular}

\textit{Notes}. (a) The units for the parameters are given in Tables~\ref{tab:syntheticfit_TIC47151245+81525800}--\ref{tab:syntheticfit_TIC356324779}. (b) The superscript of ``inf/sup'' indicates inferior vs.~superior conjunctions. (c) More explicitly, this is the angle between the orbital planes of the inner binary and the outer triple orbit.

\end{table} 

\begin{table*}
 \centering
\caption{Orbital and astrophysical parameters of TIC 47151245 and TIC 81525800 from the joint photodynamical \textit{TESS}, ETV, SED and \texttt{PARSEC} isochrone solution. Note that the orbital parameters are instantaneous, osculating orbital elements and are given for epoch $t_0$ (first row).  Therefore, the orbital periods, in particular, cannot be used for predicting the times of future eclipses; see Table \ref{tab:ephemerides} for the latter, and \citet{kostovetal21} for a more general discussion.}
 \label{tab:syntheticfit_TIC47151245+81525800}
\scalebox{0.93}{\begin{tabular}{@{}lllllll}
\hline
 & \multicolumn{3}{c}{TIC\,47151245} & \multicolumn{3}{c}{TIC\,81525800} \\
\hline
\multicolumn{7}{c}{orbital elements} \\
\hline
   & \multicolumn{3}{c}{subsystem} & \multicolumn{3}{c}{subsystem}  \\
   & \multicolumn{2}{c}{Aa--Ab} & A--B & \multicolumn{2}{c}{Aa--Ab} & A--B \\
  \hline
  $t_0$ [BJD - 2400000] & \multicolumn{3}{c}{58626.5} & \multicolumn{3}{c}{59474.0} \\
  $P$ [days] & \multicolumn{2}{c}{$1.202508_{-0.000062}^{+0.000070}$} & $284.374_{-0.568}^{+0.491}$ & \multicolumn{2}{c}{$1.6131_{-0.0065}^{+0.0043}$} & $49.75_{-0.14}^{+0.36}$  \\
  $a$ [R$_\odot$] & \multicolumn{2}{c}{$7.42_{-0.22}^{+0.16}$} & $329.0_{-9.8}^{+5.9}$ & \multicolumn{2}{c}{$8.16_{-0.09}^{+0.07}$} & $91.8_{-1.0}^{+0.8}$ \\
  $e$ & \multicolumn{2}{c}{$0.0022_{-0.0012}^{+0.0018}$} & $0.480_{-0.035}^{+0.027}$ & \multicolumn{2}{c}{$0.0202_{-0.0012}^{+0.00231}$} & $0.614_{-0.006}^{+0.017}$ \\
  $\omega$ [deg] & \multicolumn{2}{c}{$269_{-198}^{+21}$} & $218.6_{-5.7}^{+5.7}$ & \multicolumn{2}{c}{$144.1_{-3.9}^{+4.2}$} & $201.0_{-3.7}^{+1.2}$ \\ 
  $i$ [deg] & \multicolumn{2}{c}{$64.43_{-1.11}^{+0.62}$} & $90.318_{-0.046}^{+0.039}$ & \multicolumn{2}{c}{$90.63_{-3.50}^{+2.17}$} & $87.55_{-0.17}^{+0.34}$ \\
  $\mathcal{T}_0^\mathrm{inf}$ [BJD - 2400000]& \multicolumn{2}{c}{$58627.4569_{-0.0006}^{+0.0007}$} & $59368.6719_{-0.0054}^{+0.0061}$ & \multicolumn{2}{c}{$59475.5227_{-0.0047}^{+0.0025}$} & $59475.924_{-0.012}^{+0.010}$ \\
  $\tau$ [BJD - 2400000]& \multicolumn{2}{c}{$58626.648_{-0.374}^{+0.759}$} & $59355.011_{-0.882}^{+0.941}$ & \multicolumn{2}{c}{$59474.966_{-0.018}^{+0.018}$} & $59473.713_{-0.034}^{+0.037}$ \\
  $\Omega$ [deg] & \multicolumn{2}{c}{$0.0$} & $32.58_{-1.06}^{+1.11}$ & \multicolumn{2}{c}{$0.0$} & $3.14_{-2.99}^{+1.42}$ \\
  $i_\mathrm{mut}$ [deg] & \multicolumn{3}{c}{$40.88_{-0.96}^{+0.91}$} & \multicolumn{3}{c}{$4.99_{-3.77}^{+1.57}$} \\
  \hline
  mass ratio $[q=m_\mathrm{sec}/m_\mathrm{pri}]$ & \multicolumn{2}{c}{$0.881_{-0.018}^{+0.021}$} & $0.546_{-0.006}^{+0.010}$ & \multicolumn{2}{c}{$0.969_{-0.021}^{+0.057}$} & $0.498_{-0.008}^{+0.016}$ \\
 $K_\mathrm{pri}$ [km\,s$^{-1}$] & \multicolumn{2}{c}{$131.9_{-4.1}^{+2.7}$} & $23.4_{-0.5}^{+0.7}$ & \multicolumn{2}{c}{$126.5_{-1.7}^{+1.8}$} & $39.5_{-0.9}^{+0.9}$ \\ 
 $K_\mathrm{sec}$ [km\,s$^{-1}$] & \multicolumn{2}{c}{$149.7_{-4.1}^{+2.7}$} & $42.7_{-0.7}^{+1.4}$ & \multicolumn{2}{c}{$130.5_{-5.5}^{+1.5}$} & $78.5_{-0.5}^{+1.2}$ \\ 
  \hline  
\multicolumn{7}{c}{stellar parameters} \\
\hline
   & Aa & Ab &  B & Aa & Ab &  B \\
  \hline
 \multicolumn{7}{c}{Relative quantities} \\
  \hline
 fractional radius [$R/a$]  & $0.3201_{-0.0073}^{+0.0063}$ & $0.2458_{-0.0099}^{+0.0110}$ & $0.00790_{-0.00018}^{+0.00018}$ & $0.2274_{-0.0172}^{+0.0060}$ & $0.2098_{-0.0069}^{+0.0169}$ & $0.0192_{-0.0009}^{+0.0017}$ \\
 temperature relative to $(T_\mathrm{eff})_\mathrm{Aa}$ & $1$ & $0.9689_{-0.0142}^{+0.0103}$ & $0.9963_{-0.0080}^{+0.0033}$ & $1$ & $0.9983_{-0.0032}^{+0.0018}$ & $0.9999_{-0.0012}^{+0.0014}$ \\
 fractional flux [in \textit{TESS}-band] & $0.3577_{-0.0154}^{+0.0109}$ & $0.1959_{-0.0154}^{+0.0169}$ & $0.4263_{-0.0203}^{+0.0207}$ & $0.3395_{-0.0416}^{+0.0214}$ & $0.2882_{-0.0186}^{+0.0517}$ & $0.3132_{-0.0279}^{+0.0476}$ \\
 fractional flux [in $R_\mathrm{C}$-band] & $-$ & $-$ & $-$ & $0.3441_{-0.0302}^{+0.0224}$ & $0.3012_{-0.0293}^{+0.0570}$ & $0.3301_{-0.0208}^{+0.0378}$ \\
 \hline
 \multicolumn{7}{c}{Physical Quantities} \\
  \hline 
 $m$ [M$_\odot$] & $2.026_{-0.173}^{+0.119}$ & $1.778_{-0.170}^{+0.121}$ & $2.111_{-0.212}^{+0.086}$ & $1.430_{-0.093}^{+0.034}$ & $1.376_{-0.019}^{+0.025}$ & $1.407_{-0.078}^{+0.046}$ \\
 $R$ [R$_\odot$] & $2.026_{-0.173}^{+0.119}$ & $1.833_{-0.128}^{+0.100}$ & $2.586_{-0.066}^{+0.078}$ & $1.864_{-0.168}^{+0.050}$ & $1.723_{-0.062}^{+0.106}$ & $1.763_{-0.096}^{+0.157}$ \\
 $T_\mathrm{eff}$ [K]& $8569_{-383}^{+206}$ & $8220_{-207}^{+291}$ & $8563_{-485}^{+207}$ & $6332_{-112}^{+100}$ & $6326_{-141}^{+100}$ & $6333_{-114}^{+97}$ \\
 $L_\mathrm{bol}$ [L$_\odot$] & $26.66_{-5.10}^{+5.72}$ & $13.77_{-3.00}^{+3.88}$ & $32.81_{-8.01}^{+4.87}$ & $5.049_{-1.149}^{+0.557}$ & $4.305_{-0.299}^{+0.285}$ & $4.475_{-0.674}^{+1.137}$ \\
 $M_\mathrm{bol}$ & $1.21_{-0.21}^{+0.23}$ & $1.92_{-0.27}^{+0.27}$ & $0.98_{-0.15}^{+0.30}$ & $3.01_{-0.11}^{+0.28}$ & $3.19_{-0.07}^{+0.08}$ & $3.14_{-0.25}^{+0.18}$ \\
 $M_V           $ & $1.19_{-0.18}^{+0.20}$ & $1.89_{-0.24}^{+0.27}$ & $0.97_{-0.12}^{+0.27}$ & $2.99_{-0.12}^{+0.30}$ & $3.17_{-0.07}^{+0.08}$ & $3.12_{-0.24}^{+0.19}$ \\
 $\log g$ [dex] & $3.988_{-0.016}^{+0.019}$ & $4.163_{-0.027}^{+0.023}$ & $3.925_{-0.025}^{+0.021}$ & $4.055_{-0.019}^{+0.049}$ & $4.112_{-0.062}^{+0.023}$ & $4.090_{-0.060}^{+0.032}$ \\
 \hline
\multicolumn{7}{c}{Global system parameters} \\
  \hline
$\log$(age) [dex] &\multicolumn{3}{c}{$8.859_{-0.062}^{+0.130}$} &\multicolumn{3}{c}{$9.388_{-0.054}^{+0.082}$} \\
$[M/H]$  [dex]    &\multicolumn{3}{c}{$-0.069_{-0.111}^{+0.108}$} &\multicolumn{3}{c}{$0.172_{-0.052}^{+0.106}$} \\
$E(B-V)$ [mag]    &\multicolumn{3}{c}{$0.115_{-0.048}^{+0.026}$} &\multicolumn{3}{c}{$0.278_{-0.033}^{+0.024}$} \\
extra light $\ell_4$ [in \textit{TESS}-band] & \multicolumn{3}{c}{$0.019_{-0.014}^{+0.019}$} & \multicolumn{3}{c}{$0.042_{-0.030}^{+0.043}$} \\
extra light $\ell_4$ [in $R_\mathrm{C}$-band] & \multicolumn{3}{c}{$...$} & \multicolumn{3}{c}{$0.008_{-0.006}^{+0.013}$} \\
$(M_V)_\mathrm{tot}$  &\multicolumn{3}{c}{$0.09_{-0.16}^{+0.26}$} &\multicolumn{3}{c}{$1.90_{-0.12}^{+0.16}$} \\
distance [pc]           &\multicolumn{3}{c}{$983_{-52}^{+36}$} &\multicolumn{3}{c}{$1718_{-59}^{+61}$} \\  
\hline
\end{tabular}}
\end{table*}   

\begin{table*}
 \centering
\caption{The same as in Table~\ref{tab:syntheticfit_TIC47151245+81525800} above, but for TIC 99013269 and TIC 229785001.}
 \label{tab:syntheticfit_TIC99013269+229785001}
\scalebox{0.82}{\begin{tabular}{@{}llllllll}
\hline
 & \multicolumn{3}{c}{TIC\,99013269} & \multicolumn{4}{c}{TIC\,229785001} \\
\hline
\multicolumn{8}{c}{orbital elements} \\
\hline
   & \multicolumn{3}{c}{subsystem} & \multicolumn{4}{c}{subsystem}  \\
   & \multicolumn{2}{c}{Aa--Ab} & A--B & \multicolumn{2}{c}{Aa--Ab} & A--B & AB--C \\
  \hline
  $t_0$ [BJD - 2400000] & \multicolumn{3}{c}{58683.0} & \multicolumn{4}{c}{58683.0} \\
  $P$ [days] & \multicolumn{2}{c}{$6.534444_{-0.000014}^{+0.000016}$} & $604.2425_{-0.0068}^{+0.0067}$ & \multicolumn{2}{c}{$0.929762_{-0.000015}^{+0.000016}$} & $165.37_{-0.05}^{+0.05}$ & $3254_{-159}^{+136}$ \\
  $a$ [R$_\odot$] & \multicolumn{2}{c}{$21.61_{-0.09}^{+0.10}$} & $532.3_{-2.7}^{+4.8}$ & \multicolumn{2}{c}{$5.280_{-0.023}^{+0.023}$} & $200.2_{-1.0}^{+1.0}$ & $1573_{-55}^{+47}$ \\
  $e$ & \multicolumn{2}{c}{$0.00742_{-0.00066}^{+0.00069}$} & $0.4627_{-0.0053}^{+0.0052}$ & \multicolumn{2}{c}{$0.0005_{-0.0002}^{+0.0003}$} & $0.458_{-0.029}^{+0.020}$ & $0.14_{-0.03}^{+0.04}$ \\
  $\omega$ [deg] & \multicolumn{2}{c}{$259.3_{-0.9}^{+0.8}$} & $270.6_{-0.7}^{+0.6}$ & \multicolumn{2}{c}{$84_{-30}^{+64}$} & $31.1_{-3.4}^{+2.7}$ & $121_{-7}^{+10}$ \\ 
  $i$ [deg] & \multicolumn{2}{c}{$90.13_{-0.08}^{+0.07}$} & $92.33_{-0.07}^{+0.08}$ & \multicolumn{2}{c}{$87.72_{-0.63}^{+0.39}$} & $89.23_{-0.03}^{+0.04}$ & $102_{-7}^{+6}$ \\
  $\mathcal{T}_0^\mathrm{inf/sup}$ [BJD - 2400000]& \multicolumn{2}{c}{$58717.1942_{-0.0001}^{+0.0001}$} & $59422.2041_{-0.0096}^{+0.0090}$ & \multicolumn{2}{c}{$58683.4709_{-0.0012}^{+0.0009}$} & ${58740.8480_{-0.0148}^{+0.0158}}^*$ & ... \\
  $\tau$ [BJD - 2400000]& \multicolumn{2}{c}{$58716.998_{-0.016}^{+0.014}$} & $58818.49_{-0.34}^{+0.35}$ & \multicolumn{2}{c}{$58682.973_{-0.123}^{+0.100}$} & $58730.74_{-0.64}^{+0.66}$ & $57817_{-52}^{+56}$ \\
  $\Omega$ [deg] & \multicolumn{2}{c}{$0.0$} & $20.06_{-0.56}^{+0.65}$ & \multicolumn{2}{c}{$0.0$} & $2.67_{-1.23}^{+0.89}$ & $23_{-15}^{+10}$ \\
  $i_\mathrm{mut}$ [deg] & \multicolumn{3}{c}{$20.17_{-0.55}^{+0.65}$} & \multicolumn{3}{c}{$3.13_{-1.00}^{+0.83}$} & $28_{-16}^{+10}$ \\
  \hline
  mass ratio $[q=m_\mathrm{sec}/m_\mathrm{pri}]$ & \multicolumn{2}{c}{$0.926_{-0.003}^{+0.003}$} & $0.749_{-0.009}^{+0.021}$ & \multicolumn{2}{c}{$0.707_{-0.007}^{+0.007}$} & $0.724_{-0.009}^{+0.008}$ & $0.254_{-0.021}^{+0.025}$ \\
$K_\mathrm{pri}$ [km\,s$^{-1}$] & \multicolumn{2}{c}{$80.5_{-0.4}^{+0.4}$} & $21.5_{-0.2}^{+0.5}$ & \multicolumn{2}{c}{$118.9_{-0.9}^{+0.9}$} & $29.0_{-0.6}^{+0.5}$ & $4.8_{-0.3}^{+0.4}$ \\ 
$K_\mathrm{sec}$ [km\,s$^{-1}$] & \multicolumn{2}{c}{$86.9_{-0.3}^{+0.4}$} & $28.7_{-0.1}^{+0.2}$ & \multicolumn{2}{c}{$168.2_{-0.9}^{+0.8}$} & $40.0_{-0.6}^{+0.4}$ & $19.2_{-0.8}^{+0.6}$ \\ 
$\gamma$ [km\,s$^{-1}$] & \multicolumn{3}{c}{$17.81_{-0.09}^{+0.09}$} & \multicolumn{4}{c}{...} \\
  \hline  
\multicolumn{8}{c}{stellar parameters} \\
\hline
   & Aa & Ab &  B & Aa & Ab &  B & C \\
  \hline
 \multicolumn{8}{c}{Relative quantities} \\
  \hline
 fractional radius [$R/a$]  & $0.0857_{-0.0005}^{+0.0005}$ & $0.0762_{-0.0005}^{+0.0005}$ & $0.0216_{-0.0006}^{+0.0004}$ & $0.2508_{-0.0031}^{+0.0032}$ & $0.1581_{-0.0018}^{+0.0018}$ & $0.0085_{-0.0002}^{+0.0001}$ & $0.00056_{-0.00005}^{+0.00006}$ \\
 temperature relative to $(T_\mathrm{eff})_\mathrm{Aa}$ & $1$ & $0.9666_{-0.0020}^{+0.0023}$ & $0.6833_{-0.0073}^{+0.0054}$ & $1$ & $0.8178_{-0.0049}^{+0.0050}$ & $1.1330_{-0.0126}^{+0.0126}$ & $0.8494_{-0.0443}^{+0.0477}$ \\
 fractional flux [in \textit{TESS}-band] & $0.0818_{-0.0010}^{+0.0010}$ & $0.0585_{-0.0004}^{+0.0004}$ & $0.8267_{-0.0325}^{+0.0161}$ & $0.2544_{-0.0063}^{+0.0066}$ & $0.0509_{-0.0012}^{+0.0013}$ & $0.6271_{-0.0338}^{+0.0227}$ & $0.0670_{-0.0214}^{+0.0348}$ \\
 \hline
 \multicolumn{8}{c}{Physical Quantities} \\
  \hline 
 $m$ [M$_\odot$] & $1.646_{-0.019}^{+0.022}$ & $1.525_{-0.019}^{+0.021}$ & $2.371_{-0.045}^{+0.108}$ & $1.337_{-0.017}^{+0.018}$ & $0.945_{-0.015}^{+0.014}$ & $1.650_{-0.030}^{+0.032}$ & $0.999_{-0.080}^{+0.096}$ \\
 $R$ [R$_\odot$] & $1.852_{-0.011}^{+0.011}$ & $1.648_{-0.016}^{+0.014}$ & $11.509_{-0.332}^{+0.287}$ & $1.324_{-0.021}^{+0.022}$ & $0.835_{-0.013}^{+0.013}$ & $1.707_{-0.039}^{+0.037}$ & $0.890_{-0.078}^{+0.109}$ \\
 $T_\mathrm{eff}$ [K]& $6995_{-34}^{+163}$ & $6764_{-29}^{+149}$ & $4798_{-41}^{+35}$ & $6481_{-48}^{+73}$ & $5305_{-49}^{+63}$ & $7338_{-112}^{+151}$ & $5516_{-303}^{+323}$ \\
 $L_\mathrm{bol}$ [L$_\odot$] & $7.408_{-0.200}^{+0.515}$ & $5.113_{-0.156}^{+0.402}$ & $62.12_{-2.40}^{+4.76}$ & $2.791_{-0.154}^{+0.173}$ & $0.496_{-0.028}^{+0.033}$ & $7.591_{-0.531}^{+0.696}$ & $0.658_{-0.219}^{+0.382}$ \\
 $M_\mathrm{bol}$ & $2.60_{-0.07}^{+0.03}$ & $2.99_{-0.08}^{+0.03}$ & $0.29_{-0.08}^{+0.04}$ & $3.66_{-0.07}^{+0.06}$ & $5.53_{-0.07}^{+0.06}$ & $2.57_{-0.10}^{+0.08}$ & $5.22_{-0.50}^{+0.44}$ \\
 $M_V           $ & $2.53_{-0.06}^{+0.03}$ & $2.94_{-0.08}^{+0.04}$ & $0.66_{-0.12}^{+0.03}$ & $3.63_{-0.07}^{+0.06}$ & $5.70_{-0.09}^{+0.08}$ & $2.50_{-0.09}^{+0.08}$ & $5.33_{-0.56}^{+0.54}$ \\
 $\log g$ [dex] & $4.118_{-0.005}^{+0.005}$ & $4.187_{-0.004}^{+0.004}$ & $2.693_{-0.020}^{+0.021}$ & $4.319_{-0.009}^{+0.009}$ & $4.569_{-0.007}^{+0.007}$ & $4.189_{-0.013}^{+0.018}$ & $4.538_{-0.060}^{+0.043}$ \\
 \hline
\multicolumn{8}{c}{Global system parameters} \\
  \hline
$\log$(age) [dex] &\multicolumn{3}{c}{$8.972_{-0.087}^{+0.037}$} &\multicolumn{4}{c}{$8.804_{-0.052}^{+0.033}$} \\
$[M/H]$  [dex]    &\multicolumn{3}{c}{$0.322_{-0.120}^{+0.073}$} &\multicolumn{4}{c}{$0.220_{-0.063}^{+0.041}$} \\
$E(B-V)$ [mag]    &\multicolumn{3}{c}{$0.166_{-0.018}^{+0.028}$} &\multicolumn{4}{c}{$0.229_{-0.020}^{+0.030}$} \\
extra light $\ell_4$ [in \textit{TESS}-band] & \multicolumn{3}{c}{$0.033_{-0.017}^{+0.032}$} & \multicolumn{4}{c}{$-$} \\
$(M_V)_\mathrm{tot}$  &\multicolumn{3}{c}{$0.37_{-0.11}^{+0.02}$} & \multicolumn{4}{c}{$2.07_{-0.07}^{+0.07}$} \\
distance [pc]           &\multicolumn{3}{c}{$656_{-17}^{+13}$} &\multicolumn{4}{c}{$712_{-13}^{+14}$} \\  
\hline
\end{tabular}}

\textit{Notes. }{$\mathcal{T}_0^\mathrm{inf/sup}$ denotes the moment of an inferior or superior conjunction of the secondary (Ab) and the tertiary (B) along their inner and outer orbits, respectively. Superior conjunctions are noted with $^*$.}
\end{table*}  

\begin{table*}
 \centering
\caption{The same as in Table~\ref{tab:syntheticfit_TIC47151245+81525800} above, but for TIC 276162169 and TIC 280883908.}
 \label{tab:syntheticfit_TIC276162169+280883908}
\scalebox{0.93}{\begin{tabular}{@{}lllllll}
\hline
 & \multicolumn{3}{c}{TIC\,276162169} & \multicolumn{3}{c}{TIC\,280883908} \\
\hline
\multicolumn{7}{c}{orbital elements} \\
\hline
   & \multicolumn{3}{c}{subsystem} & \multicolumn{3}{c}{subsystem}  \\
   & \multicolumn{2}{c}{Aa--Ab} & A--B & \multicolumn{2}{c}{Aa--Ab} & A--B \\
  \hline
  $t_0$ [BJD - 2400000] & \multicolumn{3}{c}{58683.0} & \multicolumn{3}{c}{58983.0} \\
  $P$ [days] & \multicolumn{2}{c}{$2.549807_{-0.000017}^{+0.000016}$} & $117.27_{-0.04}^{+0.06}$ & \multicolumn{2}{c}{$5.241768_{-0.000104}^{+0.000191}$} & $184.598_{-0.021}^{+0.020}$  \\
  $a$ [R$_\odot$] & \multicolumn{2}{c}{$13.27_{-0.28}^{+0.22}$} & $186.4_{-2.7}^{+3.1}$ & \multicolumn{2}{c}{$17.57_{-0.19}^{+0.14}$} & $224.1_{-3.3}^{+2.7}$ \\
  $e$ & \multicolumn{2}{c}{$0.00108_{-0.00008}^{+0.00008}$} & $0.2682_{-0.0044}^{+0.0048}$ & \multicolumn{2}{c}{$0.0008_{-0.0003}^{+0.0006}$} & $0.2596_{-0.0012}^{+0.0011}$ \\
  $\omega$ [deg] & \multicolumn{2}{c}{$306_{-5}^{+5}$} & $243.4_{-1.3}^{+1.0}$ & \multicolumn{2}{c}{$200_{-84}^{+41}$} & $146.4_{-0.3}^{+0.3}$ \\ 
  $i$ [deg] & \multicolumn{2}{c}{$89.85_{-0.35}^{+0.33}$} & $90.20_{-0.43}^{+0.07}$ & \multicolumn{2}{c}{$89.78_{-0.60}^{+0.48}$} & $88.97_{-0.20}^{+0.53}$ \\
  $\mathcal{T}_0^\mathrm{inf/sup}$ [BJD - 2400000]& \multicolumn{2}{c}{$58685.5242_{-0.0006}^{+0.0006}$} & ${58709.4798_{-0.0120}^{+0.0128}}^*$ & \multicolumn{2}{c}{$58984.9144_{-0.0003}^{+0.0003}$} & ${59010.7402_{-0.0564}^{+0.0429}}^*$ \\
  $\tau$ [BJD - 2400000]& \multicolumn{2}{c}{$58683.234_{-0.031}^{+0.040}$} & $58636.77_{-0.29}^{+0.42}$ & \multicolumn{2}{c}{$58983.897_{-1.217}^{+0.603}$} & $58843.704_{-0.132}^{+0.138}$ \\
  $\Omega$ [deg] & \multicolumn{2}{c}{$0.0$} & $179.62_{-0.60}^{+1.48}$ & \multicolumn{2}{c}{$0.0$} & $-1.58_{-3.32}^{+2.37}$ \\
  $i_\mathrm{mut}$ [deg] & \multicolumn{3}{c}{$179.18_{-0.54}^{+0.37}$} & \multicolumn{3}{c}{$2.25_{-1.41}^{+2.78}$} \\
  \hline
  mass ratio $[q=m_\mathrm{sec}/m_\mathrm{pri}]$ & \multicolumn{2}{c}{$0.865_{-0.005}^{+0.006}$} & $0.310_{-0.008}^{+0.009}$ &\multicolumn{2}{c}{$0.931_{-0.019}^{+0.016}$} & $0.674_{-0.026}^{+0.020}$ \\
$K_\mathrm{pri}$ [km\,s$^{-1}$] & \multicolumn{2}{c}{$122.1_{-2.1}^{+1.9}$} & $19.7_{-0.2}^{+0.2}$ & \multicolumn{2}{c}{$81.6_{-1.0}^{+1.2}$} & $25.60_{-0.93}^{+0.80}$ \\ 
$K_\mathrm{sec}$ [km\,s$^{-1}$] & \multicolumn{2}{c}{$141.3_{-3.5}^{+2.6}$} & $63.8_{-1.7}^{+1.3}$ & \multicolumn{2}{c}{$87.8_{-1.1}^{+1.0}$} & $38.01_{-0.05}^{+0.05}$ \\ 
$\gamma$ [km\,s$^{-1}$] & \multicolumn{3}{c}{...} & \multicolumn{3}{c}{$-29.34_{-0.01}^{+0.01}$} \\
  \hline  
\multicolumn{7}{c}{stellar parameters} \\
\hline
   & Aa & Ab &  B & Aa & Ab &  B \\
  \hline
 \multicolumn{7}{c}{Relative quantities} \\
  \hline
 fractional radius [$R/a$]  & $0.1835_{-0.0005}^{+0.0005}$ & $0.1522_{-0.0007}^{+0.0008}$ & $0.0079_{-0.0001}^{+0.0001}$ & $0.0903_{-0.0017}^{+0.0021}$ & $0.0785_{-0.0016}^{+0.0019}$ & $0.0588_{-0.0015}^{+0.0017}$ \\
 temperature relative to $(T_\mathrm{eff})_\mathrm{Aa}$ & $1$ & $0.9252_{-0.0037}^{+0.0060}$ & $0.6918_{-0.0157}^{+0.0211}$ & $1$ & $0.9766_{-0.0074}^{+0.0061}$ & $0.7030_{-0.0058}^{+0.0077}$ \\
 fractional flux [in \textit{TESS}-band] & $0.5677_{-0.0045}^{+0.0042}$ & $0.3387_{-0.0013}^{+0.0013}$ & $0.0863_{-0.0029}^{+0.0031}$  & $0.0479_{-0.0027}^{+0.0029}$ & $0.0335_{-0.0019}^{+0.0018}$ & $0.9186_{-0.0028}^{+0.0027}$ \\
 \hline
 \multicolumn{7}{c}{Physical Quantities} \\
  \hline 
 $m$ [M$_\odot$] & $2.585_{-0.169}^{+0.134}$ & $2.234_{-0.132}^{+0.110}$ & $1.490_{-0.048}^{+0.051}$ & $1.374_{-0.048}^{+0.026}$ & $1.273_{-0.037}^{+0.035}$ & $1.781_{-0.113}^{+0.100}$ \\
 $R$ [R$_\odot$] & $2.437_{-0.054}^{+0.042}$ & $2.022_{-0.047}^{+0.040}$ & $1.471_{-0.041}^{+0.035}$ & $1.586_{-0.036}^{+0.037}$ & $1.377_{-0.034}^{+0.041}$ & $13.174_{-0.467}^{+0.479}$ \\
 $T_\mathrm{eff}$ [K]& $10087_{-517}^{+411}$ & $9334_{-419}^{+347}$ & $6972_{-145}^{+164}$ & $6688_{-60}^{+49}$ & $6523_{-34}^{+35}$ & $4704_{-34}^{+35}$ \\
 $L_\mathrm{bol}$ [L$_\odot$] & $55.12_{-12.39}^{+11.74}$ & $27.91_{-5.83}^{+5.50}$ & $4.593_{-0.605}^{+0.687}$ & $4.496_{-0.299}^{+0.344}$ & $3.090_{-0.196}^{+0.213}$ & $76.28_{-3.81}^{+3.78}$ \\
 $M_\mathrm{bol}$ & $0.42_{-0.21}^{+0.28}$ & $1.16_{-0.20}^{+0.25}$ & $3.11_{-0.15}^{+0.15}$ & $3.14_{-0.08}^{+0.07}$ & $3.55_{-0.07}^{+0.07}$ & $0.06_{-0.05}^{+0.06}$ \\
 $M_V           $ & $0.66_{-0.13}^{+0.17}$ & $1.25_{-0.13}^{+0.19}$ & $3.07_{-0.16}^{+0.16}$ & $3.11_{-0.08}^{+0.08}$ & $3.53_{-0.08}^{+0.08}$ & $0.48_{-0.06}^{+0.07}$ \\
 $\log g$ [dex] & $4.076_{-0.010}^{+0.008}$ & $4.174_{-0.006}^{+0.008}$ & $4.275_{-0.006}^{+0.010}$ & $4.172_{-0.016}^{+0.013}$ & $4.263_{-0.014}^{+0.013}$ & $2.446_{-0.020}^{+0.020}$ \\
 \hline
\multicolumn{7}{c}{Global system parameters} \\
  \hline
$\log$(age) [dex] &\multicolumn{3}{c}{$8.468_{-0.078}^{+0.096}$} &\multicolumn{3}{c}{$9.256_{-0.028}^{+0.059}$} \\
$[M/H]$  [dex]    &\multicolumn{3}{c}{$0.159_{-0.024}^{+0.027}$} &\multicolumn{3}{c}{$0.030_{-0.073}^{+0.053}$} \\
$E(B-V)$ [mag]    &\multicolumn{3}{c}{$0.648_{-0.036}^{+0.027}$} &\multicolumn{3}{c}{$0.277_{-0.025}^{+0.015}$} \\
$(M_V)_\mathrm{tot}$  &\multicolumn{3}{c}{$0.09_{-0.13}^{+0.18}$} &\multicolumn{3}{c}{$0.33_{-0.05}^{+0.06}$} \\
distance [pc]           &\multicolumn{3}{c}{$1235_{-52}^{+40}$} &\multicolumn{3}{c}{$1183_{-40}^{+40}$} \\  
\hline
\end{tabular}}

\textit{Notes. }{$\mathcal{T}_0^\mathrm{inf/sup}$ denotes the moment of an inferior or superior conjunction of the secondary (Ab) and the tertiary (B) along their inner and outer orbits, respectively. Superior conjunctions are noted with $^*$.}
\end{table*}  

\begin{table*}
 \centering
\caption{The same as in Table~\ref{tab:syntheticfit_TIC47151245+81525800} above, but for TIC 294803663 and TIC 332521671.}
 \label{tab:syntheticfit_TIC294803663+332521671}
\scalebox{0.93}{\begin{tabular}{@{}lllllll}
\hline
 & \multicolumn{3}{c}{TIC\,294803663} & \multicolumn{3}{c}{TIC\,332521671} \\
\hline
\multicolumn{7}{c}{orbital elements} \\
\hline
   & \multicolumn{3}{c}{subsystem} & \multicolumn{3}{c}{subsystem}  \\
   & \multicolumn{2}{c}{Aa--Ab} & A--B & \multicolumn{2}{c}{Aa--Ab} & A--B \\
  \hline
  $t_0$ [BJD - 2400000] & \multicolumn{3}{c}{58596.5} & \multicolumn{3}{c}{58571.0} \\
  $P$ [days] & \multicolumn{2}{c}{$2.245592_{-0.000006}^{+0.000008}$} & $153.426_{-0.047}^{+0.046}$ & \multicolumn{2}{c}{$1.247934_{-0.000059}^{+0.000055}$} & $48.5848_{-0.0038}^{+0.0062}$  \\
  $a$ [R$_\odot$] & \multicolumn{2}{c}{$11.62_{-0.10}^{+0.13}$} & $230.1_{-2.0}^{+2.5}$ & \multicolumn{2}{c}{$6.889_{-0.025}^{+0.024}$} & $94.26_{-0.47}^{+0.42}$ \\
  $e$ & \multicolumn{2}{c}{$0.0305_{-0.0018}^{+0.0019}$} & $0.0299_{-0.0099}^{+0.0087}$ & \multicolumn{2}{c}{$0.0030_{-0.0016}^{+0.0021}$} & $0.0041_{-0.0019}^{+0.0019}$ \\
  $\omega$ [deg] & \multicolumn{2}{c}{$127.1_{-3.0}^{+3.4}$} & $313_{-14}^{+9}$ & \multicolumn{2}{c}{$229_{-35}^{+17}$} & $311_{-28}^{+34}$ \\ 
  $i$ [deg] & \multicolumn{2}{c}{$86.10_{-0.24}^{+0.88}$} & $88.65_{-0.13}^{+0.08}$ & \multicolumn{2}{c}{$87.85_{-0.62}^{+0.48}$} & $85.43_{-0.17}^{+0.31}$ \\
  $\mathcal{T}_0^\mathrm{inf/sup}$ [BJD - 2400000]& \multicolumn{2}{c}{$58598.5098_{-0.0003}^{+0.0003}$} & $58629.0863_{-0.0131}^{+0.0132}$ & \multicolumn{2}{c}{$58571.3516_{-0.0002}^{+0.0003}$} & ${58583.3545_{-0.1098}^{+0.0571}}^*$ \\
  $\tau$ [BJD - 2400000]& \multicolumn{2}{c}{$58597.633_{-0.019}^{+0.021}$} & $58493.1_{-5.5}^{+3.2}$ & \multicolumn{2}{c}{$58571.210_{-0.120}^{+0.060}$} & $58568.5_{-5.8}^{+4.7}$ \\
  $\Omega$ [deg] & \multicolumn{2}{c}{$0.0$} & $-0.05_{-2.91}^{+2.77}$ & \multicolumn{2}{c}{$0.0$} & $-0.83_{-1.67}^{+1.70}$ \\
  $i_\mathrm{mut}$ [deg] & \multicolumn{3}{c}{$3.07_{-0.69}^{+1.58}$} & \multicolumn{3}{c}{$2.80_{-0.69}^{+0.84}$} \\
  \hline
  mass ratio $[q=m_\mathrm{sec}/m_\mathrm{pri}]$ & \multicolumn{2}{c}{$0.866_{-0.010}^{+0.010}$} & $0.667_{-0.006}^{+0.004}$ & \multicolumn{2}{c}{$0.640_{-0.006}^{+0.006}$} & $0.692_{-0.007}^{+0.006}$ \\
$K_\mathrm{pri}$ [km\,s$^{-1}$] & \multicolumn{2}{c}{$121.3_{-1.3}^{+1.6}$} & $30.4_{-0.4}^{+0.4}$ & \multicolumn{2}{c}{$109.0_{-0.7}^{+0.7}$} & $39.99_{-0.33}^{+0.38}$ \\ 
$K_\mathrm{sec}$ [km\,s$^{-1}$] & \multicolumn{2}{c}{$140.4_{-1.6}^{+1.4}$} & $45.6_{-0.4}^{+0.5}$ & \multicolumn{2}{c}{$170.2_{-1.0}^{+0.8}$} & $57.89_{-0.13}^{+0.28}$ \\ 
$\gamma$ [km\,s$^{-1}$] & \multicolumn{3}{c}{$-18.11_{-0.08}^{+0.08}$} & \multicolumn{3}{c}{$-15.63_{-0.04}^{+0.04}$} \\
  \hline  
\multicolumn{7}{c}{stellar parameters} \\
\hline
   & Aa & Ab &  B & Aa & Ab &  B \\
  \hline
 \multicolumn{7}{c}{Relative quantities} \\
  \hline
 fractional radius [$R/a$]  & $0.2165_{-0.0042}^{+0.0037}$ & $0.1705_{-0.0031}^{+0.0032}$ & $0.0579_{-0.0008}^{+0.0010}$ & $0.3615_{-0.0046}^{+0.0041}$ & $0.1513_{-0.0017}^{+0.0019}$ & $0.0888_{-0.0058}^{+0.0015}$ \\
 temperature relative to $(T_\mathrm{eff})_\mathrm{Aa}$ & $1$ & $0.9422_{-0.0061}^{+0.0065}$ & $0.5694_{-0.0094}^{+0.0065}$ & $1$ & $0.8643_{-0.0073}^{+0.0074}$ & $0.7091_{-0.0048}^{+0.0055}$ \\
 fractional flux [in \textit{TESS}-band] & $0.1597_{-0.0034}^{+0.0034}$ & $0.0863_{-0.0034}^{+0.0030}$ & $0.7541_{-0.0027}^{+0.0027}$ & $0.1984_{-0.0048}^{+0.0047}$ & $0.0225_{-0.0005}^{+0.0006}$ & $0.7006_{-0.0649}^{+0.0171}$ \\
 \hline
 \multicolumn{7}{c}{Physical Quantities} \\
  \hline 
 $m$ [M$_\odot$] & $2.240_{-0.067}^{+0.068}$ & $1.932_{-0.047}^{+0.071}$ & $2.776_{-0.088}^{+0.103}$ & $1.717_{-0.024}^{+0.015}$ & $1.098_{-0.012}^{+0.013}$ & $1.941_{-0.035}^{+0.037}$ \\
 $R$ [R$_\odot$] & $2.520_{-0.058}^{+0.047}$ & $1.975_{-0.040}^{+0.065}$ & $13.36_{-0.19}^{+0.19}$   & $2.490_{-0.030}^{+0.029}$ & $1.043_{-0.014}^{+0.015}$ & $8.371_{-0.580}^{+0.165}$ \\
 $T_\mathrm{eff}$ [K]& $8694_{-199}^{+98}$ & $8152_{-117}^{+129}$ & $4902_{-40}^{+91}$ & $6978_{-44}^{+55}$ & $6039_{-62}^{+41}$ & $4946_{-33}^{+43}$ \\
 $L_\mathrm{bol}$ [L$_\odot$] & $32.21_{-3.26}^{+2.83}$ & $15.66_{-1.38}^{+1.44}$ & $94.61_{-6.52}^{+5.12}$ & $13.23_{-0.39}^{+0.35}$ & $1.293_{-0.064}^{+0.069}$ & $37.41_{-3.38}^{+1.59}$ \\
 $M_\mathrm{bol}$ & $1.00_{-0.09}^{+0.12}$ & $1.78_{-0.10}^{+0.10}$ & $-0.17_{-0.06}^{+0.08}$ & $1.97_{-0.03}^{+0.03}$ & $4.49_{-0.06}^{+0.05}$ & $0.84_{-0.05}^{+0.10}$ \\
 $M_V           $ & $0.99_{-0.08}^{+0.10}$ & $1.73_{-0.09}^{+0.09}$ & $0.14_{-0.06}^{+0.08}$ & $1.91_{-0.03}^{+0.03}$ & $4.51_{-0.06}^{+0.06}$ & $1.12_{-0.05}^{+0.09}$ \\
 $\log g$ [dex] & $3.987_{-0.016}^{+0.014}$ & $4.129_{-0.010}^{+0.013}$ & $2.631_{-0.018}^{+0.015}$ & $3.878_{-0.010}^{+0.012}$ & $4.441_{-0.008}^{+0.007}$ & $2.881_{-0.012}^{+0.053}$ \\
 \hline
\multicolumn{7}{c}{Global system parameters} \\
  \hline
$\log$(age) [dex] & \multicolumn{3}{c}{$8.734_{-0.056}^{+0.064}$} &\multicolumn{3}{c}{$9.155_{-0.013}^{+0.016}$} \\
$[M/H]$  [dex]    & \multicolumn{3}{c}{$0.230_{-0.081}^{+0.056}$} &\multicolumn{3}{c}{$0.057_{-0.042}^{+0.051}$} \\
$E(B-V)$ [mag]    & \multicolumn{3}{c}{$0.631_{-0.019}^{+0.018}$} &\multicolumn{3}{c}{$0.088_{-0.016}^{+0.025}$} \\
extra light $\ell_4$ [in \textit{TESS}-band] & \multicolumn{3}{c}{...} & \multicolumn{3}{c}{$0.080_{-0.020}^{+0.061}$} \\
$(M_V)_\mathrm{tot}$  & \multicolumn{3}{c}{$-0.43_{-0.06}^{+0.09}$} &\multicolumn{3}{c}{$0.66_{-0.04}^{+0.06}$} \\
distance [pc]           & \multicolumn{3}{c}{$1739_{-26}^{+24}$} &\multicolumn{3}{c}{$799_{-47}^{+15}$} \\  
\hline
\end{tabular}}

{\textit{Notes. }{$\mathcal{T}_0^\mathrm{inf/sup}$ denotes the moment of an inferior or superior conjunction of the secondary (Ab) and the tertiary (B) along their inner and outer orbits, respectively. Superior conjunctions are noted with $^*$.}}
\end{table*}  

\begin{table*}
 \centering
\caption{The same as in Table~\ref{tab:syntheticfit_TIC47151245+81525800} above, but for TIC 356324779.}
 \label{tab:syntheticfit_TIC356324779}
\begin{tabular}{@{}llll}
\hline
\multicolumn{4}{c}{TIC\,356324779}  \\
\hline
\multicolumn{4}{c}{orbital elements} \\
\hline
   & \multicolumn{3}{c}{subsystem}  \\
   & \multicolumn{2}{c}{Aa--Ab} & A--B  \\
  \hline
  $t_0$ [BJD - 2400000] & \multicolumn{3}{c}{58816.0}  \\
  $P$ [days] & \multicolumn{2}{c}{$3.47167_{-0.00009}^{+0.00013}$} & $87.092_{-0.018}^{+0.017}$  \\
  $a$ [R$_\odot$] & \multicolumn{2}{c}{$14.66_{-0.23}^{+0.30}$} & $146.27_{-2.68}^{+3.37}$ \\
  $e$ & \multicolumn{2}{c}{$0.00032_{-0.00012}^{+0.00012}$} & $0.2836_{-0.0042}^{+0.0030}$ \\
  $\omega$ [deg] & \multicolumn{2}{c}{$66_{-13}^{+9}$} & $143.9_{-0.6}^{+0.5}$ \\ 
  $i$ [deg] & \multicolumn{2}{c}{$90.08_{-0.89}^{+0.50}$} & $89.28_{-0.08}^{+0.13}$  \\
  $\mathcal{T}_0^\mathrm{inf/sup}$ [BJD - 2400000] & \multicolumn{2}{c}{$58819.18605_{-0.00006}^{+0.00006}$} & ${58837.6027_{-0.0021}^{+0.0021}}^*$  \\
  $\tau$ [BJD - 2400000] & \multicolumn{2}{c}{$58817.213_{-0.132}^{+0.086}$} & $58757.967_{-0.143}^{+0.124}$  \\
  $\Omega$ [deg] & \multicolumn{2}{c}{$0.0$} & $-0.87_{-0.27}^{+0.39}$ \\
  $i_\mathrm{mut}$ [deg] & \multicolumn{3}{c}{$1.26_{-0.60}^{+0.41}$} \\
  \hline
  mass ratio $[q=m_\mathrm{sec}/m_\mathrm{pri}]$ & \multicolumn{2}{c}{$0.984_{-0.004}^{+0.005}$} & $0.573_{-0.015}^{+0.013}$  \\
$K_\mathrm{pri}$ [km\,s$^{-1}$] & \multicolumn{2}{c}{$106.1_{-1.6}^{+2.2}$} & $32.4_{-1.1}^{+1.3}$ \\ 
$K_\mathrm{sec}$ [km\,s$^{-1}$] & \multicolumn{2}{c}{$107.8_{-1.7}^{+2.4}$} & $56.2_{-0.5}^{+0.9}$ \\ 
  \hline  
\multicolumn{4}{c}{stellar parameters} \\
\hline
   & Aa & Ab &  B \\
  \hline
 \multicolumn{4}{c}{Relative quantities} \\
  \hline
 fractional radius [$R/a$] & $0.1372_{-0.0014}^{+0.0010}$ & $0.1330_{-0.0014}^{+0.0011}$ & $0.0198_{-0.0005}^{+0.0006}$ \\
 temperature relative to $(T_\mathrm{eff})_\mathrm{Aa}$ & $1$ & $0.9949_{-0.0027}^{+0.0024}$ & $0.9977_{-0.0425}^{+0.0214}$ \\
 fractional flux [in \textit{TESS}-band] & $0.2146_{-0.0028}^{+0.0037}$ & $0.1987_{-0.0025}^{+0.0039}$ & $0.4339_{-0.0298}^{+0.0362}$ \\
 fractional flux [in $R_C$-band] & $0.2321_{-0.0033}^{+0.0035}$ & $0.2149_{-0.0028}^{+0.0033}$ & $0.4711_{-0.0427}^{+0.0407}$ \\
 \hline
 \multicolumn{4}{c}{Physical Quantities} \\
  \hline 
 $m$ [M$_\odot$] & $1.768_{-0.082}^{+0.111}$ & $1.740_{-0.078}^{+0.108}$ & $2.022_{-0.139}^{+0.181}$ \\
 $R$ [R$_\odot$] & $2.014_{-0.047}^{+0.054}$ & $1.958_{-0.056}^{+0.052}$ & $2.909_{-0.069}^{+0.088}$  \\
 $T_\mathrm{eff}$ [K] & $7921_{-152}^{+191}$ & $7877_{-140}^{+202}$ & $7734_{-146}^{+346}$  \\
 $L_\mathrm{bol}$ [L$_\odot$] & $14.27_{-1.24}^{+1.17}$ & $13.21_{-1.15}^{+1.07}$ & $27.36_{-2.53}^{+8.23}$ \\
 $M_\mathrm{bol}$ & $1.88_{-0.09}^{+0.10}$ & $1.97_{-0.08}^{+0.10}$ & $1.18_{-0.29}^{+0.10}$ \\
 $M_V           $ & $1.83_{-0.09}^{+0.10}$ & $1.92_{-0.09}^{+0.10}$ & $1.12_{-0.29}^{+0.11}$ \\
 $\log g$ [dex] & $4.080_{-0.008}^{+0.011}$ & $4.099_{-0.007}^{+0.010}$ & $3.815_{-0.031}^{+0.038}$ \\
 \hline
\multicolumn{4}{c}{Global system parameters} \\
  \hline
$\log$(age) [dex] & \multicolumn{3}{c}{$8.966_{-0.102}^{+0.078}$} \\
$[M/H]$  [dex]    & \multicolumn{3}{c}{$0.073_{-0.240}^{+0.058}$} \\
$E(B-V)$ [mag]    & \multicolumn{3}{c}{$0.381_{-0.008}^{+0.008}$} \\
extra light $\ell_4$ [in \textit{TESS}-band] & \multicolumn{3}{c}{$0.152_{-0.040}^{+0.030}$} \\
extra light $\ell_4$ [in $R_C$-band] & \multicolumn{3}{c}{$0.081_{-0.039}^{+0.045}$} \\
$(M_V)_\mathrm{tot}$  & \multicolumn{3}{c}{$0.37_{-0.19}^{+0.09}$} \\
distance [pc]         & \multicolumn{3}{c}{$2440_{-178}^{+197}$}  \\  
\hline
\end{tabular}

\textit{Notes. }{$\mathcal{T}_0^\mathrm{inf/sup}$ denotes the moment of an inferior or superior conjunction of the secondary (Ab) and the tertiary (B) along their inner and outer orbits, respectively. Superior conjunctions are noted with $^*$.}
\end{table*}  

Virtually all of the specifics of how this code was used to analyze compact triply eclipsing triples found with {\it TESS}, were described in \citet{rappaport22}.  Therefore, we will provide a high-level overview here, highlighting inputs to the code and the parameters that are either determined or constrained by the MCMC fit.  In all, there are 26 system parameters that are output from the analysis.  Specifically, these are all the stellar parameters (including mass, radius, $T_{\rm eff}$), all of the elements of the inner and outer orbits, the system metallicity and age, as well as the distance to the source and the interstellar extinction.  

The `input' information that is used by the code to determine these 26 system parameters can be divided into two basic categories.  First, there are the `input data'.  These include the {\it TESS} lightcurve which contains (i) EB eclipse profiles, (ii) the third-body eclipse profile(s), (iii) the times of the EB eclipses which are distinct from the shape and depths of eclipses, (iv) archival SED values, (v) radial velocities (available only for four of the systems), and (vi) the outer orbital period and possibly $e_{\rm out} \cos \omega_{\rm out}$ determined from ASAS-SN, ATLAS, and other ground-based archival data.  Second, we utilize \texttt{PARSEC} model stellar evolution tracks and isochrones as well as model stellar atmospheres \citep{PARSEC}.  The former enables us to relate stellar mass to radius and $T_{\rm eff}$ for a given age and metallicity, while the latter allows us to compute magnitudes in different bands to fit the SED.  The available input information for each of the nine triples is summarized in Table \ref{tbl:input}.

The {\it TESS} lightcurves that we used for the photodynamical analysis were taken from the full-frame images (`FFI') where the photometry was done with the FITSH package \citep{pal12}. To save computational time we  binned the 10-min cadence data to 30-min cadence, and in the case of seven of the nine current systems, dropped out the out-of-eclipse sections of these lightcurves, retaining only the $\pm0\fp15$ phase-domain regions around the binary eclipses themselves.  However, during sections of the data containing the third-body (i.e., `outer') eclipses, we kept the data for an entire binary period both before and after the first and last contacts of the given third-body eclipse. 

We note that the \textit{TESS} lightcurves of TICs~47151245 and 294803663 needed some extra care. As was mentioned above, these lightcurves exhibit large stellar oscillations.  We modeled these oscillations simultaneously with the triple star lightcurve modeling, fitting harmonic functions to the residual lightcurves in each trial step. In the case of the former system we used a single frequency, while in the latter system, four\footnote{From these four frequencies, only two belong to a physical pulsation and its harmonic, while the other two lower frequencies were used to model some likely instrumental effects.}  frequencies were required.  While the frequencies were kept fixed, the corresponding coefficients of the sine and cosine terms were obtained via matrix inversion. Then the $\chi^2_\mathrm{LC}$ value was calculated for this mixed lightcurve, and used for the acceptance or rejection of the given trial step. More details of this procedure can be found in \citet{borkovitsetal18}. 
 
The system parameters that are derived from the photodynamical analyses are tabulated in Tables \ref{tab:syntheticfit_TIC47151245+81525800} through \ref{tab:syntheticfit_TIC356324779}.  The tables include 7 basic parameters describing the inner and the outer orbits, and their relative orientations; several relative values (e.g., $R/a$, $T_B/T_{Aa}$); the properties of all three stars (including $M$, $R$, $T_{\rm eff}$, $L$, and $M_V$); and 6 global system parameters such as distance, $[M/H]$, and $E(B-V)$.  The definitions of all the parameters listed in Tables \ref{tab:syntheticfit_TIC47151245+81525800} through \ref{tab:syntheticfit_TIC356324779} are given first in Table \ref{tbl:definitions}.

\section{Stellar and Orbital Parameters for the Nine Triples}
\label{sec:results}

\subsection{Individual Triples}
\label{sec:individual}

In this section we briefly discuss the bigger-picture results for each of the nine triply eclipsing triples.  Before looking at the results for the individual systems, we present a pictorial overview of all the binary orbits in Fig.~\ref{fig:orbits}. These are what the orbits would look like to a distant observer viewing the systems from the poles of the outer orbital planes.  The parameters used to generate this figure, as well as those we refer to in the discussion of each system, are taken from Tables \ref{tab:syntheticfit_TIC47151245+81525800} to \ref{tab:syntheticfit_TIC356324779} and are based on the photodynamical fits (unless otherwise specifically indicated).  

\subsubsection{TIC 47151245 = V726 Sco}
\label{sec:info_47151245}

The TIC 47151245 system was formerly classified as a likely detached EB with a 1.202-day-period \citep[see, e.g.,][]{avvakumovaetal13}. Our results confirm these findings.  In the absence of any RV data we were able to determine the stellar masses only with a moderate accuracy of 8--10\%.  To within this limiting accuracy we found that the primary and tertiary stars have similar masses ($m_\mathrm{Aa}=2.03\pm0.17\,\mathrm{M}_\odot$ and $m_\mathrm{B}=2.11\pm0.21\,\mathrm{M}_\odot$, respectively), while the secondary of the EB is a bit less massive $m_\mathrm{Ab}=1.78\pm0.17\,\mathrm{M}_\odot$. Note, however, that despite the weaker accuracy on the masses, their ratios are constrained much better, and we find $q_\mathrm{in}=0.88\pm0.02$ and $q_\mathrm{out}=0.55\pm0.01$ for the inner and outer subsystems, respectively. The triple system is less `tight' (defined as small period ratios $P_\mathrm{out}/P_\mathrm{in}$) since in this system $P_\mathrm{out}/P_\mathrm{in}=284.3/1.203=236.3$.  Hence, we cannot expect strong, and short-timescale dynamical perturbations. Despite this, what makes this triple extraordinarily interesting is that, in contrast to the previously investigated, almost flat triply eclipsing triple systems, here we have found an unusually high mutual inclination of $i_\mathrm{mut}=41\degr\pm1\degr$.

While, at the first sight, this result seems to be unexpected, an expert glance at the \textit{TESS} lightcurve (upper left panel of Fig.~\ref{fig:triples}) offers an immediate, qualitative justification of the significant non-alignment of the inner and outer orbits. As one can see, the depths of the regular eclipses of the inner EB are similar in amplitude to the strong ellipsoidal light variations of the out-of-eclipse sections of the lightcurve. This fact clearly reveals that the low amplitude of the inner eclipses cannot be explained with light dilution caused by the contaminating flux of a dominant tertiary object. The only explanation for a similar amplitude ellipsoidal light variation and eclipse depth is that the eclipses are nearly grazing, which, by itself, suggests that the inner orbit is being viewed far from a perfectly edge-on orientation. On the other hand, since the probability of outer eclipses decreases strongly with increasing separation (and hence outer orbital period) of the third component \citep[see, e.g.][Appendix~A]{borkovitsetal22b}, in order to detect third-body eclipses for such a relatively wide triple system\footnote{This triple has the second longest outer period in our sample.}, one can expect that we have an almost edge-on view of the outer orbit. Hence, one can really justify the non-alignment of the two orbits even prior to making any quantitative analysis, or saying anything about the nature of the three stellar components. 

Considering the consequences of this significant non-alignment of the two orbits, one should first realize that the observed mutual inclination is close to, but above, the well-known mutual inclination limit ($i_\mathrm{mut}=39.23\degr$) of the original, quadrupole, asymptotic ZLK theorem. In this particular system, however, due to the strong tidal distortion of the inner binary stars (the fractional radii of the EB stars are $r_\mathrm{Aa}=0.320\pm0.007$ and $r_\mathrm{Ab}=0.246\pm0.011$), the tidal forces dominate the apsidal motion.  Hence, the tidal damping prevents the inner pair from undergoing ZLK cycles, and enables the triple to remain stable in its present configuration for a very long time.

The only observable consequence of the large mutual inclination may reveal itself in the nodal precession of the orbital planes, which may lead to variations of the eclipse depths of both the inner and outer eclipses. The nodal period of the current system is $P_\mathrm{node}=580\pm40$\,years, which may seem to be quite a long time to wait for a dramatic change in the eclipse depths.  However, the grazing nature of the eclipses makes their depths and durations very sensitive to even small variations of the EB's observable inclination ($i_\mathrm{in}$). Hence, a systematic monitoring of the regular eclipses in the forthcoming years or decades would be extremely useful.

Finally we also make some cautionary notes about the ambiguity of our solution.  Out of our nine currently investigated systems, TIC~47151245 is the only one where we had neither RV data nor a good ETV curve to better constrain the system configuration. The latter resulted from the large scatter of the ETV points due to the shallow, grazing eclipses observed with \textit{TESS} (the scatter of the ETV points exceed the full amplitude of the model ETV curve.) Therefore, strictly speaking, we cannot even be certain about the type of the one observed outer eclipse, in particular, whether it occurred during an inferior or superior conjunction of the tertiary. (In other words, we cannot be certain whether the third star was the eclipser or, if it was eclipsed.)  What makes this question more ambiguous is that the surface brightnesses of stars $Aa$ and $B$ are very similar, and even star $Ab$ has only a slightly smaller surface brightness. Hence, regardless of whether star $Aa$ eclipses star $B$ or, vice versa, in the case of similar geometries the eclipses look very similar.  Hence, it was not surprising that by assuming a superior outer star conjunction for the \textit{TESS} observed third-body eclipse event, and adding 180$^\circ$ to $\omega_\mathrm{out}$ (i.e., virtually rotating the outer orbit by 180$^\circ$ to get very similar geometric conditions), we were able to obtain another, almost similarly good, solution as for the inferior conjunction case, which is tabulated in Table~\ref{tab:syntheticfit_TIC47151245+81525800}. The substantial difference, however, between the two solutions is that, in this second case, to correctly model the relative depths of the three consecutive dips of the third-body eclipse, we had to reverse the revolution of the stars along either the inner or outer orbit. (As a practical matter, we have chosen the outer orbit, setting the outer node to be $\Omega_\mathrm{out}\approx 213\degr$.) As a consequence, this solution results in a {\it retrograde} solution, with a mutual inclination of $i_\mathrm{mut}\approx139\degr$. While, these latter solutions were found to be statistically slightly weaker than the prograde solutions (Table~\ref{tab:syntheticfit_TIC47151245+81525800}), in order to make a secure choice between the two scenarios, radial velocity measurements will be critical.

\subsubsection{TIC 81525800}

With a mean third-body eclipsing period of 47.8 days, this is one of two quite short outer period systems in this set of 9 triples.  It has, by far, the most eccentric outer orbit among the set with $e = 0.61$.  The outer orbit is securely mapped out with the {\it TESS} ETV curve (see Fig.~\ref{fig:TIC81525800}) which covers one and a half outer orbital cycles.

The inner binary has a period of 1.6 days, and the period ratio of $\sim$30, puts this system quite near the dynamically unstable regime.  From estimates of what constitutes a stable triple system, the ratio of orbital periods must satisfy
\begin{equation}
P_{\rm out} \gtrsim 4.7 \left(\frac{M_{\rm AB}}{M_{\rm  B}}\right)^{1/10} \frac{(1+e_{\rm out})^{3/5}}{(1-e_{\rm out})^{9/5}} ~P_{\rm in}
\label{eqn:stability}
\end{equation}
where equation (\ref{eqn:stability}) is from \citet{rappaport13}, which in turn is based on the work of \citet{mardling01} and \citet{mikkola08}.  The numerical value of the right hand side of this equation is 58.  By comparison, both the outer eclipsing period of 47.8 and the formal osculating period given in Table \ref{tab:syntheticfit_TIC47151245+81525800} of 49.7 days are actually below the stability line. One should keep in mind however, that Eq.~(\ref{eqn:stability}) is strictly valid only for point-mass systems, and the non-negligible tidal interactions between the two components of the inner pair may act to stabilize the system. Given that the system is 2 Gyr old, and is manifestly stable, it would be worthwhile to study this problem via long-term numerical integration.

The outer and inner orbits are viewed within a couple of degrees of edge on, and the mutual inclination angle is $\simeq 5^\circ$.  So, this is a fairly flat system viewed nearly edge on.

The masses of all three stars are close to 1.4 M$_\odot$ F stars, and are just slightly evolved away from the main sequence.   

The photometric distance that we calculate is $1718 \pm 60$ pc compared to Gaia's distance of  $1979 \pm 80$ pc.  This amounts to a 2.6-$\sigma$ discrepancy, which we do not consider alarming. 

\subsubsection{TIC 99013269}
\label{sec:info_99013269}

\begin{figure}
\begin{center}
\includegraphics[width=1.00 \columnwidth]{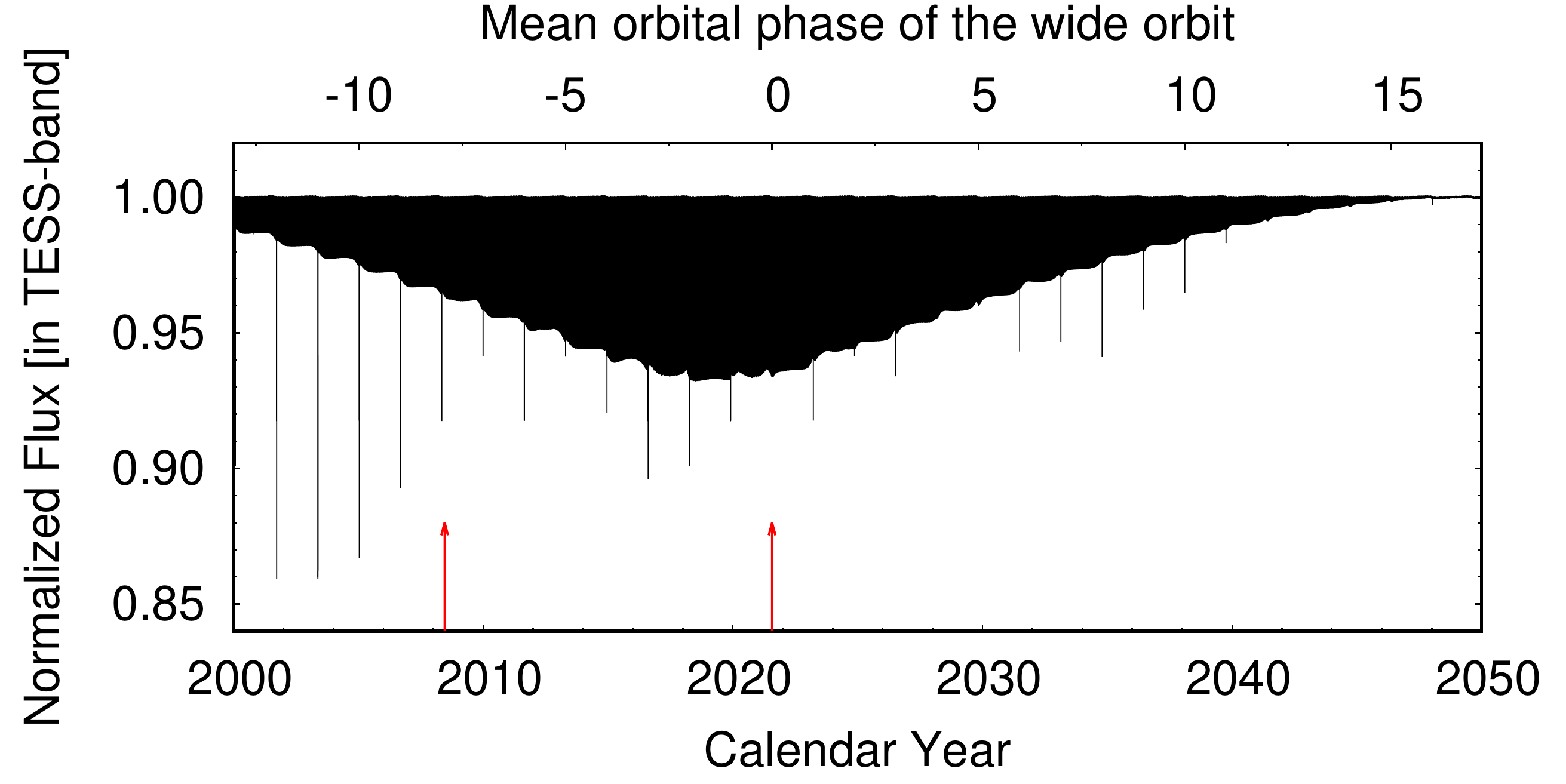}
\caption{Photodynamical model lightcurve of TIC 99013269 for the epoch 2000-2050. The red arrows denote the position of the KELT and \textit{TESS} third-body eclipses, respectively. See text for details.}
\label{fig:T99013269lc50yr}
\end{center}
\end{figure}  

This triple has the longest outer period in our current sample (604 days), and has the fifth longest outer period amongst all the securely known triply eclipsing triples \citep[see][Table 1]{borkovits22}. Despite the fact that \textit{TESS} observed only one outer eclipse, and our RV data cover only about half of a complete outer cycle, we were able to determine the outer period very robustly (with an accuracy of about 10 minutes).  This is due to the fact that the KELT survey observed one third-body outer eclipse on three subsequent nights around 1st May 2008, i.e., eight outer orbital cycles earlier than the \textit{TESS} third-body eclipse. Comparing the two third-body events (uppermost right panel of Fig.~\ref{fig:triples}) one will readily note two substantial differences.  First, the KELT third-body eclipse had a substantially longer duration than the {\it TESS} event.  Second, despite the fact that both third-body eclipses have a flat bottom, i.e. they are total eclipses, the KELT event is far deeper, substantially exceeding the depths of the regular EB eclipses.  By contrast, in the case of the \textit{TESS} event, the depth of the extra dip is in between the eclipse depths of the inner EB's primary and secondary eclipses. 

These changing outer eclipse properties can be explained by the fact that, during the KELT event, the giant tertiary star totally eclipsed both members of the inner EB, while in the case of the \textit{TESS} event, only the secondary component (star $Ab$) was eclipsed. This is a direct consequence of the remarkably inclined nature of the two orbital planes. Our results reveal robustly a mutual inclination of $i_\mathrm{mut}=20\fdg2\pm0\fdg6$ and, moreover, we obtained a relative position of the nodes ($\Delta\Omega=\Omega_\mathrm{out}=20\fdg1\pm0\fdg6$) for the epoch of (the beginning of) the \textit{TESS} observations, which leads to quite unequal eclipse probabilities for the two nearly equally massive inner stars, than for the $\Delta\Omega\approx0\degr$ case.

The mutual inclination of $i_\mathrm{mut}=20\degr$ predicts an orbital plane precession with a period of $P_\mathrm{node}\sim322\pm2$\,years. We plot the \textit{TESS}-band model lightcurve for the first half of the current century in Fig.~\ref{fig:T99013269lc50yr}. As one can see, at the very beginning of the new millennium the inner EB exhibited only very grazing eclipses, while in the same timeframe, the third-body eclipses reached their maximum depths and durations (these durations cannot be resolved in the figure).  Then, the depths of the regular eclipses increased continuously until circa 2018, when the inner orbit was seen exactly edge-on. Currently both the inner and third-body eclipse depths show a decreasing tendency\footnote{In the case of third-body eclipses we can only talk about a tendency instead of a monotonic variation. For example, the forthcoming third-body eclipse that will occur around 21-23 March 2023 is expected to be deeper than the one detected by \textit{TESS}. The reason is that next spring the third star will totally eclipse the hotter primary (i.e. $Aa$) instead of the secondary ($Ab$) of the inner pair.} and, incidentally, both kinds of eclipses will disappear at the same time around 2049.

Finally, note that this is the only system in our sample for which Gaia gives an astrometric solution. We compare two parameters, $e_{\rm out}$ and $i_{\rm out}$ between our photodynamical, and Gaia's astrometric solutions, and note that we did not utilize either of these in our fit. We find $e_{\rm out} = 0.463 \pm 0.005$ compared to Gaia's value of $e_{\rm out} = 0.458 \pm 0.026$, and $i_{\rm out} = 92.3^\circ \pm 0.08^\circ$ vs.~Gaia's value of $i_{\rm out} = 92.0^\circ \pm 0.8^\circ$.  These are both in excellent agreement.

We find masses and radii for the stars in this system of $M_{\rm Aa} = 1.65$ M$_\odot$, $M_{\rm Ab} = 1.53$ M$_\odot$, and $M_{\rm B} = 2.37$ M$_\odot$, with an evolved radius for the tertiary of $11.5$ R$_\odot$, which Gaia is not able to provide.  

\subsubsection{TIC 229785001}
\label{sec:info_229785001}

As far as the triple-system portion of TIC 229785001 goes, this is a rather flat system viewed nearly edge on, with $i_{\rm out} = 89.2^\circ$ and $i_{\rm mut} = 3^\circ$.  The three stars combined have the lowest total mass of any of our 9 current systems, with $M_{\rm Aa} = 1.3$ M$_\odot$, $M_{\rm Ab} = 0.95$ M$_\odot$, and $M_{\rm B} = 1.6$ M$_\odot$.

Perhaps the most interesting thing about this system is that the ETV curve (see Fig.~\ref{fig:TIC229785001}) clearly reveals that the triple system has another star (or stars) orbiting it with a very long period (likely more than 5 years).  This can be inferred from the overall parabolic trend in the ETV curve which is superposed on the triple orbital period of 165 days.  In fact, when this 4th star is introduced into the photodynamical analysis, the formal solution suggests an outermost period of $\sim$$8.9 \pm 0.5$ years (see Table \ref{tab:syntheticfit_TIC99013269+229785001}; but we note that the actual uncertainty in this period may be substantially larger).  If this is indeed a quadruple system with a 2+1+1 configuration, the ratio of the outermost period to the triple's period could be as low as $\sim$20, and dynamical interactions between the 4th star and the inner triple could be quite significant.    

Because of the presence of the fourth body, we thought it would be good to check the full outer orbital solution parameters (of the triple) from Gaia with those from our photodynamical fit, along with the uncertainties:  $P_{\rm out} = 166.0 \pm 0.5$, $e_{\rm out} = 0.49 \pm 0.05$, $\omega_{\rm out} = 15.9^\circ \pm 0.5^\circ$, and $K_{\rm ter} = 34.3 \pm 1.7$ km s$^{-1}$ for Gaia and $P_{\rm out} = 165.37 \pm 0.02$, $e_{\rm out} = 0.46 \pm 0.02$, $\omega_{\rm out} = 31^\circ \pm 3^\circ$, and $K_{\rm ter} = 40 \pm 0.5$ km s$^{-1}$ from our photodynamical fit.  These are in reasonable agreement, but the discrepancies in $\omega_{\rm out}$ and that of $K_{\rm ter}$ are at the 5 and 3 $\sigma$ levels.   

\subsubsection{TIC 276162169 = V493 Cyg}
\label{sec:info_276162169}

The TIC 276162169 system may be the most interesting of the current set of 9 triples.  From the photodynamical models, which incorporate a somewhat sparsely sampled, but very useful, ETV curve, the outer orbit appears to be retrograde with respect to the inner binary, with a formal value for the mutual inclination angle of $i_{\rm mut} =179\fdg2\pm0\fdg5$.  While both orbital inclination angles are within a fraction of a degree of $90^\circ$, the photodynamical fit is distinctly superior for a retrograde outer orbit than for a prograde orbit.\footnote{In contrast to TIC~47151245, in this case the ETV curve, which has contributions from both LTTE and dynamical effects that are of the same order of magnitude, clearly reveals that of the two sets of \textit{TESS} observed third-body eclipses, the first one (in Sector 14) occurred during a superior conjunction of the tertiary star, while the second one (in Sector 55) happened around its inferior conjunction. Thus, the type of ambiguity that occurred in the triple system TIC 47151245, does not exist here.}

The other unique thing about this system, at least among the current set of 9 triples, in that the tertiary star is distinctly lower in mass than either of the two stars in the binary: $M_{\rm Aa} = 2.6\pm0.2$ M$_\odot$, $M_{\rm Ab} = 2.2\pm0.1$ M$_\odot$, and $M_{\rm B} = 1.49\pm0.05$ M$_\odot$.  The system is also relatively young at 300 Myr.

Due to the potentially important finding of an outer retrograde orbit, we would like to see this confirmed with a good set of radial velocity measurements.  At G = 12 this should not be particularly challenging.  

Finally, we note that before our current analysis, which is based on the high-precision \textit{TESS} observations, this system was misclassified as a semi-detached, classic Algol-type EB with a period of exactly half the actual eclipsing period \citep[see, e.~g.][]{buddingetal04,samusetal17}. The origin of this misclassification might have come from the fact that the primary and secondary eclipses have nearly the same depths, which made them largely indistinguishable in the lesser quality ground-based observations. What is more interesting, however, is that in these catalogs, in addition to the 0.5\,mag primary eclipse depth, a shallow 0.1\,mag secondary eclipse amplitude is also reported. In our interpretation, this fact might indicate an historical observation of a third-body event which was probably thought to be a secondary eclipse -- most likely reported by Eva Ahnert-Rohlfs about 70 years ago \citep{hoffmeisteretal54}.

\subsubsection{TIC 280883908}

TIC  280883908 has one of the most robustly determined sets of system parameters among our set of 9 systems thanks to the availability of RV data, a meaningful ETV curve, and the detection of both types of outer eclipses in the archival data.  The outer orbital period is 184.2 days, with a medium-level outer eccentricity 0.260.  The EB period is 5.4218 days.  The tertiary in the system is also one of the two largest among our collection, with $R_B = 13.2 \pm 0.5$\,R$_\odot$. The mutual inclination angle is $i_{\rm mut} = 2.3^\circ$, but with uncertainties that make it still consistent with $0^\circ$.  

The tertiary, with a mass of 1.78 M$_\odot$, has evolved substantially off the main sequence while the EB primary, at 1.37 M$_\odot$, is only slightly evolved.  The SED fit is in agreement with the full photodynamical fit, and it yields $R_B = 14.3 \pm 2.3$\,R$_\odot$, which is quite consistent with the photodynamical analysis value for $R_B$. All of the masses deduced from the SED alone are in excellent agreement with the photodynamical results.  

Because this system is so well diagnosed, we compare the full outer orbital solution parameters from Gaia with those from our photodynamical fit with the uncertainties:  $P_{\rm out} = 184.12 \pm 0.07$, $e_{\rm out} = 0.256 \pm 0.004$, $\omega_{\rm out} = 142^\circ \pm 1^\circ$, and $K_{\rm ter} = 38.06 \pm 0.17$ km s$^{-1}$ for Gaia and $P_{\rm out} = 184.35 \pm 0.02$\footnote{Note that this is the outer eclipsing period (see Table~\ref{tab:ephemerides})}, $e_{\rm out} = 0.260 \pm 0.001$, $\omega_{\rm out} = 146.4^\circ \pm 0.3^\circ$, and $K_{\rm ter} = 38.01 \pm 0.05$ km s$^{-1}$ from our photodynamical fit.  These are in substantial agreement, but the uncertainties on the photodynamical parameters are several times smaller than for Gaia. 

The distance from the full photodynamical fit is $1183\pm40$ pc, and $1275 \pm 180$ pc from the SED fit alone.  These are to be compared to the Gaia distance of $3072^{+2200}_{-900}$ pc, but Gaia reports a very large RUWE value of 9.8, so we are rather more confident of the photometric distance.   The fitted $E(B-V)$ is close to 0.30, while we had no cataloged working value to start with.  The age of the system is found to be 1800 Myr.

\subsubsection{TIC 294803663}

TIC 294803663 is one of two of our systems (along with TIC 280883908) that has a quite evolved tertiary star with $R_B = 13.4\pm0.2$\,R$_\odot$.  The masses are all comparable at: $M_{\rm Aa} = 2.24\pm0.07$ M$_\odot$, $M_{\rm Ab} = 1.93\pm0.07$ M$_\odot$, and $M_{\rm B} = 2.8\pm0.1$ M$_\odot$.  However, the extra few tenths of a solar mass are just sufficient to have allowed the tertiary to become substantially evolved, while the EB stars are still near the main sequence.  

The system is quite flat ($i_{\rm mut} = 3\fdg1_{-0.7}^{+1.6}$) and viewed nearly edge on with $i_{\rm out} = 88\fdg7\pm0\fdg1$.  The outer orbital eccentricity is fairly small at $e_{\rm out} = 0.030 \pm 0.010$.  The outer period of $153.43\pm0.05$ days and the outer eccentricity are in good agreement with the corresponding Gaia values of $P_{\rm out} = 153.1 \pm 0.2$ d and $e_{\rm out} = 0.047 \pm 0.01$.  However, our value for the outer argument of periastron of $\omega_{\rm out} = 313 \pm 14$ is not in complete accord with Gaia's value of $\omega_{\rm out} = 252 \pm 18$, even given the large uncertainty in the Gaia value.  Our photodynamical distance and Gaia's compare very favorably at $1739\pm26$\,pc and $1770\pm55$\,pc, respectively.

One of more unusual features of this system, among our collection, is that there are prominent stellar pulsations (see Fig.~\ref{fig:294803663_oscillation}) which we believe are from the giant tertiary star.  We reach this conclusion because (i) the EB is completely blocked during the {\it TESS} third body event and the pulsations are still present, and (ii) the giant tertiary comprises 75\% of the system flux, and only this is likely to accommodate the observed large pulsation amplitude.

\subsubsection{TIC 332521671}
\label{Sect:TIC 332521671}

TIC 332521671 is the second of our systems with a short outer orbital period of 48.5 days, similar to that of TIC 81525800.  But, in this case, the tertiary star is quite evolved with R = 8.4 R$_\odot$, compared to that of TIC 81525800 where the tertiary star is still only barely evolved away from the MS.  

For this system, we have a nicely sampled RV curve with 11 points around the outer orbit, as well as an ETV curve with a reasonable fraction of the outer orbit sampled on each of two occasions separated by two years.  The remarkable thing about this system is the extremely small outer orbital eccentricity of $e_{\rm out} = 0.004 \pm 0.002$.  It seems likely that the combination of a short outer orbital period and a large radius for the tertiary has led to tidal circularization.

While the mutual inclination angle is small ($i_{\rm mut} = 3^\circ$), i.e., the system is flat, we are actually viewing the outer orbit $5^\circ$ from edge on.  It is the large radius of the tertiary which still allows the outer eclipses to be observed.

Here we briefly discuss the reliability of the rapid dynamically-forced apsidal motion that is suggested by our photodynamical solution (see bottom panel of Fig.~\ref{fig:TIC332521671}). One should be cautious about accepting this at face value after noticing that during both sections of the ETV curves---where actual observations are available---one does not observe divergent ETV curves, but rather, the primary and secondary ETV curves overlap each other. In order to investigate this issue, we carried out further numerical integrations by setting the osculating inner eccentricity strictly to zero\footnote{One should keep in mind, however, that for a perturbed triple system the osculating eccentricity cannot remain zero for all times. The same is true for tidally distorted stars and, especially for the present combination of them, as was discussed, e.g., in \citet{kiselevaetal98} and \citet{borkovitsetal02}.} at epoch $t_0$. The only parameter that we modified in the different new runs was the apsidal motion parameter ($k_2$) of each star, thereby controlling the magnitude of the tidal contribution to the motion of the stars. This included the ($k_2=0$) case as well, which literally means that we omitted the tidal contribution during the integration of the motion of the three stars, considering them to be point masses. What we found is that all MCMC runs converged to solutions with very similar system parameters, and the ETV curves fit well the observed ETV points.  But the fits exhibited different kinds of ``apsidal motion'' behaviour during the non-constrained sections (i.e., where there are no data). Hence, our conclusion is that the apparent low amplitude, but rapid, dynamically-forced apsidal motion is only some numerical artefact which is the consequence of the insufficient data coverage in the case of this practically circular eclipsing binary ($e \simeq 0.003$).

\subsubsection{TIC 356324779}

TIC 356324779 yielded one of the two most useful outer orbit folds based on the ASAS-SN and ATLAS archival data (the other being TIC 47151245; see Fig.~\ref{fig:outer_fold}).  For TIC 356324779 we clearly see both primary and secondary outer eclipses during the 86.65-day outer orbit.  From the eclipse spacing we find that $e_{\rm out} \cos \omega_{\rm out} = 0.22$.  This is in terrific accord with our photodynamical solution of $e_{\rm out} = 0.284$ and $\omega_{\rm out} = 144^\circ$.

The system is rather flat ($i_{\rm mut} = 1.3^\circ \pm 0.5^\circ$) and viewed very close to exactly edge on with $i_{\rm in} = 90.1^\circ$ and $i_{\rm out} = 89.3^\circ$.  The masses are all fairly similar, with the tertiary being more massive by only 0.25 M$_\odot$: $M_{\rm Aa} = 1.8$ M$_\odot$, $M_{\rm Ab} = 1.7$ M$_\odot$, and $M_{\rm B} = 2.0$ M$_\odot$.

\subsection{Common Properties}
\label{sec:common}

The nine triply eclipsing triples we report on here have a number of properties in common, though they differ in some important details.  Here we mention briefly some of their shared characteristics and differences.

The outer orbital periods of the 9 triples range from 47.8 days to 604 days, with two shorter than 50 days and two longer than 250 days.  The outer orbital eccentricities range from 0.004 to 0.61, with no obvious correlation between $P_{\rm out}$ and $e_{\rm out}$.  Since the outer orbital semi-major axes range from $\sim$1/2 to 2 AU, we consider all of them to be `close' systems, but they are not particularly `tight' in the sense of having a small value of $P_{\rm out}/P_{\rm in}$, with ratios in the range of 25-230.   

All of the 9 triples are observed rather edge-on with respect to the outer orbit with values of $i_{\rm out}$ differing from $90^\circ$ by between $0.2^\circ$ and $4.5^\circ$.  The latter, of course, is something of a selection effect since we were searching for outer orbit eclipses.  Seven of the nine triples are rather flat, with $i_{\rm mut} \lesssim 5^\circ$.  However, the remaining two systems have distinctly misaligned orbits with $i_{\rm m} = 20^\circ, 40^\circ$.  One of the flat systems, TIC 276162169, however, presents the exciting possibility of a compact triple with a retrograde outer orbit, as the preferred photodynamical solution gives mutual inclination of $i_\mathrm{mut}=179\degr$. Note, while we are fairly certain of this result, we would like to see it confirmed by further measurements of TIC 276162169. 

All the tertiary masses lie in the range of 1.4-2.8 M$_\odot$, with a mean of 1.95 M$_\odot$.  Four of the tertiary stars are substantially evolved away from the main sequence, with radii of 8, 11, 13, and 13 R$_\odot$ with $T_{\rm eff}$ in the range of 4700-4950 K\footnote{In a small minority of our systems, we cannot exclude the possibility that the tertiary might have ascended the RGB and then returned to near the horizontal branch (HB).  However, this is possible only in those cases where the tertiary does not overflow its Roche lobe at the tip of the RGB, which is allowed in only three of the four systems where the tertiary happens to be located near the HB. A detailed investigation of this question is outside the scope of this paper, and whether one or more of the tertiaries is actually on the HB is largely inconsequential for the determinations of our stellar and orbital parameters.}.  Three other tertiaries are mildly evolved off the MS.  In all cases except for TIC 276162169, the tertiary is either the most massive of three stars, or close to the mass of binary star $Aa$.  By contrast, in TIC 276162169, the tertiary mass is only 1.49 M$_\odot$, while the EB stars are considerably more massive at 2.2 and 2.6 M$_\odot$.  The fact that a fair fraction of the tertiary stars are evolved is a selection effect in that larger tertiary stars make it more likely to exhibit third-body eclipses, while at the same time the EB stars cannot grow too large without overflowing their Roche lobes.

The range of the 18 stellar masses in the 9 EBs is 0.95-2.6 M$_\odot$, which is not so different from the range of tertiary masses.  However, as mentioned above, in all but one case, the  tertiary manages to be the most massive star in the system, even if not by much. 

We find that the ages of the 9 triples lie in the range of 0.3-2.5 Gyr, with 7 of the systems older than 1 Gyr.  Thus, the systems are nearly as old as is required for the most massive star to start its ascent of the giant branch, thereby enhancing the detection probability, while avoiding Roche lobe overflow in the EB.  This is consistent with the intermediate mass stars we are observing.

Finally, we note that eight of the triples in this collection show no evidence for any stars other than the three stellar components of the triple.  This includes looking for non-linear behavior in our ETV curves beyond those induced by the tertiary star on the EB, and by checking Gaia for any bound companion over a separation range of 0.5$''$ to 20$''$, and down to 4 magnitudes fainter than the triple.  On the other hand, TIC 229785001 clearly includes at least one other bound star, making this a somewhat rare 2+1+1 hierarchical quadruple.

\section{Summary and Discussion}
\label{sec:summary}

In this work we have presented 9 compact triply eclipsing triples with a full evaluation of their system parameters -- both stellar and orbital.  We found all of these by searching through millions of {\it TESS} photometric lightcurves looking for third body eclipses in binary systems (see e.g., \citealt{kristiansen22,rappaport22}).  While we were carrying out the research for this paper, the outer eclipses in four of them were also found independently by \citet{zasche22}, but no further analyses of them were carried out in that work.  For four of the systems, Gaia has measured the radial velocities of the outer orbit of the tertiary (in three cases), or the photocenter of the system (in one case).  No other analyses were done, nor was there any way to use the Gaia data to even know that one of the components in the system was a binary.  

\begin{figure*}
\begin{center}
\includegraphics[width=0.323 \textwidth]{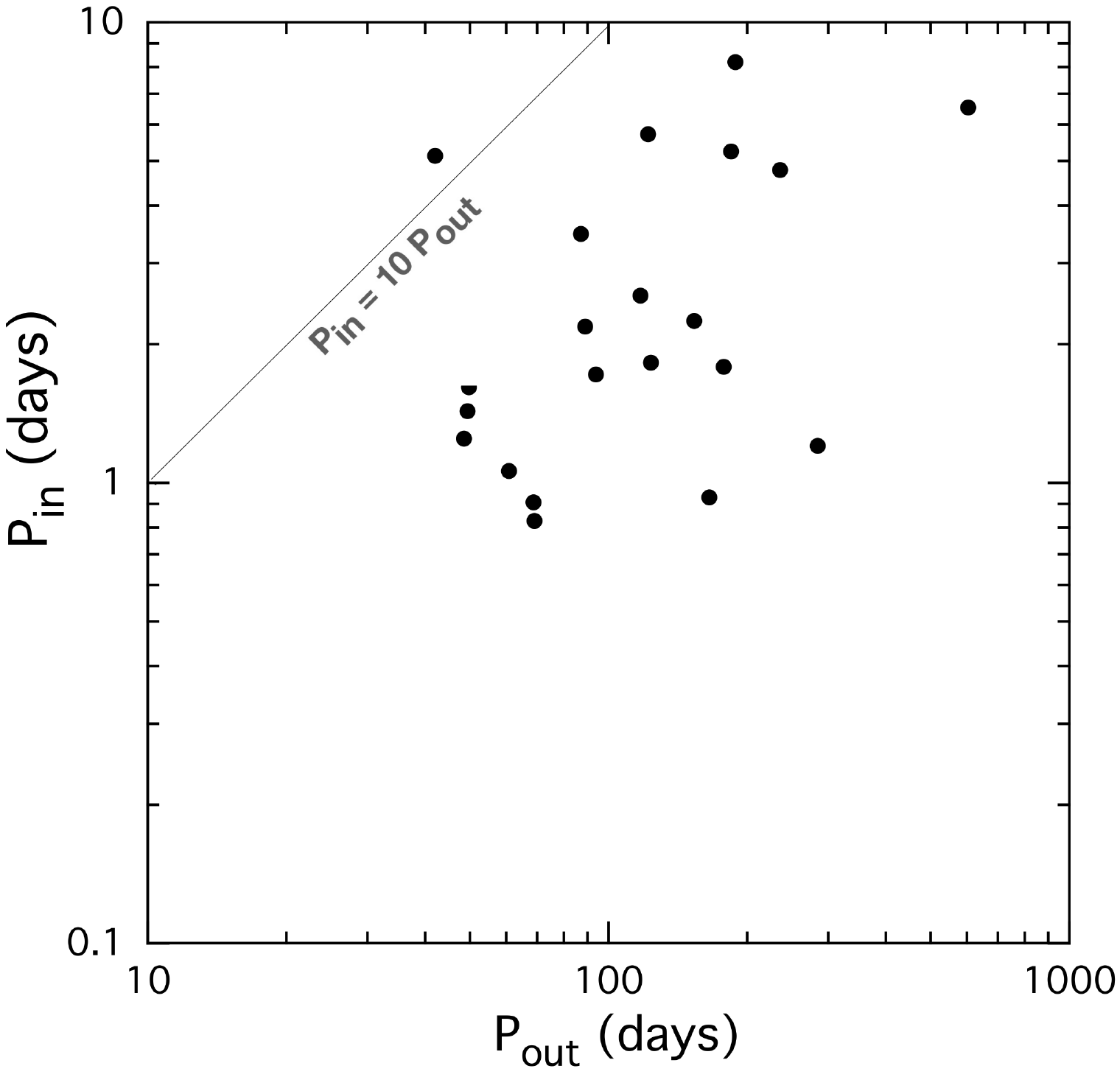}
\includegraphics[width=0.329 \textwidth]{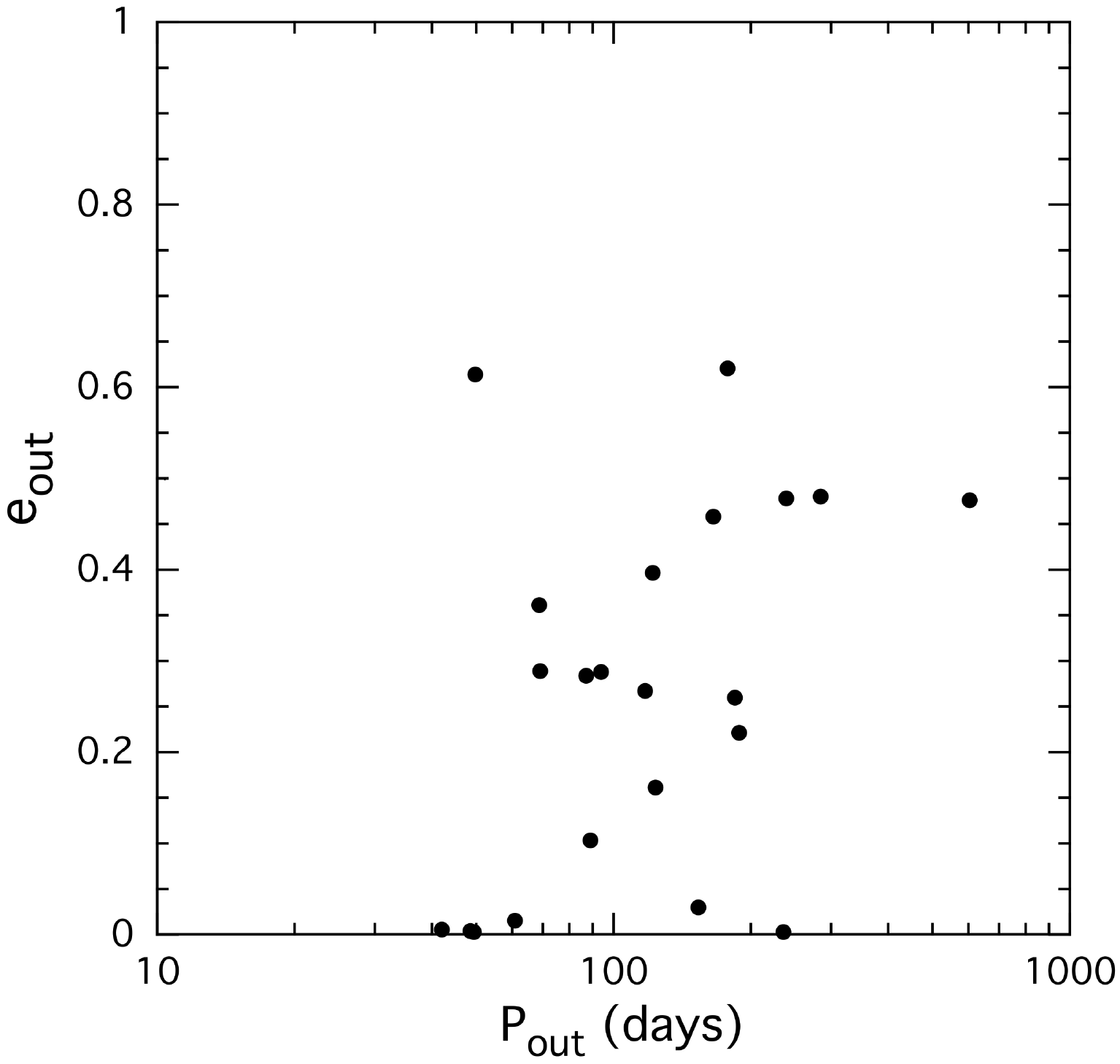} \hglue-0.15cm
\includegraphics[width=0.32 \textwidth]{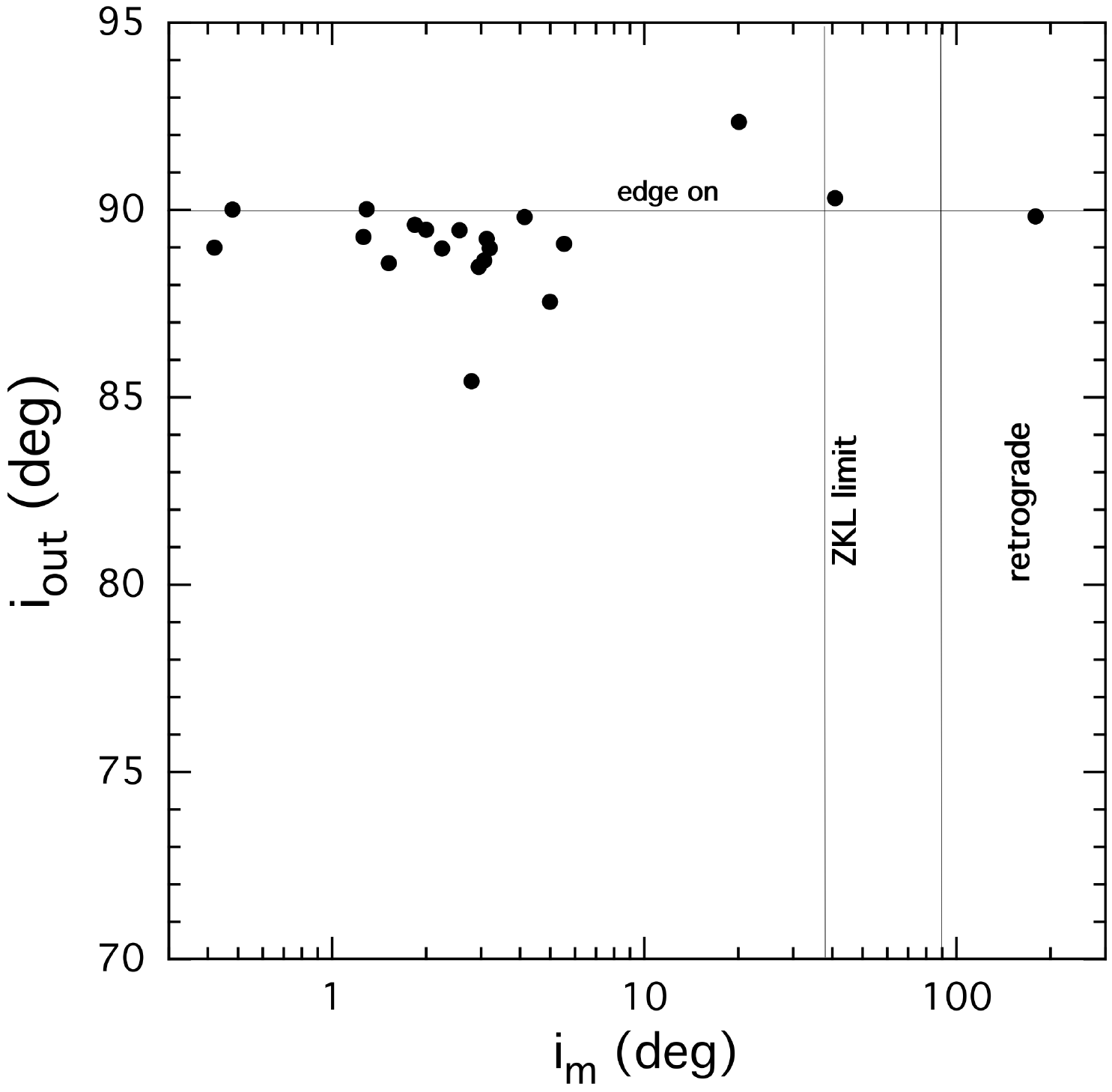} \hglue0.1cm  \hglue-0.10cm
\includegraphics[width=0.318 \textwidth]{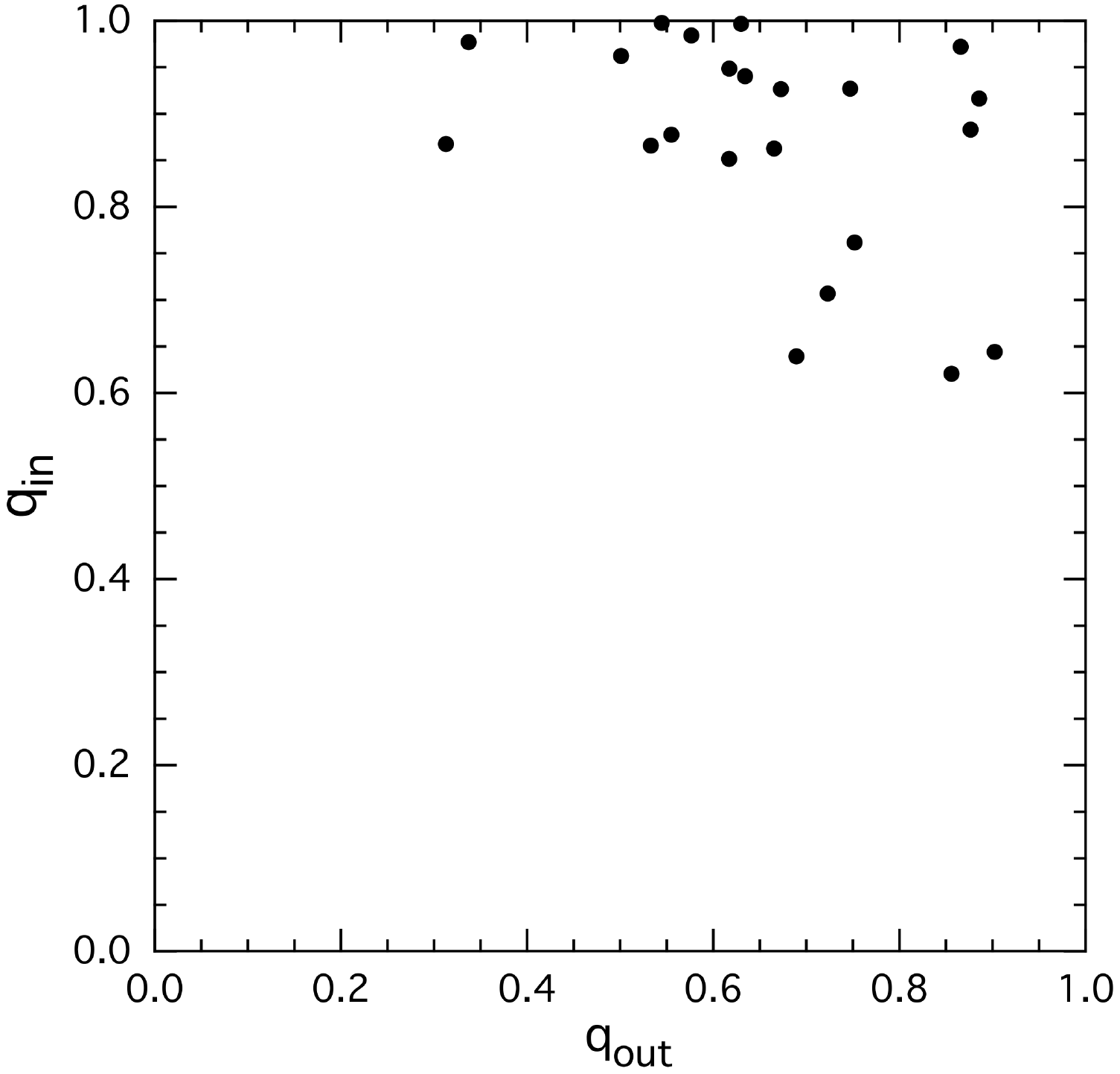} \hglue0.15cm
\includegraphics[width=0.316 \textwidth]{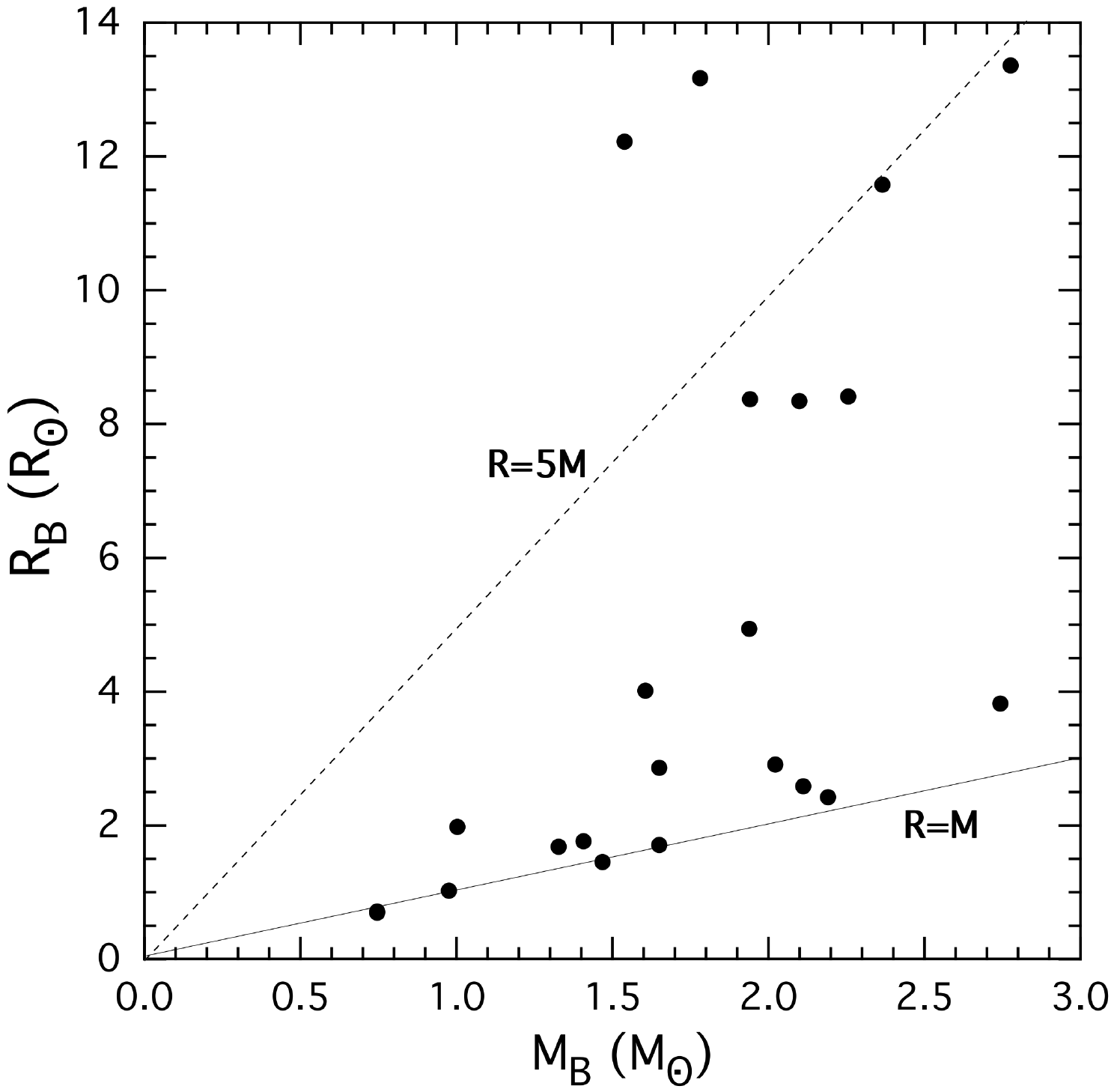} \hglue0.2cm
\includegraphics[width=0.305 \textwidth]{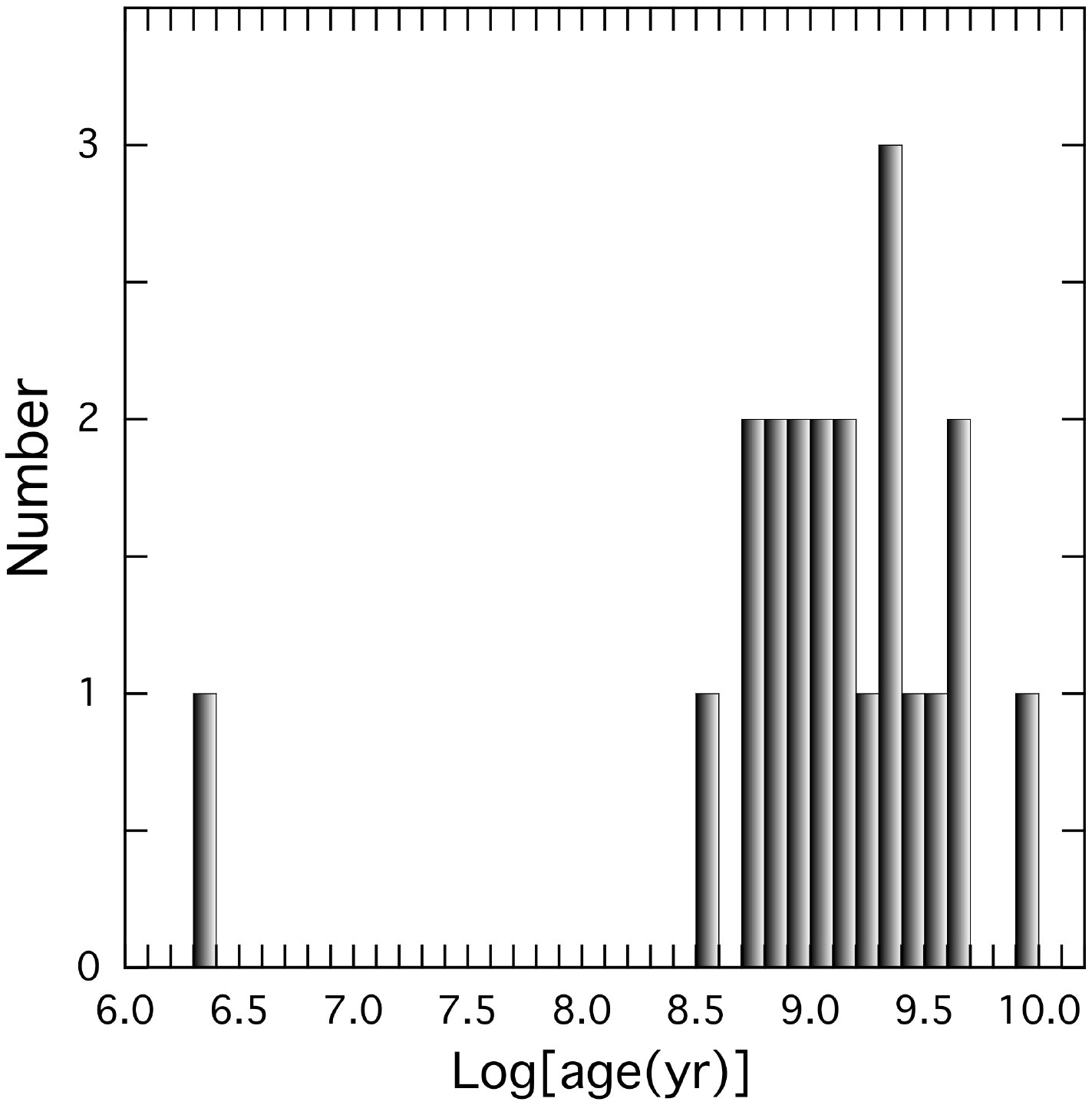}   

\caption{Statistical plots for properties of 22 triply eclipsing triples uniformly analyzed (see text for references).  {\it Top-row panels:} $P_{\rm in}$ vs.~$P_{\rm out}$, $e_{\rm out}$ vs.~$P_{\rm out}$, and $i_{\rm out}$ vs.~ $i_{\rm mut}$. The diagonal dotted line in the first plot is a rough estimate of the dividing point between stable and unstable orbits for mildly eccentric outer orbits.  In the upper rightmost panel the vertical lines denote the transition to the ZLK cycles, and to retrograde orbits, respectively. {\it Bottom-row panels:} $q_{\rm in}$ vs.~$q_{\rm out}$, $R_{\rm B}$ vs.~$M_{\rm B}$, and a 1D distribution of log system ages. The sloped dashed lines in the middle panel are for $R_{\rm B}$ =1\,M$_{\rm B}$ and = 5\,M$_{\rm B}$, as rough guides of unevolved and quite evolved stars, respectively.}
\label{fig:statistics}
\end{center}
\end{figure*} 

We utilized the {\it TESS} photometric lightcurve, the ETV points derived from the lightcurve, archival SED data, archival photometry from ASAS-SN and ATLAS, and in some cases, ground-based follow up RV observations as well as eclipse photometry.  These were combined in a complex photodynamical analysis where we solve for all the system parameters, as well as the distance to the source.  Typical uncertainties on the masses and radii are in the range of a couple of per cent to not much more than $\sim$5 per cent.  Uncertainties on the angles associated with the orbital planes (e.g., $i_{\rm out}$ and $i_{\rm mut}$) range from a fraction of a degree to about a degree.  All of the many system parameters that we have solved for, and the associated uncertainties, are given in Tables \ref{tab:syntheticfit_TIC47151245+81525800} through \ref{tab:syntheticfit_TIC356324779}.

In Fig.~\ref{fig:statistics} we present a set of correlation plots for some of the physically interesting parameters associated with our collection of triply eclipsing triples.  To the nine sources we have studied in this work, we add the six triples from our previous closely related paper \citep{rappaport22}, as well as seven triples studied in \citet{borkovitsetal19a,borkovitsetal20b,borkovitsetal22,borkovitsetal22b}, and \citet{mitnyanetal20} using largely the same selection criteria and methods of analysis.

The plot of $P_{\rm in}$ vs $P_{\rm out}$ shows that the bulk of the outer periods range from 48-285 days, while the inner periods span $\sim$0.8-7 days. The dotted diagonal line represents the dynamical stability limit taken from Eqn.~(\ref{eqn:stability}) for a nominal case where $e_{\rm out} \simeq 0.3$. Most systems stay comfortably away from this limit.

The top middle panel in Fig.~\ref{fig:statistics} shows how $e_{\rm out}$ varies with $P_{\rm out}$.  In general, there is little correlation, except to note that for the most circular outer orbits, the outer periods tend to be relatively short.  The circularization is likely brought about by tidal damping with an evolved tertiary.  

The top right panel of Fig.~\ref{fig:statistics} shows the inclination angle of the outer orbit vs. the mutual inclination angle of the inner binary vs.~the outer orbit.  The fact that most of the values of $i_{\rm out}$ hover near $90^\circ$ is due to selection effects since these triples were discovered by searching for third-body eclipses. To a lesser extent, the same selection also holds for the alignment of the inner binary, otherwise third body eclipses would be somewhat more difficult to detect, but this is mitigated by the smaller values of orbital separations in the EB more so than for the outer orbit.  Two of the systems have large enough $i_{\rm mut}$ (20$^\circ$ and 40$^\circ$) to undergo serious precession of their orbital planes, where the one with the larger tilt (TIC 47151245) is, in principle, potentially subject to ZLK cycles (but see the discussion in Sect.~\ref{sec:info_47151245}).  Finally, there is one triple (TIC 276162169) in a nearly flat system, but where the outer orbit is {\it retrograde} with respect to the inner EB.  These are rare systems (see Sect.~\ref{sec:info_276162169}).  

The $q_{\rm in}$ vs.~$q_{\rm out}$ plot (lower left panel in Fig.~\ref{fig:statistics}) shows two noteworthy features.  First, most of the inner EBs have mass ratios not far from unity, with a mean of 0.87.  This may say more about the intrinsic mass ratio distribution in EBs, than anything necessarily to do with triple stars.  On the other hand, it may be predictable from the accretion scenario for forming triple-star systems \citep{tokovininmoe20}.  Second, and more interesting, is that the ratio of $q_{\rm out} \equiv M_{\rm B}/M_{\rm EB}$ never exceeds unity (and not just by definition) and has a mean value of 0.66.  This strongly favors the accretion formation scenario of triples \citep[see][for a recent review]{offneretal22}, and is not obviously caused by any selection effects.

The relation between the tertiary mass and its radius is shown in the bottom middle panel of Fig.~\ref{fig:statistics}.  Some 8 of the stars are fairly well clustered not too far from the main sequence, crudely represented by the $R=M$ line shown in the figure.  However, the other 14 of the tertiary stars are distinctly evolved away from the main sequence.  That is due both to an observational selection effect (larger stellar radii lead to more probable third body eclipses), as well as to the physical limitation whereby only the tertiary can grow this large (the EB stars are confined to their Roche lobes in relatively close orbits).

The age distribution of the systems is shown in the bottom right panel of Fig.~\ref{fig:statistics}.  All but three of the systems have ages between 1/2 and 5 Gyr.  The biggest exception is TIC 52041148 for which we found only a pre-main-sequence solution \citep{borkovitsetal22}.  Thus, 21 of these triples are manifestly long-term dynamically stable and will last until the tertiary overflows its Roche lobe.  In principle, the EB stars could also evolve to mass exchange, but the tertiaries in these systems are sufficiently more massive (in all but a couple of cases), so that they will fill their Roche lobes first, before the EB stars have reached Roche-lobe overflow.  (The tertiaries' Roche lobes range in size from 27 to 187 R$_\odot$).  On the other hand just 2/3 of the tertiaries will reach their Roche lobe radii on the ascent of the red giant branch.  However, they would all do so on their ascent of the asymptotic giant branch.  

Here we briefly follow up on two facts noted above: (i) the mass of all tertiaries is less than the mass of the inner EB, and (ii) in half of the 22 systems, the tertiary stars have evolved to at least twice their main sequence radius.  The latter point is something of an observational selection effect since larger stars have a greater probability of eclipsing the other star(s) in a wide orbit.  And, the tertiary is more free to expand than the EB stars which are confined in much closer orbits.  In our photodynamical analyses we have adopted the assumption that the tertiary star is coeval with the inner EB stars and, more specifically, that no mass has been lost or transferred among the stars.  This implicitly includes the assumption that the tertiary star itself was always a single star, and not the merger product of another binary star, e.g., starting life as a member of a quadruple system, where one of the binaries evolved and merged into a single star.

This scenario can be substantially ruled against by the fact that the mass of the current tertiary star is always (at least in our sample of 22 systems) less than the mass of the inner binary star.  This demonstrates that the tertiary is unlikely to be the merger product of a binary that was more massive than the current inner EB\footnote{We note, however, that there is actually a small region of parameter space for a binary B to be less massive than its companion binary A, but still contain the most massive star in the system.  Example: $M_{\rm Aa} = 1.2\,M_\odot; M_{\rm Ab} = 1.0\,M_\odot; M_{\rm Ba} = 1.5\,M_\odot; M_{\rm Bb} =0.5\,M_\odot$. So, while $M_A > M_B$, the B binary would still evolve to merger first. In addition, mergers may also result in mass loss, thereby rendering the initially more massive binary, the less massive one.}.  However, there is still the possibility that the tertiary was a merger product of a lower mass binary that evolved to merger first due to magnetic braking (see, e.g., \citealt{verbunt81}); though, the four masses would have to be tuned just right.  We might also mention that all but four of the systems in Fig.~\ref{fig:statistics} have outer periods $< 200$ days.  Since there are currently no published papers claiming any quadruple with an outer period this short, we conclude that triples arise from a distinct evolutionary channel than that of the quadruples.

Now that the 22 triply eclipsing triples discussed above (see Fig.~\ref{fig:statistics}) are known, and their basic parameters determined, the group of them would make for an excellent follow-up ground-based eclipse timing project.  This is especially true for modest-size amateur telescopes since the vast majority of the 22 objects have G magnitudes of $\lesssim 13$.  The ETV data from {\it TESS} itself was often instrumental in determining some of the parameters found from the photodynamical analyses. Thus, future timing observations of the ordinary EB eclipses in these systems would be quite helpful in significantly improving the parameter determinations.  The dynamical delays in these systems as well as the LTTE delays are typically in the minute range, so readily within the realm of amateur observations.  There are a sufficient number of these triply eclipsing triples so that anyone interested in pursuing such timing observations would have one of them to observe nearly every night.  

For the purpose of predicting accurate timings for future eclipse events, the periods and epochs (conjunction times) given in Tables~\ref{tab:syntheticfit_TIC47151245+81525800}--\ref{tab:syntheticfit_TIC356324779}, however, cannot be used. The reason, as explained earlier, is that the periods tabulated there are instantaneous, osculating ones that differ substantially from the average eclipsing period.  The latter can be deduced, e.g., from ETV curves or the folds of archival data accumulated over nearly a decade. Hence, similar to Table~6 of \citet{rappaport22} we give ephemerides for planning future eclipse observation of the nine triples in Table~\ref{tab:ephemerides}.

Finally, we note that \citet{{czavalinga22}} have presented an alternative method for finding substantial numbers of new compact hierarchical triples to the approach taken in this work.  They matched known eclipsing binaries with the orbital solutions that Gaia is now including (\citealt{babusiaux22}; \citealt{gaia22}) with their distance and kinematic findings. Thus far, this study has already potentially increased the number of compact triples by some 50\%.  Going forward, this approach is a good complement to the study of triples presented in this work.  In addition, the approach present by \citet{{czavalinga22}} can also lead directly to the discovery of new triply eclipsing systems---and indeed three new triply eclipsing systems were reported in that work.  And, of course, the outer period is automatically known once a triple has been found by matching known EBs with Gaia orbital solutions.

\begin{table*}
\centering 
\caption{Derived ephemerides for the nine triple systems to be used for planning future observations.}
 \label{tab:ephemerides}
 \begin{tabular}{llllllllll}
 \hline 
TIC ID               & 47151245       & 81525800  & 99013269 & 229785001       & 276162169  & 280883908 & 294803663 & 332521671 & 356324779 \\
\hline
&\multicolumn{9}{c}{Inner binary} \\
\hline
$P$  & 1.201783 &  1.64935   & 6.534695 & 0.92951 &  2.55018   &  5.24795 & 2.246005 & 1.247149 & 3.47683\\
     &          &            &          &         &            &          & 2.246025 &          &        \\
$\mathcal{T}_0$ & 8\,627.453 & 9\,475.556 & 8\,717.199  & 8\,683.4830 & 8\,685.522 & 8\,984.923 & 8\,598.513 & 8\,571.352 & 8\,819.194\\
                &            &            &             &             &            &            & 8\,598.543 &            &           \\
$\mathcal{A}_\mathrm{ETV}$  & 0.003 & 0.013 & 0.012 & 0.0015 & 0.003 & 0.010 & 0.0025 & 0.001 & 0.010\\
$D$  & 0.150 & 0.247 & 0.327  & 0.132 & 0.281 & 0.252 & 0.265 & 0.224 & 0.278\\
\hline
&\multicolumn{9}{c}{Wide binary (third body eclipses)} \\
\hline
$P$ & 284.90 & 47.85 & 604.05 & 165.25 & 117.10 &  184.35 & 153.20 & 48.51 & 86.88 \\
$\mathcal{T}_0^\mathrm{inf}$  &  9\,368.4 &  9\,523.7 & 9\,421.0 &  (9\,686.1) & 9\,813.1 &  9\,629.0 & 8\,629.2 & 9\,335.1: & 9\,648.9:\\
$D^\mathrm{inf}$ & 1.1 & 0.6 &  2.0 & (0.4) & 1.6 & 7.2  & 4.2 & 1.6 & 3.3\\
$\mathcal{T}_0^\mathrm{sup}$ &  9\,294.9: &  $-$ &  $-$ &  9\,731.9 & 8\,709.0 &  9\,748.1 & 9\,317.8: & 9\,310.8 & 8\,837.5\\
$D^\mathrm{sup}$ & 2.2 & $-$ &  $-$ & 0.8 & 2.8 &  5.5  & 4.1 & 1.3 & 1.3\\
\hline
\end{tabular}

\textit{Notes.} (a) For the inner pairs: $P$, $\mathcal{T}_0$, $\mathcal{A}_\mathrm{ETV}$, $D$ are the period, reference time of a primary minimum, half-amplitude of the ETV curve, and the full duration of an eclipse, respectively. $\mathcal{T}_0$ is given in BJD -- 2\,450\,000, while the other quantities are in days. As all but one of the inner eccentricities are very small and, hence, the shifts of the secondary eclipses relative to phase 0.5 are negligible (quantitatively, they are much smaller than the full durations of the individual eclipses), the same reference times and periods can be used to predict the times of the secondary eclipses. The only exception is the inner EB of TIC~294803663. For this system we give a separate period and reference time for the secondary eclipses, listing them below the primary ephemerides. (b) For the outer orbits we give separate reference times for the third body eclipses around the inferior and superior conjunctions of the tertiary component. The eclipse durations, $D$, of the third-body eclipses do not give the extent of any specific third body events.  Rather $D$ represents the time difference corresponding to the very first and last moments around a given third-body conjunction when the first/last contact of a third-body event may occur). Double dots (:) call attention to the less certain superior/inferior conjunction times at those types of third-body events (i.e., primary vs.~secondary outer eclipses) because they were not observed by either \textit{TESS} or KELT. The absence of a reference time at a given type of conjunction indicates the absence of a third-body eclipse at those conjunctions. Note, the superior conjunction data of TIC~229785001, given in parentheses, indicate that only very shallow third-body eclipses may occur which can hardly be observed with ground-based instruments.

\end{table*} 

\section*{Data availability}

The \textit{TESS} data underlying this article were accessed from MAST (Barbara A. Mikulski Archive for Space Telescopes) Portal (\url{https://mast.stsci.edu/portal/Mashup/Clients/Mast/Portal.html}). The ASAS-SN archival photometric data were accessed from \url{https://asas-sn.osu.edu/}. The ATLAS archival photometric data were accessed from \url{https://fallingstar-data.com/forcedphot/queue/}. A part of the data were derived from sources in the public domain as given in the respective footnotes. The derived data generated in this research and the code used for the photodynamical analysis will be shared upon a reasonable request to the corresponding author.

\section*{Acknowledgments}

We thank an anonymous referee for some very insightful comments and suggestions.

We are grateful to Allan R. Schmitt for making his lightcurve examining software tools LcTools freely available.

TB and ZG acknowledge the support of the Hungarian National Research, Development and Innovation Office (NKFIH) grant K-125015, a PRODEX Experiment Agreement No. 4000137122 between the ELTE E\"otv\"os Lor\'and University and the European Space Agency (ESA-D/SCI-LE-2021-0025), and the support of the city of Szombathely. ZG acknowledges the VEGA grant of the Slovak Academy of Sciences No. 2/0031/22, and the Slovak Research and Development Agency contract No. APVV-20-0148.

VBK is thankful for support from NASA grants 80NSSC21K0351. 

AP acknowledges the financial support of the Hungarian National Research, Development and Innovation Office -- NKFIH Grant K-138962.

We thank David Latham for facilitating the TRES observations of TIC 280883909 at the F.\,L.\,Whipple Observatory.  The TRES observations were obtained with the able help of P.\,Berlind, M.\,Calkins, and G.\,Esquerdo. We also thank J.\,Mink for maintaining the echelle database at the CfA.

The operation of the BRC80 robotic telescope of Baja Astronomical Observatory has been supported by the project ``Transient Astrophysical Objects'' GINOP 2.3.2-15-2016-00033 of the National Research, Development and Innovation Office (NKFIH), Hungary, funded by the European Union.

T. Mitnyan gratefully acknowledge observing grant support from the Institute of Astronomy and National Astronomical Observatory, Bulgarian Academy of Sciences. We thank Mitko Churalski for obtaining spectra of TIC 99013269 and TIC 280883908 at the NAO Rozhen Observatory.

This paper includes data collected by the \textit{TESS} mission. Funding for the \textit{TESS} mission is provided by the NASA Science Mission directorate. Some of the data presented in this paper were obtained from the Mikulski Archive for Space Telescopes (MAST). STScI is operated by the Association of Universities for Research in Astronomy, Inc., under NASA contract NAS5-26555. Support for MAST for non-HST data are provided by the NASA Office of Space Science via grant NNX09AF08G and by other grants and contracts.

We have made extensive use of the All-Sky Automated Survey for Supernovae archival photometric data.  See \citet{shappee14} and \citet{kochanek17} for details of the ASAS-SN survey.

We also acknowledge use of the photometric archival data from the Asteroid Terrestrial-impact Last Alert System (ATLAS) project.  See \citet{tonry18} and \citet{heinze18} for specifics of the ATLAS survey.

This work has made use  of data  from the European  Space Agency (ESA)  mission {\it Gaia}\footnote{\url{https://www.cosmos.esa.int/gaia}},  processed  by  the {\it   Gaia}   Data   Processing   and  Analysis   Consortium   (DPAC)\footnote{\url{https://www.cosmos.esa.int/web/gaia/dpac/consortium}}.  Funding for the DPAC  has been provided  by national  institutions, in  particular the institutions participating in the {\it Gaia} Multilateral Agreement.

This publication utilized data products from the Wide-field Infrared Survey Explorer, which is a joint project of the University of California, Los Angeles, and the Jet Propulsion Laboratory/California Institute of Technology, funded by the National Aeronautics and Space Administration. 

This work also utilized data products from the Two Micron All Sky Survey, which is a joint project of the University of Massachusetts and the Infrared Processing and Analysis Center/California Institute of Technology, funded by the National Aeronautics and Space Administration and the National Science Foundation.

We  used the  Simbad  service  operated by  the  Centre des  Donn\'ees Stellaires (Strasbourg,  France) and the ESO  Science Archive Facility services (data  obtained under request number 396301).   

Finally, we acknowledge the use of the VizieR catalogue access tool, CDS, Strasbourg, France (DOI : 10.26093/cds/vizier). The original description of the VizieR service was published in \citet{ochsenbein00}.






\appendix

\section{Supplementary Material -- RV data}
\label{app:RV_data}

In this appendix, we tabulate the radial velocity data obtained for four tertiary components out of our nine triples (Tables~\ref{tab:TIC99013269_RVdata}-\ref{tab:TIC332521671_RVdata}).

\begin{table}
\centering
\caption{Measured radial velocities of the tertiary component of TIC~99013269. The date is given as BJD -- 2\,450\,000, while the RVs and their uncertainties are in km\,s$^{-1}$.}
\label{tab:TIC99013269_RVdata}
\begin{tabular}{lrrlrr}
\hline
\hline
Date & RV$_\mathrm{B}$ & $\sigma_\mathrm{B}$ & Date & RV$_\mathrm{B}$ & $\sigma_\mathrm{B}$ \\  
\hline
$9564.20623^a$ & $ 39.49 $ & $ 0.34$ & $9782.39505^a$ & $ 10.65 $ & $ 0.17$ \\
$9575.21922^a$ & $ 38.06 $ & $ 0.45$ & $9785.51202^a$ & $ 10.70 $ & $ 0.26$ \\
$9583.21267^a$ & $ 35.94 $ & $ 0.29$ & $9824.46323^a$ & $  4.28 $ & $ 0.77$ \\
$9711.53417^b$ & $ 18.97 $ & $ 0.28$ & $9827.53333^a$ & $  5.06 $ & $ 0.19$ \\
$9718.53900^b$ & $ 18.08 $ & $ 0.26$ & $9830.39248^c$ & $  3.81 $ & $ 0.29$ \\
$9745.55576^a$ & $ 15.31 $ & $ 0.38$ & $9831.44676^c$ & $  3.92 $ & $ 0.26$ \\
$9746.54710^a$ & $ 14.38 $ & $ 0.30$ & $9857.29764^a$ & $  0.69 $ & $ 0.18$ \\
$9748.55289^a$ & $ 15.47 $ & $ 0.74$ & $9896.25484^a$ & $ -4.55 $ & $ 0.22$ \\
$9749.54329^a$ & $ 13.94 $ & $ 0.16$ & $9897.25350^a$ & $ -4.98 $ & $ 0.26$ \\
$9750.54483^a$ & $ 14.16 $ & $ 0.21$ & $9898.25200^a$ & $ -5.34 $ & $ 0.22$ \\
\hline
\end{tabular}

Notes. Instruments: $^a$ Konkoly; $^b$ Skalnat\'e Pleso; $^c$ Rozhen
\end{table}  

\begin{table}
\centering
\caption{Measured radial velocities of the tertiary component of TIC~280883908. The date is given as BJD -- 2\,450\,000, while the RVs and their uncertainties are in km\,s$^{-1}$.}
\label{tab:TIC280883908_RVdata}
\begin{tabular}{lrrlrr}
\hline
\hline
Date & RV$_\mathrm{B}$ & $\sigma_\mathrm{B}$ & Date & RV$_\mathrm{B}$ & $\sigma_\mathrm{B}$ \\  
\hline
$9473.008411$ & $-14.43$ & $0.18$ & $9567.743108$ & $-45.90$ & $0.19$ \\ 
$9479.909811$ & $-10.02$ & $0.16$ & $9585.735308$ & $-74.11$ & $0.16$ \\
$9489.986911$ & $ -5.21$ & $0.14$ & $9596.41174^a$ & $-74.43$ & $0.22$\\
$9500.912210$ & $ -1.48$ & $0.15$ & $9600.792707$ & $-70.85$ & $0.16$  \\
$9519.953010$ & $  0.12$ & $0.19$ & $9605.803607$ & $-65.95$ & $0.14$ \\
$9522.827010$ & $ -0.30$ & $0.16$ & $9615.820607$ & $-53.68$ & $0.19$ \\
$9524.803410$ & $ -0.37$ & $0.15$ & $9622.745707$ & $-45.20$ & $0.18$ \\
$9527.826110$ & $ -1.58$ & $0.15$ & $9635.636206$ & $-31.53$ & $0.15$ \\
$9531.770109$ & $ -2.87$ & $0.12$ & $9658.648005$ & $-13.66$ & $0.23$ \\
$9548.802309$ & $-15.56$ & $0.13$ & $9915.745996$ & $-13.94$ & $0.15$ \\
\hline
\end{tabular}

Notes. All but one RV data were gathered with TRES spectrograph. The only exception is superscripted with an $^a$ and was obtained with the 2m Rozhen telescope (Bulgaria).  The Rozhen point and the last two TRES RV data were not used for the analysis.

\end{table}  

\begin{table}
\centering
\caption{Measured radial velocities of the tertiary component of TIC~294803663. The date is given as BJD -- 2\,450\,000, while the RVs are in km\,s$^{-1}$.}
\label{tab:TIC294803663_RVdata}
\begin{tabular}{lrlr}
\hline
\hline
Date & RV$_\mathrm{B}$ & Date & RV$_\mathrm{B}$  \\  
\hline
$9700.700109$ & $-19.367$ & $9742.572900$ & $ 28.245$ \\ 
$9720.616709$ & $ 15.841$ & $9792.486631$ & $-43.705$ \\
$9725.621810$ & $ 21.267$ & $9837.471313$ & $-48.825$ \\
\hline
\end{tabular}

Notes. All RV data were obtained with CHIRON instrument.

\end{table}  

\begin{table}
\centering
\caption{Measured radial velocities of the tertiary component of TIC~332521671. The date is given as BJD -- 2\,450\,000, while the RVs are in km\,s$^{-1}$.}
\label{tab:TIC332521671_RVdata}
\begin{tabular}{lrlr}
\hline
\hline
Date & RV$_\mathrm{B}$ & Date & RV$_\mathrm{B}$  \\  
\hline
$9650.749645032$ & $-19.111$ & $9692.612050781$ & $ 25.653$ \\ 
$9656.684773510$ & $-58.279$ & $9699.574493482$ & $-21.489$ \\
$9659.753177934$ & $-70.110$ & $9702.580351819$ & $-42.815$ \\
$9671.652089117$ & $-36.776$ & $9716.612527877$ & $-58.393$ \\
$9677.611086097$ & $  6.692$ & $9725.529218565$ & $  2.654$ \\
$9684.681348653$ & $ 40.296$ & & \\
\hline
\end{tabular}

Notes. All RV data were obtained with CHIRON instrument. 
\end{table}  

\onecolumn

\section{Tables of determined times of minima for all the nine systems}
\label{app:ToMs}

In this appendix, we tabulate the individual mid-minima times of the primary and secondary eclipses, including mostly \textit{TESS}, and a few ground-based observed ones, for the inner EBs of the triples considered in this study (Tables \ref{Tab:V726_Sco_(TIC_047151245)_ToM}-\ref{Tab:TIC_356324779_ToM}).

The complete Appendix~\ref{app:ToMs} is available supplementary material to the journal, and also in this arXiv version.

\begin{table*}
\caption{Times of minima of TIC 047151245 (V726 Sco)}
 \label{Tab:V726_Sco_(TIC_047151245)_ToM}
\scalebox{0.92}{
\begin{tabular}{@{}lrllrllrllrl}
\hline
BJD & Cycle  & std. dev. & BJD & Cycle  & std. dev. & BJD & Cycle  & std. dev. & BJD & Cycle  & std. dev. \\ 
$-2\,400\,000$ & no. &   \multicolumn{1}{c}{$(d)$} & $-2\,400\,000$ & no. &   \multicolumn{1}{c}{$(d)$} & $-2\,400\,000$ & no. &   \multicolumn{1}{c}{$(d)$} & $-2\,400\,000$ & no. &   \multicolumn{1}{c}{$(d)$} \\ 
\hline
58626.858995 &   -0.5 & 0.000927 & 58640.670145 &   11.0 & 0.000923 & 59362.346333 &  611.5 & 0.000186 & 59377.365321 &  624.0 & 0.000232  \\ 
58627.454056 &    0.0 & 0.000550 & 58641.273052 &   11.5 & 0.000366 & 59362.941691 &  612.0 & 0.000181 & 59377.969436 &  624.5 & 0.000315  \\ 
58628.044977 &    0.5 & 0.000816 & 58641.873918 &   12.0 & 0.000757 & 59363.552102 &  612.5 & 0.000298 & 59378.573030 &  625.0 & 0.000249  \\ 
58628.657908 &    1.0 & 0.000442 & 58642.467826 &   12.5 & 0.000371 & 59364.153877 &  613.0 & 0.000173 & 59379.172045 &  625.5 & 0.000186  \\ 
58629.259583 &    1.5 & 0.000660 & 58643.071932 &   13.0 & 0.002422 & 59364.748281 &  613.5 & 0.000160 & 59379.769220 &  626.0 & 0.000215  \\ 
58629.855647 &    2.0 & 0.000322 & 58643.675185 &   13.5 & 0.000377 & 59365.349327 &  614.0 & 0.000133 & 59380.370227 &  626.5 & 0.000253  \\
\hline
\end{tabular}}

{\it Notes.} Integer and half-integer cycle numbers refer to primary and secondary eclipses, respectively. All eclipses were observed by the \textit{TESS} spacecraft.
\end{table*}  

\begin{table*}
\caption{Times of minima of TIC 81525800}
 \label{Tab:TIC_081525800_ToM}
\scalebox{0.92}{\begin{tabular}{@{}lrllrllrllrl}
\hline
BJD & Cycle  & std. dev. & BJD & Cycle  & std. dev. & BJD & Cycle  & std. dev. & BJD & Cycle  & std. dev. \\ 
$-2\,400\,000$ & no. &   \multicolumn{1}{c}{$(d)$} & $-2\,400\,000$ & no. &   \multicolumn{1}{c}{$(d)$} & $-2\,400\,000$ & no. &   \multicolumn{1}{c}{$(d)$} & $-2\,400\,000$ & no. &   \multicolumn{1}{c}{$(d)$} \\ 
\hline
59474.728608 &   -0.5 & 0.000097 & 59496.168409 &   12.5 & 0.000064 & 59515.947921 &   24.5 & 0.000084 & 59534.109947 &   35.5 & 0.000079  \\ 
59475.561386 &    0.0 & 0.000147 & 59496.995804 &   13.0 & 0.000068 & 59516.775969 &   25.0 & 0.000077 & 59534.938437 &   36.0 & 0.000062  \\ 
59476.384883 &    0.5 & 0.000132 & 59497.816054 &   13.5 & 0.000060 & 59517.596762 &   25.5 & 0.000075 & 59535.758567 &   36.5 & 0.000080  \\ 
59477.214604 &    1.0 & 0.000112 & 59498.643803 &   14.0 & 0.000152 & 59518.425218 &   26.0 & 0.000089 & 59536.586067 &   37.0 & 0.000053  \\ 
59478.035069 &    1.5 & 0.000130 & 59500.292489 &   15.0 & 0.000100 & 59519.245915 &   26.5 & 0.000058 & 59537.406589 &   37.5 & 0.000055  \\ 
59478.864782 &    2.0 & 0.000072 & 59501.111780 &   15.5 & 0.000078 & 59520.076806 &   27.0 & 0.000070 & 59540.701995 &   39.5 & 0.000099  \\
\hline
\end{tabular}}

{\it Notes.} Integer and half-integer cycle numbers refer to primary and secondary eclipses, respectively. All but one eclipses were observed by the \textit{TESS} spacecraft. The very last time of minimum was determined from ground-based BAO80 observations.
\end{table*}  

\begin{table*}
\caption{Times of minima of TIC 99013269}
 \label{Tab:TIC_099013269_ToM}
\scalebox{0.92}{\begin{tabular}{@{}lrllrllrllrl}
\hline
BJD & Cycle  & std. dev. & BJD & Cycle  & std. dev. & BJD & Cycle  & std. dev. & BJD & Cycle  & std. dev. \\ 
$-2\,400\,000$ & no. &   \multicolumn{1}{c}{$(d)$} & $-2\,400\,000$ & no. &   \multicolumn{1}{c}{$(d)$} & $-2\,400\,000$ & no. &   \multicolumn{1}{c}{$(d)$} & $-2\,400\,000$ & no. &   \multicolumn{1}{c}{$(d)$} \\ 
\hline
58684.513594 &   -5.0 & 0.000208 & 58720.461635 &    0.5 & 0.000136 & 59426.222237 &  108.5 & 0.000082 & 59801.948299 &  166.0 & 0.000064   \\ 
58687.787001 &   -4.5 & 0.000283 & 58723.724083 &    1.0 & 0.000092 & 59429.487117 &  109.0 & 0.000064 & 59805.219247 &  166.5 & 0.000081   \\ 
58691.048878 &   -4.0 & 0.000248 & 58726.996785 &    1.5 & 0.000132 & 59436.023888 &  110.0 & 0.000069 & 59808.483194 &  167.0 & 0.000069   \\ 
58694.322099 &   -3.5 & 0.000277 & 58730.258331 &    2.0 & 0.000109 & 59439.295332 &  110.5 & 0.000073 & 59811.754445 &  167.5 & 0.000078   \\ 
58697.583534 &   -3.0 & 0.000299 & 58733.532437 &    2.5 & 0.000136 & 59442.559063 &  111.0 & 0.000066 & 59815.017709 &  168.0 & 0.000061   \\ 
58704.118271 &   -2.0 & 0.000301 & 58736.793434 &    3.0 & 0.000101 & 59445.831342 &  111.5 & 0.000070 & 59818.289028 &  168.5 & 0.000075   \\ 
58713.926899 &   -0.5 & 0.000126 & 59422.950663 &  108.0 & 0.000064 & 59798.684472 &  165.5 & 0.000076 & 59821.551830 &  169.0 & 0.000079   \\ 
58717.188773 &    0.0 & 0.000107 &  \\ 
\hline
\end{tabular}}

{\it Notes.} Integer and half-integer cycle numbers refer to primary and secondary eclipses, respectively. All eclipses were observed by the \textit{TESS} spacecraft.
\end{table*}  

\begin{table*}
\caption{Times of minima of TIC 229785001}
 \label{Tab:TIC_229785001_ToM}
\scalebox{0.92}{\begin{tabular}{@{}lrllrllrllrl}
\hline
BJD & Cycle  & std. dev. & BJD & Cycle  & std. dev. & BJD & Cycle  & std. dev. & BJD & Cycle  & std. dev. \\ 
$-2\,400\,000$ & no. &   \multicolumn{1}{c}{$(d)$} & $-2\,400\,000$ & no. &   \multicolumn{1}{c}{$(d)$} & $-2\,400\,000$ & no. &   \multicolumn{1}{c}{$(d)$} & $-2\,400\,000$ & no. &   \multicolumn{1}{c}{$(d)$} \\ 
\hline
58683.483148 &    0.0 & 0.000168 & 58854.049357 &  183.5 & 0.000283 & 58999.054886 &  339.5 & 0.000391 & 59627.400615 & 1015.5 & 0.000164  \\ 
58683.947729 &    0.5 & 0.000304 & 58854.514901 &  184.0 & 0.000157 & 58999.519334 &  340.0 & 0.000175 & 59627.865729 & 1016.0 & 0.000077  \\ 
58684.412838 &    1.0 & 0.000164 & 58856.837483 &  186.5 & 0.000391 & 58999.984097 &  340.5 & 0.000366 & 59628.330658 & 1016.5 & 0.000181  \\ 
58684.876945 &    1.5 & 0.000382 & 58857.303168 &  187.0 & 0.000180 & 59000.448813 &  341.0 & 0.000172 & 59628.795175 & 1017.0 & 0.000091  \\ 
58685.342210 &    2.0 & 0.000129 & 58857.767945 &  187.5 & 0.000375 & 59000.913961 &  341.5 & 0.000376 & 59629.259757 & 1017.5 & 0.000162  \\ 
58685.806781 &    2.5 & 0.000301 & 58858.232903 &  188.0 & 0.000131 & 59001.378173 &  342.0 & 0.000139 & 59629.724858 & 1018.0 & 0.000087  \\
\hline
\end{tabular}}
\end{table*}  

\addtocounter{table}{-1}
\begin{table*}
\caption{(\textit{continued})}
\scalebox{0.92}{\begin{tabular}{@{}lrllrllrllrl}
\hline
BJD & Cycle  & std. dev. & BJD & Cycle  & std. dev. & BJD & Cycle  & std. dev. & BJD & Cycle  & std. dev. \\ 
$-2\,400\,000$ & no. &   \multicolumn{1}{c}{$(d)$} & $-2\,400\,000$ & no. &   \multicolumn{1}{c}{$(d)$} & $-2\,400\,000$ & no. &   \multicolumn{1}{c}{$(d)$} & $-2\,400\,000$ & no. &   \multicolumn{1}{c}{$(d)$} \\ 
\hline
58709.508594 &   28.0 & 0.000136 & 58882.398358 &  214.0 & 0.000142 & 59026.010273 &  368.5 & 0.000283 & 59654.357363 & 1044.5 & 0.000163  \\ 
58709.973641 &   28.5 & 0.000311 & 58882.863143 &  214.5 & 0.000347 & 59026.474785 &  369.0 & 0.000149 & 59654.822222 & 1045.0 & 0.000080  \\ 
58711.832077 &   30.5 & 0.000474 & 58883.328108 &  215.0 & 0.000149 & 59026.939800 &  369.5 & 0.000302 & 59655.286938 & 1045.5 & 0.000156  \\ 
58712.297086 &   31.0 & 0.000127 & 58883.792659 &  215.5 & 0.000338 & 59027.404150 &  370.0 & 0.000179 & 59655.751699 & 1046.0 & 0.000096  \\ 
58712.761190 &   31.5 & 0.000285 & 58885.186563 &  217.0 & 0.000160 & 59027.869058 &  370.5 & 0.000384 & 59656.215859 & 1046.5 & 0.000155  \\ 
58713.226529 &   32.0 & 0.000132 & 58885.650874 &  217.5 & 0.000365 & 59028.333843 &  371.0 & 0.000165 & 59656.681111 & 1047.0 & 0.000076  \\
\hline
\end{tabular}}
\end{table*}  

\addtocounter{table}{-1}
\begin{table*}
\caption{(\textit{continued})}
\scalebox{0.92}{\begin{tabular}{@{}lrllrllrllrl}
\hline
BJD & Cycle  & std. dev. & BJD & Cycle  & std. dev. & BJD & Cycle  & std. dev. & BJD & Cycle  & std. dev. \\ 
$-2\,400\,000$ & no. &   \multicolumn{1}{c}{$(d)$} & $-2\,400\,000$ & no. &   \multicolumn{1}{c}{$(d)$} & $-2\,400\,000$ & no. &   \multicolumn{1}{c}{$(d)$} & $-2\,400\,000$ & no. &   \multicolumn{1}{c}{$(d)$} \\ 
\hline
58747.152757 &   68.5 & 0.000309 & 58919.578223 &  254.0 & 0.000177 & 59415.937026 &  788.0 & 0.000069 & 59689.212598 & 1082.0 & 0.000094  \\ 
58747.617051 &   69.0 & 0.000127 & 58920.042914 &  254.5 & 0.000309 & 59416.401202 &  788.5 & 0.000170 & 59689.677076 & 1082.5 & 0.000162  \\ 
58748.082377 &   69.5 & 0.000295 & 58920.507708 &  255.0 & 0.000208 & 59416.866683 &  789.0 & 0.000083 & 59690.142317 & 1083.0 & 0.000080  \\ 
58748.546142 &   70.0 & 0.000164 & 58920.972518 &  255.5 & 0.000334 & 59417.331430 &  789.5 & 0.000157 & 59690.607204 & 1083.5 & 0.000174  \\ 
58749.011152 &   70.5 & 0.000292 & 58921.437214 &  256.0 & 0.000219 & 59417.796065 &  790.0 & 0.000082 & 59691.071731 & 1084.0 & 0.000078  \\ 
58749.476297 &   71.0 & 0.000145 & 58921.901623 &  256.5 & 0.000244 & 59418.260822 &  790.5 & 0.000161 & 59693.395860 & 1086.5 & 0.000153  \\
\hline
\end{tabular}}
\end{table*}  

\addtocounter{table}{-1}
\begin{table*}
\caption{(\textit{continued})}
\scalebox{0.92}{\begin{tabular}{@{}lrllrllrllrl}
\hline
BJD & Cycle  & std. dev. & BJD & Cycle  & std. dev. & BJD & Cycle  & std. dev. & BJD & Cycle  & std. dev. \\ 
$-2\,400\,000$ & no. &   \multicolumn{1}{c}{$(d)$} & $-2\,400\,000$ & no. &   \multicolumn{1}{c}{$(d)$} & $-2\,400\,000$ & no. &   \multicolumn{1}{c}{$(d)$} & $-2\,400\,000$ & no. &   \multicolumn{1}{c}{$(d)$} \\ 
\hline
58810.361726 &  136.5 & 0.000299 & 58955.829846 &  293.0 & 0.000412 & 59584.176809 &  969.0 & 0.000081 & 59725.926160 & 1121.5 & 0.000155  \\ 
58810.826899 &  137.0 & 0.000128 & 58956.294921 &  293.5 & 0.000306 & 59584.641670 &  969.5 & 0.000185 & 59726.390665 & 1122.0 & 0.000088  \\ 
58811.291215 &  137.5 & 0.000303 & 58956.759920 &  294.0 & 0.000163 & 59585.106526 &  970.0 & 0.000094 & 59726.855468 & 1122.5 & 0.000145  \\ 
58811.756001 &  138.0 & 0.000157 & 58957.225011 &  294.5 & 0.000307 & 59585.571290 &  970.5 & 0.000157 & 59727.320292 & 1123.0 & 0.000075  \\ 
58812.220792 &  138.5 & 0.000297 & 58957.689301 &  295.0 & 0.000197 & 59586.035968 &  971.0 & 0.000086 & 59727.784840 & 1123.5 & 0.000154  \\ 
58812.685948 &  139.0 & 0.000120 & 58958.154654 &  295.5 & 0.000298 & 59586.500395 &  971.5 & 0.000185 & 59728.249580 & 1124.0 & 0.000076  \\
\hline
\end{tabular}}
\end{table*}  

\addtocounter{table}{-1}
\begin{table*}
\caption{(\textit{continued})}
\scalebox{0.92}{\begin{tabular}{@{}lrllrllrllrl}
\hline
BJD & Cycle  & std. dev. & BJD & Cycle  & std. dev. & BJD & Cycle  & std. dev. & BJD & Cycle  & std. dev. \\ 
$-2\,400\,000$ & no. &   \multicolumn{1}{c}{$(d)$} & $-2\,400\,000$ & no. &   \multicolumn{1}{c}{$(d)$} & $-2\,400\,000$ & no. &   \multicolumn{1}{c}{$(d)$} & $-2\,400\,000$ & no. &   \multicolumn{1}{c}{$(d)$} \\ 
\hline
58846.148886 &  175.0 & 0.000144 & 58990.224015 &  330.0 & 0.000147 & 59618.570653 & 1006.0 & 0.000078 & 59761.712551 & 1160.0 & 0.000082  \\ 
58846.614054 &  175.5 & 0.000279 & 58990.688828 &  330.5 & 0.000299 & 59619.034741 & 1006.5 & 0.000167 & 59762.176882 & 1160.5 & 0.000147  \\ 
58847.078816 &  176.0 & 0.000154 & 58991.153144 &  331.0 & 0.000134 & 59619.500094 & 1007.0 & 0.000088 & 59762.642524 & 1161.0 & 0.000087  \\ 
58847.543469 &  176.5 & 0.000284 & 58991.618479 &  331.5 & 0.000336 & 59619.964014 & 1007.5 & 0.000178 & 59763.106835 & 1161.5 & 0.000159  \\ 
58848.007966 &  177.0 & 0.000144 & 58992.082865 &  332.0 & 0.000149 & 59620.429869 & 1008.0 & 0.000100 & 59763.571985 & 1162.0 & 0.000082  \\ 
58848.472845 &  177.5 & 0.000276 & 58992.548056 &  332.5 & 0.000298 & 59620.893937 & 1008.5 & 0.000174 & 59764.036271 & 1162.5 & 0.000159  \\
\hline
\end{tabular}}

{\it Notes.} Integer and half-integer cycle numbers refer to primary and secondary eclipses, respectively. All eclipses were observed by the \textit{TESS} spacecraft.
\end{table*}  

\begin{table*}
\caption{Times of minima of TIC~276162169 (V493 Cyg)}
 \label{Tab:V493_Cyg_(TIC_276162169)_ToM}
\scalebox{0.92}{\begin{tabular}{@{}lrllrllrllrl}
\hline
BJD & Cycle  & std. dev. & BJD & Cycle  & std. dev. & BJD & Cycle  & std. dev. & BJD & Cycle  & std. dev. \\ 
$-2\,400\,000$ & no. &   \multicolumn{1}{c}{$(d)$} & $-2\,400\,000$ & no. &   \multicolumn{1}{c}{$(d)$} & $-2\,400\,000$ & no. &   \multicolumn{1}{c}{$(d)$} & $-2\,400\,000$ & no. &   \multicolumn{1}{c}{$(d)$} \\ 
\hline
58684.246260 &   -0.5 & 0.000252 & 58721.222486 &   14.0 & 0.000070 & 59439.097560 &  295.5 & 0.000042 & 59796.122515 &  435.5 & 0.000190 \\ 
58685.521975 &    0.0 & 0.000245 & 58722.496413 &   14.5 & 0.000066 & 59440.373326 &  296.0 & 0.000039 & 59797.399406 &  436.0 & 0.000049 \\ 
58686.796686 &    0.5 & 0.000260 & 58723.772578 &   15.0 & 0.000050 & 59441.648017 &  296.5 & 0.000048 & 59798.673885 &  436.5 & 0.000058 \\ 
58688.072057 &    1.0 & 0.000255 & 58726.322720 &   16.0 & 0.000043 & 59442.923824 &  297.0 & 0.000039 & 59799.949733 &  437.0 & 0.000042 \\ 
58689.346606 &    1.5 & 0.000279 & 58727.596846 &   16.5 & 0.000054 & 59444.198456 &  297.5 & 0.000044 & 59801.224408 &  437.5 & 0.000052 \\ 
58690.622134 &    2.0 & 0.000232 & 58728.873088 &   17.0 & 0.000045 & 59445.474552 &  298.0 & 0.000040 & 59802.500610 &  438.0 & 0.000047 \\
\hline
\end{tabular}}

{\it Notes.} Integer and half-integer cycle numbers refer to primary and secondary eclipses, respectively. Most of the eclipses (cycle nos. $-0.5$ to $446.5$) were observed by the \textit{TESS} spacecraft. The last 3 times of minima were determined from the ground-based GAO80 observations.
\end{table*}  

\begin{table*}
\caption{Times of minima of TIC 280883908}
 \label{Tab:TIC_280883908_ToM}
\scalebox{0.92}{\begin{tabular}{@{}lrllrllrllrl}
\hline
BJD & Cycle  & std. dev. & BJD & Cycle  & std. dev. & BJD & Cycle  & std. dev. & BJD & Cycle  & std. dev. \\ 
$-2\,400\,000$ & no. &   \multicolumn{1}{c}{$(d)$} & $-2\,400\,000$ & no. &   \multicolumn{1}{c}{$(d)$} & $-2\,400\,000$ & no. &   \multicolumn{1}{c}{$(d)$} & $-2\,400\,000$ & no. &   \multicolumn{1}{c}{$(d)$} \\ 
\hline
58984.913027 &    0.0 & 0.000211 & 59005.904767 &    4.0 & 0.000115 & 59029.532528 &    8.5 & 0.000130 & 59737.995390 &  143.5 & 0.000109  \\ 
58987.539927 &    0.5 & 0.000185 & 59008.531699 &    4.5 & 0.000194 & 59032.154316 &    9.0 & 0.000119 & 59740.617096 &  144.0 & 0.000083  \\ 
58990.160151 &    1.0 & 0.000139 & 59013.781797 &    5.5 & 0.000122 & 59034.782130 &    9.5 & 0.000204 & 59745.865507 &  145.0 & 0.000154  \\ 
58992.785349 &    1.5 & 0.000183 & 59016.403673 &    6.0 & 0.000445 & 59719.626354 &  140.0 & 0.000130 & 59753.742490 &  146.5 & 0.000098  \\ 
58995.408270 &    2.0 & 0.000516 & 59019.031229 &    6.5 & 0.000167 & 59724.873615 &  141.0 & 0.000113 & 59758.993159 &  147.5 & 0.000169  \\ 
58998.034108 &    2.5 & 0.000141 & 59021.655780 &    7.0 & 0.000180 & 59727.499237 &  141.5 & 0.000112 & 59761.616627 &  148.0 & 0.000084  \\ 
59000.656182 &    3.0 & 0.000226 & 59024.283264 &    7.5 & 0.000126 & 59730.120787 &  142.0 & 0.002262 & 59764.245467 &  148.5 & 0.000086  \\ 
59003.283139 &    3.5 & 0.000161 & 59026.905888 &    8.0 & 0.000111 & 59732.748670 &  142.5 & 0.000102 & 59766.867410 &  149.0 & 0.000131  \\ 
\hline
\end{tabular}}

{\it Notes.} Integer and half-integer cycle numbers refer to primary and secondary eclipses, respectively. All eclipses were observed by the \textit{TESS} spacecraft.
\end{table*}  

\begin{table*}
\caption{Times of minima of TIC 294803663}
 \label{Tab:TIC_294803663_ToM}
\scalebox{0.92}{\begin{tabular}{@{}lrllrllrllrl}
\hline
BJD & Cycle  & std. dev. & BJD & Cycle  & std. dev. & BJD & Cycle  & std. dev. & BJD & Cycle  & std. dev. \\ 
$-2\,400\,000$ & no. &   \multicolumn{1}{c}{$(d)$} & $-2\,400\,000$ & no. &   \multicolumn{1}{c}{$(d)$} & $-2\,400\,000$ & no. &   \multicolumn{1}{c}{$(d)$} & $-2\,400\,000$ & no. &   \multicolumn{1}{c}{$(d)$} \\ 
\hline
58597.414393 &   -0.5 & 0.001380 & 58625.464654 &   12.0 & 0.000234 & 59335.199971 &  328.0 & 0.000101 & 59363.308963 &  340.5 & 0.000133  \\ 
58598.511084 &    0.0 & 0.000176 & 58626.612731 &   12.5 & 0.000692 & 59336.353685 &  328.5 & 0.000152 & 59364.399049 &  341.0 & 0.000118  \\ 
58599.661625 &    0.5 & 0.000177 & 58631.108275 &   14.5 & 0.000558 & 59337.443872 &  329.0 & 0.000123 & 59365.554127 &  341.5 & 0.000121  \\ 
58600.759152 &    1.0 & 0.000139 & 58632.204460 &   15.0 & 0.000151 & 59338.600152 &  329.5 & 0.000172 & 59366.646276 &  342.0 & 0.000107  \\ 
58601.907966 &    1.5 & 0.000195 & 58633.353822 &   15.5 & 0.000158 & 59339.690501 &  330.0 & 0.000125 & 59367.800004 &  342.5 & 0.000142  \\ 
58603.004370 &    2.0 & 0.000139 & 58634.448397 &   16.0 & 0.000126 & 59340.844364 &  330.5 & 0.000139 & 59368.891494 &  343.0 & 0.000099  \\
\hline
\end{tabular}}

{\it Notes.} Integer and half-integer cycle numbers refer to primary and secondary eclipses, respectively. All eclipses were observed by the \textit{TESS} spacecraft.
\end{table*}  

\begin{table*}
\caption{Times of minima of TIC 332521671}
 \label{Tab:TIC_332521671_ToM}
\scalebox{0.92}{\begin{tabular}{@{}lrllrllrllrl}
\hline
BJD & Cycle  & std. dev. & BJD & Cycle  & std. dev. & BJD & Cycle  & std. dev. & BJD & Cycle  & std. dev. \\ 
$-2\,400\,000$ & no. &   \multicolumn{1}{c}{$(d)$} & $-2\,400\,000$ & no. &   \multicolumn{1}{c}{$(d)$} & $-2\,400\,000$ & no. &   \multicolumn{1}{c}{$(d)$} & $-2\,400\,000$ & no. &   \multicolumn{1}{c}{$(d)$} \\ 
\hline
58571.351631 &    0.0 & 0.000622 & 58585.069584 &   11.0 & 0.000222 & 59307.792721 &  590.5 & 0.000260 & 59322.135649 &  602.0 & 0.000194  \\ 
58571.975838 &    0.5 & 0.000425 & 58585.693745 &   11.5 & 0.000435 & 59308.415519 &  591.0 & 0.000166 & 59322.759532 &  602.5 & 0.000309  \\ 
58572.598164 &    1.0 & 0.001552 & 58586.316751 &   12.0 & 0.000219 & 59309.038454 &  591.5 & 0.000211 & 59323.382946 &  603.0 & 0.000199  \\ 
58573.222659 &    1.5 & 0.000853 & 58586.941261 &   12.5 & 0.000305 & 59309.662771 &  592.0 & 0.000161 & 59324.006946 &  603.5 & 0.000272  \\ 
58573.845601 &    2.0 & 0.000566 & 58587.564477 &   13.0 & 0.000462 & 59312.157410 &  594.0 & 0.000156 & 59324.630096 &  604.0 & 0.000198  \\ 
58574.468385 &    2.5 & 0.000672 & 58588.188971 &   13.5 & 0.000476 & 59312.781035 &  594.5 & 0.000154 & 59325.253236 &  604.5 & 0.000263  \\
\hline
\end{tabular}}

{\it Notes.} Integer and half-integer cycle numbers refer to primary and secondary eclipses, respectively. All eclipses were observed by the \textit{TESS} spacecraft.
\end{table*}  

\begin{table*}
\caption{Times of minima of TIC 356324779}
 \label{Tab:TIC_356324779_ToM}
\scalebox{0.92}{\begin{tabular}{@{}lrllrllrllrl}
\hline
BJD & Cycle  & std. dev. & BJD & Cycle  & std. dev. & BJD & Cycle  & std. dev. & BJD & Cycle  & std. dev. \\ 
$-2\,400\,000$ & no. &   \multicolumn{1}{c}{$(d)$} & $-2\,400\,000$ & no. &   \multicolumn{1}{c}{$(d)$} & $-2\,400\,000$ & no. &   \multicolumn{1}{c}{$(d)$} & $-2\,400\,000$ & no. &   \multicolumn{1}{c}{$(d)$} \\ 
\hline
58817.450745 &   -0.5 & 0.000204 & 58826.139648 &    2.0 & 0.000254 & 58834.832035 &    4.5 & 0.000417 & 59542.378549 &  208.0 & 0.000066  \\ 
58819.188471 &    0.0 & 0.000287 & 58827.878404 &    2.5 & 0.000463 & 58836.570878 &    5.0 & 0.000577 & 59587.572095 &  221.0 & 0.000059  \\ 
58820.925547 &    0.5 & 0.000243 & 58829.615688 &    3.0 & 0.000493 & 58838.311296 &    5.5 & 0.000534 & 59596.260941 &  223.5 & 0.000063  \\ 
58822.664508 &    1.0 & 0.000233 & 58831.354344 &    3.5 & 0.000496 & 58840.050670 &    6.0 & 0.000582 & 59615.378465 &  229.0 & 0.000084  \\ 
58824.402098 &    1.5 & 0.000229 & 58833.093073 &    4.0 & 0.000516 & 59535.414985 &  206.0 & 0.000074 & 59662.328637 &  242.5 & 0.000111  \\ 
\hline
\end{tabular}}

{\it Notes.} Integer and half-integer cycle numbers refer to primary and secondary eclipses, respectively. Most of the eclipses (cycle nos. $-0.5$ to $6.0$) were observed by the \textit{TESS} spacecraft. The last 6 times of minima were determined from the ground-based BAO80 (nos. $206$, $242.5$) and GAO80 (nos. $208$ to $229$) observations.
\end{table*}  

\end{document}